\documentclass[floatfix,amsmath,amssymb,aps, prfluids]{revtex4-2}
\usepackage{amsmath}
\usepackage{bm}
\usepackage{appendix}

\usepackage{graphicx}
\usepackage{epstopdf}
\usepackage{float}
\usepackage{subcaption}
\usepackage{cleveref}
\usepackage{fullpage}
\usepackage{comment}
\usepackage{amsfonts}

\newcommand{\Wi}{\mbox{Wi}}
\newcommand{\De}{\mbox{De}}
\newcommand{\Reynolds}{\mbox{Re}}

\begin{document}


\title{On the unsteady and lineal   translation of a sphere through a viscoelastic fluid}

\author{Mary A. Joens}
\author{James W. Swan}
 \email{jswan@mit.edu}
\affiliation{  Department of Chemical Engineering Massachusetts Institute of Technology, 
  \\ Cambridge, MA 02139, USA
}

\date{\today}

\begin{abstract}
The unsteady, lineal translation of a solid spherical particle through viscoelastic fluids described by the Johnson-Segalman and Giesekus models is studied analytically. Solutions for the pressure and velocity fields as well as the force on the particle are expanded as a power series in the Weissenberg number.  The momentum balance and constitutive equation are solved asymptotically for a steadily translating particle up to second order in the particle velocity, and rescaling of the pressure and velocity in the frequency domain is used to relate the solutions for steady lineal translation to those for unsteady lineal translation. The unsteady force at third order in the particle velocity is then calculated through application of the Lorentz reciprocal theorem, and it is shown that this weakly nonlinear contribution to the force can be expressed as part of a Volterra series. Through a series of examples, it is shown that a truncated representation of this Volterra series, which can be manipulated to describe the velocity in terms of an imposed force, is useful for analyzing specific time-dependent particle motions. Two examples studied using this relationship are the force on a particle suddenly set into motion and the velocity of a particle in response to a suddenly imposed steady force.  Additionally, the weakly nonlinear response of particle captured by a harmonic trap moving lineally through the fluid is computed.  This is an analog to active microrheology experiments, and can be used to explain how weakly nonlinear responses manifest in active microrheology experiments with spherical probes.
\end{abstract}

\maketitle

\section{Introduction} \label{intro}

Flows of viscoelastic fluids around immersed objects have long been a subject of interest in both experimental and theoretical studies. Understanding such flows around spheres, particularly, has held appeal for its broad range of potential applications, from use as a benchmarking problem for validation of complicated numerical methods and models \cite{Housiadas2016}, to better understanding sedimentation of slurries \cite{McKinley2001} and suspensions with non-Newtonian matrix fluids \cite{Housiadas2011a}.  A variety of analytical solutions have been determined for the steady flows around solid spheres in viscoelastic fluids. In one early example, Giesekus used a retarded motion expansion to solve the problem of steady rotation and translation in a viscoelastic fluid \cite{Giesekus1963}. Since, solutions have been determined for steady flows of various description using different constitutive models via perturbation expansions \cite{Housiadas2016, Housiadas2011a, Housiadas2011,  Housiadas2012, Gkormpatsis2020, Becker1996}.  And though some studies have used numerical simulations to understand unsteady flows of viscoelastic fluids around spheres \cite{Bodart1994, Becker1994, Chilcott1988}, comparatively few have attempted an analytical description of such flows. In work by Moore and Shelley, analytical solutions were determined for some specific flows in the so-called ``weak-coupling" limit, where viscoelastic stresses are treated as an asymptotically small contribution to the overall stress in the fluid \cite{Moore2012}, providing an analytical description of the forces exerted on a sphere moving in response to a suddenly imposed force and subject to oscillatory forcing. However, we still lack a general theoretical understanding of these types of unsteady particle motions and flows outside of the weak-coupling limit and without restriction to simple temporal protocols.  

Having such a general theoretical understanding would be useful for a variety of reasons. Perhaps most obviously, it would serve to unite the theoretical and experimental work that has already been done. Experimental studies have identified flow phenomena like the sustained oscillation in the settling velocity of a falling particle \cite{Mollinger1999, Zhang2018} whose underlying causes are still not agreed upon. While numerical studies have captured similar behavior \cite{Becker1994}, a thorough theoretical exploration of this behavior or identification of criteria for its onset has not been achieved. Additionally, one particularly under-explored potential application for a more general understanding of dynamic flows of viscoelastic fluids around spherical probe particles is in the field of nonlinear microrheology. 

Microrheology is a set of experimental techniques used to measure the rheological properties of non-Newtonian fluids. In microrheology experiments, microscale probes, typically inert spherical particles \cite{Wilson2011} -- though some measurements have been carried out using other particle shapes such as nanorods \cite{Dhar2010a} -- are embedded in a sample of the viscoelastic fluid of interest. Unlike traditional macroscale rheology techniques, which require sample volumes on the order of milliliters, microrheology can be used for samples with volumes as small as a few nanoliters \cite{Gardel2005}, or for unconventional samples like the interior of living cells \cite{Guo2013}. This is a major advantage over traditional rheological techniques, particularly if the measurements are performed on precious or difficult to produce materials for which a small sample volume may drastically reduce costs per experiment, or on fluids like cytoplasm encased in a fragile structure like the cell that cannot be transferred to a regular rheometer.

There are two main categories of microrheology techniques: passive and active. In \textit{passive microrheology}, the thermal fluctuations of a Brownian probe are the driving force for its movement, and the generalized Stokes-Einstein relation can be used to infer the linear rheological properties of the fluid from the mean-squared displacement of the probe particle \cite{Squires2010}. This particle tracking technique is limited to determining the linear response of materials.  Conversely, \textit{active microrheology} is not subject to this limitation. In active microrheology, the probe particle is impelled by some known force perhaps through the use of optical \cite{Guo2013, Robertson-Anderson2018, Meyer2006, Brau2007, Yao2009, Gupta2017} or magnetic \cite{Amblard1996a, Bausch1998a, Rich2011, Schmidt1996} tweezers, and the relationship between the applied force and the probe displacement is used to infer a micro-rheological property of the fluid.

Currently, one of the major disadvantages of active microrheology is the lack of a general understanding of how to interpret these micro-rheological properties when the viscoelastic material is subject to strong deformations \cite{Squires2010, Robertson-Anderson2018}. For small deformations, a known linear relationship between the applied force and the probe displacement can be used determine the macroscopic linear viscoleastic properties of the fluid \cite{Guo2013}. Typically, the reported property  is the complex modulus, $G^*(\omega)$, as this is most familiar from macroscale rheology. 

While nonlinear microrheology measurements have been made \cite{Meyer2006, Rich2011, Armstrong2020, Cappallo2007, Sriram2009}, the lack of a generalized relationship between force and displacement has made it difficult to compare measurements across materials or experiments. Additionally, theoretical work developing a fundamental understanding of nonlinear microrheology has been mainly limited to analysis or simulations of rheological probes in colloidal suspensions \cite{Khair2005, DePuit2011, Gazuz2013, Squires2005} or a few other model materials \cite{Winter2012}. The range of real materials to which such theories are applicable is not broad. 

In recent work by Lennon, et. al. \cite{Lennon2020, Lennon2020a}, a similar problem in the general representation of nonlinear responses in traditional macroscale rheology was addressed via a Volterra series expansion of the shear stress developed in response to arbitrary time-dependent, viscometric deformations. For weakly nonlinear flows, the expansions derived in that work reveal a transfer function called the third-order complex modulus, that is a material property characterizing nonlinear mechanical responses. In this work, we develop an equivalent framework for interpretation of microrheological measurements in the weakly nonlinear regime by formulating another Volterra series expansion for the time-dependent force on a spherical probe undergoing lineal translation in a viscoelastic fluid.

Here, we solve for the force exerted on a sphere undergoing lineal translation through a viscoelastic fluid.  We employ two models for polymeric fluids: the Johnson-Segalman and Giesekus models. The force is computed using a perturbation expansion of the governing equations of these flows in the limit of small deformation amplitudes (small Weissenberg numbers), as detailed in Sections \ref{definiton} and \ref{methods}. We solve for the pressure and velocity fields in the fluid around the sphere to second order in deformation amplitude and then show that the third-order contribution to the force on the particle can be computed easily using the Lorentz reciprocal theorem.  When the force on the particle is expressed as a Volterra series expansion, a transfer function appearing at third order in deformation amplitude, which we call the third-order resistivity, can be evaluated explicitly and characterizes the nonlinear viscoelasticity of the fluid much like the third-order complex modulus in viscometric flows. Section \ref{generalize} details the implications of this Volterra series representation of the force and describes some visualization strategies that can be used to understand how constitutive model parameters affect the third-order resistivity. Finally, in Section \ref{examples} this Volterra series is applied to some specific  particle motions and flows in the limit of weak deformations: start-up of lineal motion, and active microrheology controlled by the motion of a harmonic trap.

\section{Problem definition}\label{definiton}

\begin{figure}[H]
\centering
\includegraphics[width=0.5\linewidth]{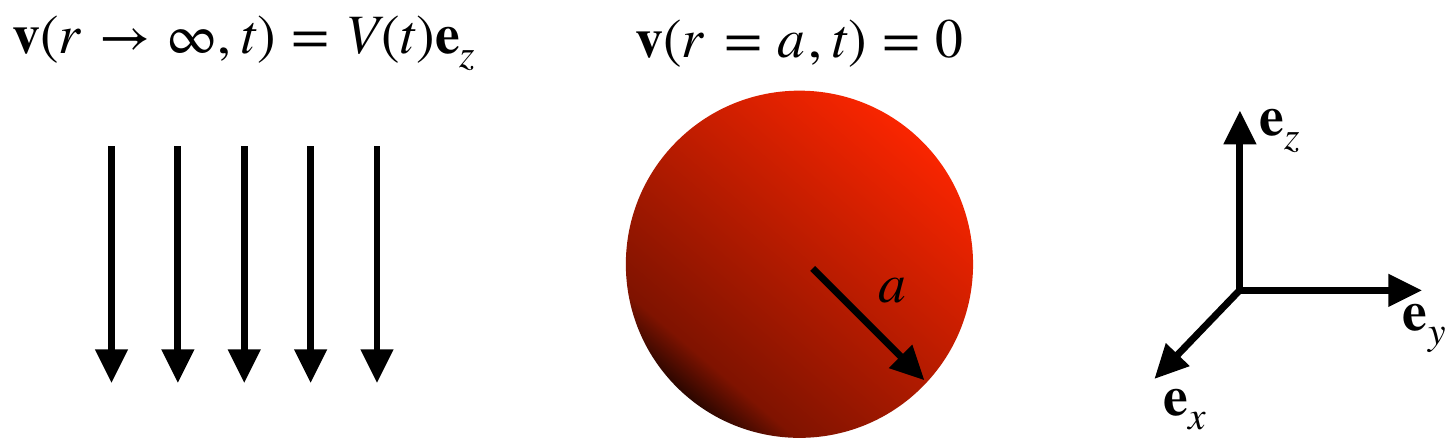}
\caption{Schematic depicting the flow geometry and boundary conditions for the velocity field.}
\label{fig:system}
\end{figure}

A spherical particle with radius $a$ executes an arbitrary unsteady, lineal motion through a viscoelastic fluid. A schematic representation of this scenario is found in Figure \ref{fig:system}. The motion of the fluid and particle is assumed to be inertialess ($\Reynolds \ll 1$) and isothermal. The flow will be described in a Cartesian coordinate system, with unit vectors $\mathbf{e}_{x}, \mathbf{e}_{y}, \mathbf{e}_{z}$.  The boundary conditions are specfied from the frame of reference moving with the particle.  Far from the sphere, the flow field is assumed to have a known, uniform, time-dependent profile $V(t)\mathbf{e}_{z}$. The viscoelastic fluid is characterized by a relaxation time $\lambda$ and a total zero-shear viscosity $\eta_{0} = \eta_{s} + \eta_{p}$, where $\eta_{s}$ is a Newtonian solvent viscosity and $\eta_{p}$ is a polymeric zero-shear viscosity. 

The governing and constitutive equations are made dimensionless by rescaling the variables carefully. The spatial position, $\mathbf{r} $, which is centered on the spherical particle is scaled by the particle radius $a$.  We denote the distance from the particle center, $ r = | \mathbf{r} | $.  Time, $ t $, is scaled by a characteristic time $t_{c}$, which reflects the timescale on which the prescribed flow rate, $ V(t) $, is changing. $ t_c $ can be mathematically defined in a variety of ways depending on the temporal protocol. For most cases, an obvious $t_c$ will arise.  For example, in a flow field with a time periodic velocity, a natural definition for $t_{c}$ is the period.  In a steady flow field, $t_{c}$ would be the duration of observation and could tend to infinity.  For a more complicated or even stochastically varying flow field, $t_c$ could be defined using features of the time autocorrelation function of $ V( t ) $.  The fluid velocity, $ \mathbf{v}( \mathbf{r}, t ) $, is scaled on $v_{c}$, which is taken to be the maximum value of $|V(t)|$.  Because we will solve the for the fluid motion asymptotically, the velocity must be bounded in magnitude.  The pressure, $p( \mathbf{r}, t )$, is scaled on $\eta_{0}v_{c}/a$.  The Newtonian solvent stress, $\bm{\tau}_{s}( \mathbf{r}, t ) $, is scaled on $\eta_{s}v_{c}/a$.  The polymeric or viscoelastic stress, $\bm{\tau}_{p}( \mathbf{r}, t )$, is scaled on $\eta_{p}v_{c}/a$.  The force exerted on the particle, $ \mathbf{F}(t) $, is scaled on $ \eta_0 v_c a $. 

When these scaling relationships are applied, the conservation of mass and momentum in the fluid can be written as: 
\begin{subequations}
   \begin{equation}
     \nabla \cdot \mathbf{v}( \mathbf{r}, t ) = 0, 
     \label{eq:continuity}
   \end{equation}
   \begin{equation}
     -\nabla p( \mathbf{r}, t ) + \beta \nabla \cdot \bm{\tau}_{s}( \mathbf{r}, t ) + (1-\beta) \nabla \cdot \bm{\tau}_{p}( \mathbf{r}, t ) = 0,
     \label{eq:momentumbal}
    \end{equation}
\end{subequations}
where the Newtonian, $ \bm{\tau}_s( \mathbf{r}, t ) $, and polymeric, $ \bm{\tau}_p( \mathbf{r}, t ) $, stresses are simply superimposed to give the total stress in the fluid.  The quantity, $\beta = \eta_s / \eta_0$, denotes the fractional contribution of the solvent to the zero shear viscosity.  The quantity $1-\beta = \eta_p / \eta_0$ is the similarly scaled polymeric viscosity. When the constitutive model used to describe the polymeric stress is the Johnson-Segalman model with a Newtonian solvent, its representation in dimensionless form is: 
\begin{subequations}
   \begin{equation}
     \bm{\tau}_{s}( \mathbf{r}, t ) = 2\mathbf{e}( \mathbf{r}, t ),
     \label{eq:solventstress}
    \end{equation}
    \begin{equation}
    \bm{\tau}_{p}( \mathbf{r}, t ) +   \De \frac{\partial}{\partial t}\bm{\tau}_p( \mathbf{r}, t ) + \Wi \left[\mathbf{v}( \mathbf{r}, t ) \cdot \nabla\bm{\tau}_{p}( \mathbf{r}, t ) - \frac{1}{2} (\mathbf{A}( \mathbf{r}, t ) \cdot \bm{\tau}_{p}( \mathbf{r}, t ) + \bm{\tau}_{p}( \mathbf{r}, t ) \cdot \mathbf{A}( \mathbf{r}, t )^T)\right] = 2\mathbf{e}( \mathbf{r}, t )
     \label{eq:polymerstress}
    \end{equation}
\end{subequations}
where $\mathbf{e}( \mathbf{r}, t )$ is the rate-of-strain tensor: 
\begin{equation}
    \mathbf{e}( \mathbf{r}, t ) = \frac{1}{2} \left( \nabla \mathbf{v}( \mathbf{r}, t ) + \nabla \mathbf{v}( \mathbf{r}, t )^{T} \right),
    \label{eq:rosdef}
\end{equation}
and
\begin{equation}
        \mathbf{A}( \mathbf{r}, t ) = (b-1) \nabla \mathbf{v}( \mathbf{r}, t ) + (1+b) \nabla \mathbf{v}( \mathbf{r}, t )^T .
\end{equation}
In this model, \textit{b} is an adjustable \textit{slip parameter}, which can take on values between -1 and 1 and is a measure of the contribution of non-affine motion to the stress tensor. When $b = 1$, the Oldroyd-B model is recovered.  When $b=1$ and $\beta = 0$, the upper-convected Maxwell model is recovered.  With $ \beta = 0 $, when $ b = 0 $ and $ b = -1 $, the corotational and lower-convected Maxwell models are recovered, respectively. 

The two additional dimensionless groups that emerge from these scaling arguments are: $ \De = \lambda/t_{c} $ and $ \Wi = \lambda v_{c}/a $.  The Deborah number, $\De$, is a measure of how quickly the lineal velocity is changing relative to the relaxation time of the fluid.  The Weissenberg number, $\Wi$, can be interpreted as a ratio of the rate of lineal motion, $ v_c / a $, to the stress relaxation rate in the fluid, $ \lambda^{-1} $, or as a ratio of elastic, $ \lambda \eta_0 v_c ^2  $, to viscous, $ \eta_0 v_c a $, forces.  Finally, the boundary conditions applied to the problem are a no-slip and no-penetration condition at the surface of the sphere, $ \mathbf{v}( r = 1, t ) = 0 $, the far-field pressure approaching zero, $ p( r \rightarrow \infty, t ) \rightarrow 0 $, and the imposition of a time varying, uniform flow profile in the far-field, $ \mathbf{v}( r \rightarrow \infty, t ) \rightarrow V(t)\mathbf{e}_{z}$.

Up to this point, we have described the viscoelastic fluid as if it has a single relaxation time, $\lambda$. However, in many real viscoelastic fluids, there is a distribution of relaxation times, characterized by the probability density function: $ P( \lambda )$. To more accurately model such a system, we introduce an averaged polymeric stress: 
\begin{equation}
    \left< \bm{\tau}_{p}( \mathbf{r}, t, \lambda ) \right>_\lambda = \int_{0}^{\infty} \bm{\tau}_{p}(\mathbf{r}, t, \lambda) P(\lambda) \, d\lambda
    \label{eq:avgstress},
\end{equation}
where $ \bm{\tau}_p( \mathbf{r}, t, \lambda ) $ is the stress in a polymer mode with relaxation time $ \lambda $.  The subscript on the angle brackets indicates that the average is taken with respect to the variable $ \lambda $.  In a fluid with a distribution of relaxation times, $ \eta_p $ is now defined as the difference between the zero shear viscosity of the fluid and the Newtonian solvent viscosity: $ \eta_p = \eta_0 - \eta_s $.  In other literature, the quantity: $ \eta_p P( \lambda ) $ is sometimes used as a descriptor of the `stress relaxation spectrum' \cite{Tschoegl1989}.

Numerous methods have been developed and used in the literature for estimation of both discrete \cite{Smail1996} and continuous \cite{Malkin2006,Shanbhag2019} relaxation time distributions for viscoelastic materials. Most of these methods use information about the linear responses of these materials to deformation in order to infer a relaxation time spectrum using either theoretical models of these spectra or numerical optimization routines. Many of these calculated distributions could be applied directly to the present calculations to estimate the effect of having a broad relaxation time distribution. This can be done analytically for discrete spectra, and in most cases could be computed numerically for continuous spectra, so we aim to generalize the range of modeled responses and leave open that possibility.

The introduction of a range of relaxation times also calls for a redefinition of the Deborah and Weissenberg numbers. Rather than being defined using the singular relaxation time $\lambda$, the dimensionless groups will be defined on a characteristic relaxation time $\Bar{\lambda}$.  This value should be chosen carefully.  Good choices include either the longest relaxation time in a discrete distribution or the mean relaxation time: $ \left< \lambda \right>_\lambda $, in the fluid, which will ensure that dimensionless groups defined upon this quantity are representative for all polymer stress modes. Regardless of how $ \Bar \lambda $ is specified, for a fluid with a distribution of relaxation times, we can redefine the Weissenberg and Deborah numbers as: 
\begin{equation}
    \De = \frac{\Bar{\lambda}}{t_{c}}, \qquad \Wi = \frac{\Bar{\lambda} v_{c}}{a}.
\end{equation}
With these definitions, the scaled expression for a mode of the polymer stress in Equation \ref{eq:polymerstress} having relaxation time, $ \lambda $, is: 
\begin{align}
\bm{\tau}_{p}( \mathbf{r}, t, \lambda ) &+ \De \left( \frac{\lambda}{\Bar{\lambda}} \right) \frac{\partial }{\partial t} \bm{\tau}_p( \mathbf{r}, t, \lambda ) \nonumber \\
&+ \Wi \left( \frac{\lambda}{\Bar{\lambda}}\right) \left[\mathbf{v}( \mathbf{r}, t ) \cdot \nabla\bm{\tau}_{p}( \mathbf{r}, t, \lambda ) - \frac{1}{2} (\mathbf{A}( \mathbf{r}, t ) \cdot \bm{\tau}_{p}( \mathbf{r}, t, \lambda ) + \bm{\tau}_{p}( \mathbf{r}, t, \lambda ) \cdot \mathbf{A}( \mathbf{r}, t )^T)\right] = 2\mathbf{e}( \mathbf{r}, t ) .
\label{eq:avg_polstress}
\end{align}
As is often done in calculations of this sort, we are assuming that the stresses from these different polymer modes can be linearly superimposed and that the modes only interact indirectly through local gradients in the velocity field \cite{Larson1987, Hulsen1991, Oztekin1994}.  The inertialess momentum balance in a fluid of this sort is then:
\begin{equation}
    -\nabla p( \mathbf{r}, t ) + \beta \nabla \cdot \bm{\tau}_{s}( \mathbf{r}, t ) + (1-\beta) \nabla \cdot \left< \bm{\tau}_{p}( \mathbf{r}, t, \lambda ) \right>_\lambda = 0,
    \label{eq:momentumbalavg}
\end{equation}
where we have assumed that the distribution of relaxation times is spatially and temporally invariant.  Finally, the dimensionless force exerted on the particle is simply:
\begin{equation}
    \mathbf{F}( t ) = \int_{\partial \Omega} \mathbf{n} \cdot \left( -p( \mathbf{r}, t ) \mathbf{I} + \beta \bm{\tau}_{s}( \mathbf{r}, t ) + (1-\beta) \left< \bm{\tau}_p( \mathbf{r}, t, \lambda ) \right>_\lambda \right) \, d \mathbf{r},
    \label{eq:forcedef}
\end{equation}
where $ \partial \Omega $ is the surface with of the spherical particle with outward pointing unit normal, $ \mathbf{n} $.

\section{Method of solution} \label{methods}

We aim to calculate the force exerted on a spherical particle immersed in a viscoelastic fluid with a time-dependent, uniform flow in the far-field.  Equivalently, this is the force exerted on a spherical particle executing a time-dependent lineal motion through a viscoelastic fluid.  In this section, we briefly describe the nature of the solution we will produce and the solution method.  Any reader can understand the basis of the solution and method from this brief description and move onto the results presented in Section \ref{generalize}.  A detailed derivation is provided in the following sections for the more interested reader.

For weak imposed flows, we expect the time-dependent force, $\mathbf{F}(t)$, on the particle to align with the flow direction.  Reversing the direction of the imposed flow should reverse the force. Therefore, we also expect that the force, when represented in dimensional form, has a regular expansion in odd powers of the characteristic velocity, $ v_c $.   In dimensionless form, the leading order nonlinear contributions to the velocity and pressure fields with respect to a small deformation amplitude will be $O( \Wi )$.  Similarly, the leading order nonlinear contribution to the dimensionless force will be $ O(\Wi^2) $.  We compute all of these leading order perturbations to the well-known linear response solution, which is briefly re-derived in Section \ref{firstorder}.  The nonlinear contributions to the velocity and pressure fields are derived directly through an asymptotic expansion at small $ \Wi $ in Section \ref{secondorder}.  The nonlinear contribution to the force is derived from application of the Lorentz reciprocal theorem in Section \ref{thirdorder}.  

Because the far-field boundary conditions are time-dependent, the governing equations, their asymptotic expansions at small $ \Wi $, solutions for the velocity and pressure fields, and the force are represented more conveniently in the frequency domain through use of Fourier transformations. In Sections \ref{firstorder} and \ref{secondorder}, a clever rescaling of variables allows the linear response and leading nonlinearities in the velocity and pressure fields in frequency space to be expressed in terms of the velocity and pressure fields that would be calculated when the imposed far-field flow is steady and of unit magnitude.  Therefore, we compute steady-state solutions explicitly, and through these scaling relationships, we determine the velocity and pressure fields in response to an unsteady, far-field flow.

It should be noted for conceptual clarity that the application of an asymptotic expansion implies the solution is only valid for vanishingly small Weissenberg numbers, so that the amplitude of deformation is very weak.  However, the Deborah number can be specified arbitrarily, so that the imposed flow can change with arbitrary rapidity.  In principle, it can be nice to close such an asymptotic analysis by estimating the truncation error associated with terms neglected in the asymptotic expansion or even computing the radius of convergence of the expansion if one exists.  We do not consider that problem here, which is quite difficult in general.  However, results for the radius of convergence of such an asymptotic expansion with respect to $ \Wi $ for the steady lineal motion of a sphere in an Oldroyd-B fluid exist \cite{Housiadas2016}.

\subsection{Asymptotic expansion of variables and governing equations}\label{expansion}

For a small Wi, we suppose that all dimensionless variables can be written as a power series in the Weissenberg number as follows: 
\begin{subequations}
\begin{align}
    \mathbf{v}( \mathbf{r}, t ) &= \mathbf{v}^{(1)}( \mathbf{r}, t ) + \Wi \mathbf{v}^{(2)}( \mathbf{r}, t ) + \Wi^2 \mathbf{v}^{(3)}( \mathbf{r}, t ) + \ldots,
    \\
    p( \mathbf{r}, t )& = p^{(1)}( \mathbf{r}, t ) + \Wi p^{(2)}( \mathbf{r}, t ) + \Wi^2 p^{(3)}( \mathbf{r}, t ) + \ldots,
\\
    \boldsymbol{\tau}_s( \mathbf{r}, t ) &= \boldsymbol{\tau}_{s}^{(1)}( \mathbf{r}, t ) + \Wi \boldsymbol{\tau}_{s}^{(2)}( \mathbf{r}, t ) + \Wi^2 \boldsymbol{\tau}_{s}^{(3)}( \mathbf{r}, t ) + \ldots,
\\
    \boldsymbol{\tau}_p( \mathbf{r}, t, \lambda ) &= \boldsymbol{\tau}_{p}^{(1)}( \mathbf{r}, t, \lambda ) + \Wi \boldsymbol{\tau}_{p}^{(2)}( \mathbf{r}, t, \lambda ) + \Wi^2 \boldsymbol{\tau}_{p}^{(3)}( \mathbf{r}, t, \lambda ) + \ldots,
\\
    \mathbf{F}( t )& = \mathbf{F}^{(1)}( t ) + \Wi^2 \mathbf{F}^{(3)}( t ) + \ldots.
\end{align}
\end{subequations}
In dimensional terms, $ \mathbf{v}^{(1)}( \mathbf{r}, t ) $ contributes a term to the velocity that is linear in the deformation amplitude, $ v_c $.  Similarly, $ \Wi^2 \mathbf{v}^{(3)}( \mathbf{r}, t ) $ contributes a term that is cubic in the deformation amplitude.  With these expansions for the variables, Equations \ref{eq:continuity} and \ref{eq:momentumbalavg} can be rewritten by matching terms of $O(\Wi^{n-1})$ for $ n \le 3 $: 
\begin{subequations}
\begin{equation}
    \nabla \cdot \mathbf{v}^{(n)}( \mathbf{r}, t ) = 0,
    \label{eq:expandedcontinuity}
\end{equation}
\begin{equation}
     - \nabla p^{(n)} ( \mathbf{r}, t )+ \beta \nabla \cdot  \bm{\tau}_{s}^{(n)}( \mathbf{r}, t )  + (1-\beta) \nabla \cdot \left< \bm{\tau}_{p}^{(n)}( \mathbf{r}, t, \lambda ) \right>_\lambda  = 0, 
     \label{eq:expandedmombal}
\end{equation}
\label{eq:expandgov}
\end{subequations}

while the constitutive model in Equations \ref{eq:solventstress} and \ref{eq:polymerstress} matched at the same order is: 
 \begin{subequations}
\begin{equation}
    \bm{\tau}_{s}^{(n)}( \mathbf{r}, t ) = 2 \bm{e}^{(n)}( \mathbf{r}, t ),
    \label{eq:expandsolvstress}
\end{equation}
\begin{align}
        &\bm{\tau} _{p}^{(n)}( \mathbf{r}, t, \lambda ) + \De \left(\frac{\lambda}{\Bar{\lambda}}\right) \frac{\partial}{\partial t} \bm{\tau}_p^{(n)}( \mathbf{r}, t, \lambda ) =  2\mathbf{e}^{(n)}( \mathbf{r}, t )  \label{eq:expandpolstress}  \\
        & \qquad- \frac{\lambda}{\Bar{\lambda}} \sum_{m=1}^{n-1}  \left[\mathbf{v}^{(n-m)}( \mathbf{r}, t ) \cdot \nabla \bm{\tau}_{p}^{(m)}( \mathbf{r}, t , \lambda) - \frac{1}{2} (\mathbf{A}^{(n-m)}( \mathbf{r}, t ) \cdot \bm{\tau}_{p}^{(m)}( \mathbf{r}, t, \lambda )
 + \bm{\tau}_p^{(m)}( \mathbf{r}, t, \lambda ) \cdot {\mathbf{A}^{(n-m)}}( \mathbf{r}, t )^T)\right].
\nonumber
\end{align}
\label{eq:expandconstitutive}
\end{subequations}

Here, $ \mathbf{e}^{(n)}( \mathbf{r}, t ) $ and $ \mathbf{A}^{(n)}( \mathbf{r}, t ) $ have analogous definitions to $ \mathbf{e}( \mathbf{r}, t ) $ and $ \mathbf{A}( \mathbf{r}, t ) $ but using the $ O( \Wi^{n-1} ) $ term from the expansion of the velocity.  For the dimensionless force on the particle:
\begin{equation}
    \mathbf{F}^{(n)}(t) = \int_{\partial \Omega} \mathbf{n} \cdot \left[ -p^{(n)}( \mathbf{r}, t ) \mathbf{I} + \beta \bm{\tau}_s^{(n)}( \mathbf{r}, t ) + (1-\beta) \left< \bm{\tau}_p^{(n)}( \mathbf{r}, t, \lambda ) \right>_\lambda \right] \, d \mathbf{r}. \label{eq:forcedefn}
\end{equation}

There are two specific representations of the variables and governing equations with which we will work simultaneously in what follows: the steady-state case and the frequency-domain representation, with frequency denoted by $ \omega $. For the steady-state case, the time derivative of the polymeric stress in Equation \ref{eq:expandpolstress} is zero.  Variables solving the steady state governing equations will be indicated with a tilde, for example: $\tilde{\bm{\tau}}_{p}^{(n)}( \mathbf{r}, \lambda )$ .  

The frequency-domain representation is formulated by rewriting all the variables in terms of their Fourier transformation, which is indicated with a caret, \emph{e.g.} 
\begin{equation}
\bm{\tau}_{p}^{(n)}( \mathbf{r}, t, \lambda ) = \frac{1}{2 \pi} \int_{-\infty}^{\infty} e^{i \omega t} \bm{\hat {\tau}}_{p}^{(n)}( \mathbf{r}, \omega, \lambda ) \, d\omega.
\end{equation}
When substituted into the mass and momentum balance (Equations \ref{eq:expandedcontinuity} and \ref{eq:expandedmombal}), orthogonality of Fourier modes can be used to simply replace time-domain variables (no caret) with frequency-domain variables (caret).  When substituted into the constitutive model, the $ O( \Wi^{n-1} ) $ contribution to the frequency-domain polymeric stress for $ n \le 3 $ can be written as:
\begin{align}
     \hat{\bm{\tau}} _{p}^{(n)}( \mathbf{r}, \omega, \lambda )  &=  \chi(\omega, \lambda) \Big( 2\hat{\mathbf{e}}^{(n)}( \mathbf{r}, \omega )
     \label{eq:expandedpolstress_FT} \\
     & \qquad -  \frac{\lambda}{\Bar{\lambda}} \sum_{m=1}^{n-1} \mathcal{F}\Big[\mathbf{v}^{(n-m)} \cdot\nabla \bm{\tau} _{p}^{(m)} - \frac{1}{2} \Big(\mathbf{A}^{(n-m)} \cdot \bm{\tau} _{p}^{(m)} + \bm{\tau} _{p}^{(m)} \cdot {\mathbf{A} ^{(n-m)}}^T \Big) \Big] \Big), \nonumber
\end{align}
where the Fourier transformation is:
\begin{equation}
    \mathcal{F}[ * ] = \int_{-\infty}^\infty e^{-i \omega t} (*) \, d t.
\end{equation} 
The frequency-domain transfer function: $\chi(\omega, \lambda) = (1+i\omega\De\lambda/\Bar{\lambda})^{-1} $, will used frequently throughout the remainder of this work.  It is the contribution of a polymer mode with relaxation time $\lambda$ to the complex viscosity of the viscoelastic fluid: $\eta^*(\omega) = \beta + (1-\beta)\left< \chi(\omega, \lambda)\right>_\lambda $, made dimensionless on $ \eta_0 $.

\subsection{Linear solution for the velocity, pressure and force} \label{firstorder}

At leading order, the frequency-domain governing equations and boundary conditions are:
\begin{equation}
\begin{gathered}
    \nabla \cdot \hat{\mathbf{v}}^{(1)}( \mathbf{r}, \omega ) = 0, \qquad - \nabla \hat{p}^{(1)}( \mathbf{r}, \omega ) + \beta \nabla \cdot \hat{\bm{\tau}}_{s}^{(1)}( \mathbf{r}, \omega ) + (1-\beta)\nabla \cdot \left< \hat{\bm{\tau}}_{p}^{(1)}( \mathbf{r}, \omega, \lambda ) \right>_\lambda = 0, \\
    \hat{\mathbf{v}}^{(1)}(r \rightarrow \infty, \omega) \rightarrow \hat V(\omega)\mathbf{e}_{z}, \qquad \hat{\mathbf{v}}^{(1)} (r = 1, \omega) = 0, \qquad \hat{p}^{(1)} (r \rightarrow \infty, \omega) \rightarrow 0,
    \label{eq:order1_freqspace}
\end{gathered}
\end{equation}
where $\hat{V}(\omega)$ is the frequency-domain, far-field velocity.  The frequency-domain constitutive model at leading order is: 
\begin{subequations}
\begin{equation}
    \hat{\bm{\tau}}_{s}^{(1)}( \mathbf{r}, \omega ) = 2 \hat{\bm{e}}^{(1)}( \mathbf{r}, \omega ), 
\end{equation}
\begin{equation}
    \hat{\bm{\tau}}_{p}^{(1)}( \mathbf{r}, \omega, \lambda )  = 2 \chi(\omega, \lambda) \hat{\mathbf{e}}^{(1)}( \mathbf{r}, \omega ).
     \label{eq:order1_constitutive}
\end{equation}
\end{subequations}
The variables in the frequency-domain (caret) can be related to their steady-state counterparts (tilde) via the following relationships: 
\begin{equation}
    \begin{gathered}
    \hat{\mathbf{v}}^{(1)}( \mathbf{r}, \omega ) = \hat{V}(\omega)\tilde{\mathbf{v}}^{(1)}( \mathbf{r} ), \hspace{12pt}  \hat{p}^{(1)}( \mathbf{r}, \omega ) = \hat{V}(\omega)  \eta^*( \omega ) \tilde{p}^{(1)}( \mathbf{r} ),
    \\
    \hat{\bm{\tau}}_{p}^{(1)}( \mathbf{r}, \omega, \lambda ) = \hat{V}(\omega)\chi(\omega, \lambda)\tilde{\bm{\tau}}_{p}^{(1)}( \mathbf{r} ), \hspace{12pt} \hat{\bm{\tau}}_{s}^{(1)}( \mathbf{r}, \omega ) = \hat V(\omega)\tilde{\bm{\tau}}_{s}^{(1)}( \mathbf{r} ).
    \label{eq:o1scaling}
    \end{gathered}
\end{equation}

Substituting these relationships into the frequency domain equations recovers exactly the steady-state form of the governing equations and constitutive model. The first-order velocity and pressure profiles are given by the well-known solution for Stokes flow of a Newtonian fluid around a sphere.  In index notation:
\begin{subequations}
\begin{equation}
\tilde{v}_{i}^{(1)}( \mathbf{r} ) = \frac{3}{4r} \left(\delta_{i 3}+ \frac{r_{i}r_{3}}{r^2}\right)  - \frac{1}{4r^3} \left(\delta_{i3}- \frac{3r_{i}r_{3}}{r^2}\right),
\label{eq:SSFOS}
\end{equation}
\begin{equation}
    \tilde p^{(1)}( \mathbf{r} ) = \frac{3r_3}{2r^3},
\end{equation}
\end{subequations}

where the index $3$ corresponds to the direction of the lineal motion, $ \mathbf{e}_z $.

At leading order, the steady state force is given simply the dimensionless Stokes drag: $ \tilde{\mathbf{F}}^{(1)} = 6 \pi \mathbf{e}_{z} $.   The unsteady force in frequency space is obtained by applying the scaling relationships for the steady state variables to the definition of the force at leading order (Equation \ref{eq:forcedefn}): $ \hat{\mathbf{F}}^{(1)} (\omega) = 6 \pi \eta^*( \omega ) \hat{V}(\omega) \mathbf{e}_{z}$.    For convenience, we can define a dimensional quantity: $\zeta^{*}_{1}(\omega) = 6 \pi \eta_0 a [ \beta + ( 1 - \beta ) \left< \chi( \omega, \lambda ) \right>_\lambda ] $, called the first-order complex resistivity.  It is the linear, frequency-domain transfer function between the dimensional force and the dimensional imposed velocity.

\subsection{Leading-order nonlinearities in the velocity and pressure} \label{secondorder}

At the next order in $\Wi$, the governing equations and boundary conditions are: 
\begin{equation}
\begin{gathered}
    \nabla \cdot \hat{\mathbf{v}}^{(2)}( \mathbf{r}, \omega ) = 0, \qquad - \nabla \hat{p}^{(2)}( \mathbf{r}, \omega ) + \beta \nabla \cdot \hat{\bm{\tau}}_{s}^{(2)}( \mathbf{r}, \omega ) + (1-\beta) \nabla \cdot \left< \hat{\bm{\tau}}_{p}^{(2)}( \mathbf{r}, \omega, \lambda ) \right>_\lambda  = 0, \\
    \hat{\mathbf{v}}^{(2)}(r \rightarrow \infty, \omega) \rightarrow 0, \qquad \hat{\mathbf{v}}^{(2)} (r = 1, \omega) = 0, \qquad \hat{p}^{(2)} (r \rightarrow \infty, \omega) \rightarrow 0,
    \label{eq:order2BC}
 \end{gathered}
\end{equation}
and the constitutive equation at this order is: 
\begin{subequations}
\begin{equation}
    \hat{\bm{\tau}}_{s}^{(2)}( \mathbf{r}, \omega ) = 2 \hat{\mathbf{e}}^{(2)}( \mathbf{r}, \omega ),
\end{equation}
\begin{align}
    \bm{\hat \tau}_{p}^{(2)}( \mathbf{r}, \omega, \lambda )  &= \chi(\omega, \lambda) \Big( 2\mathbf{\hat e}^{(2)}( \mathbf{r}, \omega ) 
      \label{eq:order2polstress} \\
    &-  \frac{\lambda}{\Bar{\lambda}}\Big[\hat{\mathbf{v}}^{(1)}( \mathbf{r}, \omega ) * \nabla \hat{\bm{\tau}}_{p}^{(1)}( \mathbf{r}, \omega, \lambda ) - \frac{1}{2} \Big( \hat{\mathbf{A}}^{(1)}( \mathbf{r}, \omega ) *  \hat{\bm{\tau}}_{p}^{(1)}( \mathbf{r}, \omega, \lambda ) + \hat{\bm{\tau}}_{p}^{(1)}( \mathbf{r}, \omega, \lambda ) * {\hat{\mathbf{A}}^{(1)}} ( \mathbf{r}, \omega )^ {T} \Big) \Big] \Big).\nonumber
\end{align}
\end{subequations}
Here the $*$ in Equation \ref{eq:order2polstress} indicates a scaled convolution (and a dot product for vector/tensor arguments) of two terms in the frequency domain:
\begin{equation}
    \mathcal{F} \left[ f(t) g(t) \right] = \hat f( \omega ) * \hat g( \omega ) = \frac{1}{2\pi} \iint_{-\infty}^{\infty}   \hat f( \omega_1 ) \hat g( \omega_2 ) \delta(\omega - \omega_1 - \omega_2) d\omega_1 d\omega_2 ,
\end{equation}
and appears due to application of the convolution theorem in evaluating the Fourier transform in Equation \ref{eq:expandedpolstress_FT}. It should be noted that while the symbol $*$ is typically used to denote convolution, here it includes the scaling factor introduced due to the definition of the Fourier transform being used.

The relationships between the steady-state and frequency-domain terms at first order were already specified and can be applied directly.  Additional relationships between terms at second order can also be determined: 
\begin{equation}
\begin{gathered}
    \hat{\mathbf{v}}^{(2)}( \mathbf{r}, \omega ) = \frac{1}{\eta^*( \omega ) \Bar \lambda} \Big< \lambda \chi( \omega, \lambda ) \Big[ \hat V(\omega) * \Big( \hat V(\omega) \chi (\omega, \lambda) \Big) \Big] \Big>_\lambda \tilde{\mathbf{v}}^{(2)}( \mathbf{r} ),
    \\ \hat{p}^{(2)}( \mathbf{r} ) = \frac{1}{\Bar \lambda} \Big< \lambda\chi(\omega, \lambda) \Big[ \hat V(\omega) * \Big( \hat V(\omega) \chi (\omega, \lambda) \Big) \Big] \Big>_\lambda \tilde{p}^{(2)}( \mathbf{r} ).
    \label{eq:SOSRelations}
\end{gathered}
\end{equation}
 It is important that the terms in angle brackets exist.  Recall that the quantity, $ \chi( \omega, \lambda ) $, scales as $ \lambda^{-1} $ for large $ \lambda $ when $ \omega \ne 0 $ and that $ P( \lambda ) $ must decay faster than $ \lambda^{-1} $ to be integrable.  Therefore, the scaling factor: $ \left< \lambda \chi( \omega, \lambda ) [ \hat V( \omega ) * ( \hat V( \omega ) \chi( \omega, \lambda ) ] \right>_\lambda $ is finite for all $ \omega \ne 0 $.  If this scaling factor must be finite when $ \omega = 0 $, then $ P( \lambda ) $ must decay faster than $ \lambda^{-2} $ for large values of $ \lambda $, because $ \chi( 0, \lambda ) = 1 $.  An equivalent requirement in this case is that $ P( \lambda ) $ must have a finite mean value: $ \left< \lambda \right>_\lambda $.

Substitution of these scaled, second order terms into the frequency-domain governing and constitutive equations recovers the steady-state momentum balance at second order for a single mode Johnson-Segalman fluid: 
\begin{equation}
         \nabla^{2} \tilde{\mathbf{v}}^{(2)}( \mathbf{r})  - \nabla \tilde{p}^{(2)}( \mathbf{r}) = (1-\beta) \nabla \cdot \left[ \tilde{\mathbf{v}}^{(1)}( \mathbf{r}) \cdot \nabla \tilde{\bm{\tau}}_{p}^{(1)}( \mathbf{r}) - \frac{1}{2} \left( \tilde{\mathbf{A}}^{(1)}( \mathbf{r}) \cdot  \tilde{\bm{\tau}}_{p}^{(1)}( \mathbf{r}) + \tilde{\bm{\tau}}_{p}^{(1)}( \mathbf{r}) \cdot {\tilde{\mathbf{A}}^{(1)} }( \mathbf{r})^ {T} \right) \right].
         \label{SOS_SS_mombal}
\end{equation}

From conservation of mass, $ \mathbf{\tilde v}^{(2)}( \mathbf{r}) $ is also known to be divergence free.  Using this in conjunction with the steady-state momentum balance, the second-order steady-state velocity and pressure profiles can be solved for analytically. We do this in tensorial form through use of the computer algebra system Mathematica and the plugin EinS \cite{Klioner1998}: 
\begin{subequations}
\begin{multline}
    \tilde{v}^{(2)}_i( \mathbf{r} ) = b (1-\beta)  \left[ \frac{3}{8r^{3}} \left( 1 - \frac{3}{r} + \frac{3}{r^2} - \frac{1}{r^3 }  \right) r_{i} - \frac{9}{8r^{4}} \left( 1 - \frac{2}{r} + \frac{1}{r^{2}} \right) \delta_{i3}r_{3}   \right. \\ \left. - \frac{9}{8r^{5}} \left( 1 - \frac{4}{r} + \frac{5}{r^{2}} - \frac{2}{r^3} \right)r_{i} r_{3} r_{3}  \right],
    \label{eq:SOS_SS}
\end{multline}

\begin{equation}
    \tilde{p}^{(2)}( \mathbf{r} ) =  \frac{3}{2r^3} \left(  1 - \frac{3}{4 r} - \frac{3}{r^2} + \frac{5}{4 r^3} + \frac{3}{r^5} - \frac{3}{2 r^7} \right) + \frac{3b}{4r^3}\left( 1 - \frac{3}{2 r} - \frac{3}{r^2} + \frac{12}{r^3} - \frac{15}{r^5} + \frac{6}{r^7} \right).
    \label{eq:SOS_SS_P}
\end{equation}
\end{subequations}

These profiles, when converted to spherical coordinates with $b=1$, corresponding to the Oldroyd-B model, match exactly those determined at second order by Housiadas and Tanner for the same model \cite{Housiadas2016}. These steady-state analytical solutions can then be related to their frequency-domain counterparts using the relationships in Equation \ref{eq:SOSRelations} to describe the time-dependent response. The contribution of these fields to the force can also be calculated at second order. However, due to the spatial symmetry of the pressure and stress at second order, this contribution to the force is exactly zero.

\subsection{Leading order nonlinearity in the force} \label{thirdorder}

In order to determine the force at third order in deformation amplitude, we must be able to evaluate the surface integral of the third order traction, $\mathbf{n} \cdot \left( -\hat p^{(3)}( \mathbf{r}, \omega ) \mathbf{I} + \bm{\hat \tau}_{s}^{(3)}( \mathbf{r}, \omega ) + \left< \bm{\hat \tau}_{p}^{(3)}( \mathbf{r}, \omega, \lambda ) \right>_\lambda \right) $. Equation \ref{eq:expandpolstress} allows one to specify part of $ \bm{\hat \tau}_p^{(3)}( \mathbf{r}, \omega, \lambda ) $ in terms of the first and second order velocity profiles.  However, without solving the system of partial differential equations for the third order fields, we cannot know $\hat{ \textbf{e}}^{(3)}( \mathbf{r}, \omega ) $, which is needed to compute the solvent stress, or $\hat p^{(3)}( \mathbf{r}, \omega )$.  Yet, by making use of the Lorentz reciprocal theorem, we can avoid having to explicitly solve for the pressure and velocity fields at this next order. 

We first define an auxiliary velocity field, $\mathbf{\hat v}^\prime( \mathbf{r} )$, having pressure profile $\hat p^\prime( \mathbf{r}, \omega )$.  These auxiliary velocity and pressure fields satisfy the Stokes equations:
\begin{equation}
    \eta^*( \omega ) \nabla^2 \mathbf{\hat v}^\prime( \mathbf{r} ) = \nabla \hat p^\prime( \mathbf{r}, \omega ), \qquad \nabla \cdot \mathbf{\hat v}^\prime( \mathbf{r} ) = 0,
\end{equation}
with boundary conditions $\mathbf{\hat v}^\prime(r = 1 ) = \mathbf{u}^\prime$, $\mathbf{\hat v}^\prime(r \rightarrow \infty ) \rightarrow 0$, and $\hat p^\prime(r \rightarrow \infty, \omega) \rightarrow 0$. The constant vector, $ \mathbf{u}^\prime $, is chosen arbitrarily.  The deviatoric stress in this auxiliary flow is $\bm{\hat \tau}^\prime( \mathbf{r}, \omega ) = 2 \eta^*(\omega) \mathbf{\hat e}^\prime( \mathbf{r} ) $, where $\eta^*(\omega)$ is just the complex viscosity of the viscoelastic fluid under study at frequency, $ \omega $, normalized on its zero shear viscosity.  The rate of strain in the auxiliary flow is: $\mathbf{e}^\prime( \mathbf{r} ) = \mathbf{u}^\prime \cdot \mathbf{R}(\mathbf{r}) $, with
\begin{equation}
    R_{ijk}( \mathbf{r} ) = \frac{3}{4r^3} \left( -r_k \delta_{ij}  + \frac{1}{r^2} \left( 3r_i r_j r_k + r_i \delta_{jk} + r_j \delta_{ik} + r_k \delta_{ij} \right) - \frac{1}{r^4}5r_i r_j r_k \right)
\end{equation}

Following the usual construction of reciprocal theorem type arguments, we know that
\begin{align}
    &\int_{\Omega}  \nabla \cdot \left[ -\hat p^{(3)}( \mathbf{r}, \omega ) \mathbf{I} + \beta \bm{\hat \tau}_s^{(3)}( \mathbf{r}, \omega ) + (1-\beta) \left< \bm{\hat \tau}_p^{(3)}( \mathbf{r}, \omega, \lambda ) \right>_\lambda \right] \cdot \mathbf{v}^\prime( \mathbf{r} ) \, d\mathbf{r} \nonumber \\
    & \qquad \qquad = \int_{\Omega}  \nabla \cdot \left[ -p^\prime( \mathbf{r}, \omega ) \mathbf{I} + \bm{\tau}^\prime( \mathbf{r}, \omega ) \right] \cdot \hat{\mathbf{v}}^{(3)}( \mathbf{r}, \omega ) \, d\mathbf{r},
    \label{eq:recipthm}
\end{align}
with $\Omega$ the volume of the fluid around the particle, because the divergence of the total stress in each flow is zero. Applying the product rule and the divergence theorem along with some tedious algebra leads to a surprisingly simple expression for the third order contribution to the force on the particle: 
\begin{align}
      \mathbf{u}^\prime \cdot \mathbf{\hat F}^{(3)}( \omega ) &=\mathbf{u}^\prime \cdot \int_{\partial \Omega} \mathbf{n} \cdot \left[ -\hat p^{(3)}( \mathbf{r}, \omega ) \mathbf{I} + \beta \bm{\hat \tau}_s^{(3)}( \mathbf{r}, \omega ) + (1-\beta) \left< \bm{\hat \tau}_p^{(3)}( \mathbf{r}, \omega, \lambda ) \right>_\lambda \right] \, d \mathbf{r} \nonumber \\
      &=  (1-\beta) \int_{\Omega}  \mathbf{e}^\prime( \mathbf{r} ) : \left[ \left< \hat{\bm{\tau}}^{(3)}_p( \mathbf{r}, \omega, \lambda ) \right>_\lambda -2\left< \chi (\omega)\right>\hat{\mathbf{e}}^{(3)}( \mathbf{r}, \omega ) \right] \, d \mathbf{r},
      \label{eq:forcedev3}
\end{align}
  Recalling the definition of $\mathbf{e}'( \mathbf{r} )$ and recognizing that the heterogeneity in the auxiliary flow field, $\mathbf{u}^\prime$, is arbitrary, reduces the expression for the third order force to: 
\begin{equation}
      \hat{\mathbf{F}}^{(3)}( \omega ) = (1-\beta) \int_{\Omega}  \mathbf{R}(\mathbf{r}) : \left(\left<\hat{\bm{\tau}}_p^{(3)}( \mathbf{r}, \omega, \lambda ) \right>_\lambda - 2\left< \chi(\omega, \lambda)\right> \hat{\mathbf{e}}^{(3)}( \mathbf{r}, \omega ) \right) \, d\mathbf{r}.
      \label{eq:thirdorderforce}
\end{equation}
Finally, we can close this expression for the force by computing the average over modes for the polymeric stress constitutive model at third order:
\begin{align}
    \left< \hat{\bm{\tau}}^{(3)}_p( \mathbf{r}, \omega, \lambda ) \right>_\lambda &-2 \left< \chi (\omega, \lambda) \right>_\lambda \hat{\mathbf{e}}_{3}( \mathbf{r}, \omega ) = \label{eq:tau3def}
    - \Big< \chi(\omega, \lambda) \left(\frac{\lambda}{\Bar{\lambda}} \right) \Big[ \hat{\mathbf{v}}^{(1)}( \mathbf{r}, \omega ) * \nabla \hat{\bm{\tau}}_{p}^{(2)}( \mathbf{r}, \omega, \lambda ) \\ &+ \hat{\mathbf{v}}_{2}( \mathbf{r}, \omega ) * \nabla \hat{\bm{\tau}}_{p}^{(1)}( \mathbf{r}, \omega, \lambda ) \nonumber - \frac{1}{2} \Big( \hat{\mathbf{A}}^{(1)}( \mathbf{r}, \omega ) * \hat{\bm{\tau}}_{p}^{(2)}( \mathbf{r}, \omega, \lambda ) + \hat{\mathbf{A}}^{(2)}( \mathbf{r}, \omega ) * \hat{\bm{\tau}}_{p}^{(1)}( \mathbf{r}, \omega, \lambda )  \\
    &  + \hat{\bm{\tau}}_{p}^{(2)}( \mathbf{r}, \omega, \lambda ) * \hat{\mathbf{A}}^{(1)}( \mathbf{r}, \omega )^T + \hat{\bm{\tau}}_{p}^{(1)}( \mathbf{r}, \omega, \lambda ) * \hat{\mathbf{A}}^{(2)}( \mathbf{r}, \omega )^T \Big) \Big] \Big>_\lambda.
    \nonumber
\end{align}
As illustrated in the previous sections, there are scaling relationships between the frequency-domain terms in Equation \ref{eq:tau3def} (carets) and the steady-state solutions for those values (tildes).  Therefore, one can use the steady-state solutions for the velocity field at first and second order to determine the third order force.  One caveat when formulating the solution this way is that the volume integral over the fluid domain in Equation \ref{eq:thirdorderforce} must converge.  We know that $ \mathbf{R}(\mathbf{r}) \sim O(r^{-2})$, and $\left< \hat{\bm{\tau}}_{p}^{(3)}( \mathbf{r}, \omega, \lambda ) \right>_\lambda - 2 \eta^*(\omega)\hat{\mathbf{e}}^{(3)}( \mathbf{r}, \omega )  \sim O(r^{-4})$ in the large $ r $ limit.  Therefore, this volume integral is absolutely convergent.  Additionally, the relaxation time distribution must decay fast enough that the average in Equation \ref{eq:thirdorderforce} is finite.  As with the calculation of the leading order nonlinearity in the velocity and pressure fields, this always happens when the relaxation time distribution has a finite mean value.

Once the third-order force is re-expressed in terms of the steady-state solutions, three groupings of terms emerge as the products of volume integrals over the steady state solutions with distinct frequency domain coefficients. These coefficients depend on averages over the relaxation time distribution and convolutions of the time-dependent flow, $ \hat V( \omega ) $, with the transfer function $ \chi( \omega, \lambda ) $.  Here, we take advantage of the fact that averaging and convolution are integral operations whose order can be interchanged freely so long as the integrals are absolutely convergent:

\begin{align}
    \mathbf{\hat F}^{(3)}( \omega ) &= \mathbf{C}_1 (1-\beta) \frac{1}{\Bar \lambda^2} \Big< \Big< \lambda_1 \lambda_2 \chi( \omega, \lambda_1 ) \Big( \Big[ \hat V( \omega ) \chi( \omega, \lambda_1 ) \Big] * \Big[ \frac{\chi( \omega, \lambda_2 )}{\eta^*( \omega )}  \Big( \hat V( \omega ) * \Big[ \hat V( \omega ) \chi( \omega, \lambda_2 )  \Big] \Big) \Big] \Big) \Big>_{\lambda_1} \Big>_{\lambda_2} \nonumber \\
    &+ \mathbf{C}_2 ( 1 - \beta ) \frac{1}{\Bar \lambda^2} \Big< \Big< \lambda_1 \lambda_2 \chi( \omega, \lambda_1 ) \Big( \hat V( \omega ) * \Big[ \frac{\chi( \omega, \lambda_1 ) \chi( \omega, \lambda_2 )}{\eta^*( \omega )}  \Big( \hat V( \omega ) * \Big[ \hat V( \omega ) \chi( \omega, \lambda_2 ) \Big]  \Big) \Big] \Big)  \Big>_{\lambda_1} \Big>_{\lambda_2} \nonumber \\
    &+ \mathbf{C}_3 ( 1 - \beta ) \frac{1}{\Bar \lambda^2} \Big< \lambda^2 \chi( \omega, \lambda ) \Big( \hat V( \omega ) * \Big[ \chi( \omega, \lambda ) \Big( \hat V( \omega ) * \Big[ \hat V( \omega ) \chi( \omega, \lambda ) \Big] \Big) \Big] \Big>_\lambda, 
    \label{eq:forceform}
\end{align}

with
\begin{subequations}
\begin{align}
    \mathbf{C}_{1} &=  -\int_{\Omega} \mathbf{R}(\mathbf{r}):\left[  \tilde{\mathbf{v}}^{(2)}(\mathbf{r}) \cdot \nabla \tilde{\bm{\tau}}_{p}^{(1)}(\mathbf{r}) - \frac{1}{2} \left( \tilde{\mathbf{A}}^{(2)}(\mathbf{r}) \cdot \tilde{\bm{\tau}}_{p}^{(1)} (\mathbf{r}) + \tilde{\bm{\tau}}_{p}^{(1)}(\mathbf{r}) \cdot \tilde{\mathbf{A}}^{(2)}(\mathbf{r})^{T} \right) \right] \, d\mathbf{r} \nonumber \\
    & = -(1-\beta)\frac{3\pi b^2}{175} \mathbf{e}_z,
\end{align}
\begin{align}
    \mathbf{C}_{2} &= -2 \int_{\Omega}\mathbf{R}(\mathbf{r}): \left[ \tilde{\mathbf{v}}^{(1)} (\mathbf{r})\cdot \nabla \tilde{\mathbf{e}}^{(2)}(\mathbf{r}) - \frac{1}{2} \left( \tilde{\mathbf{A}}^{(1)}(\mathbf{r}) \cdot \tilde{\mathbf{e}}^{(2)}(\mathbf{r})  + \tilde{\mathbf{e}}^{(2)}(\mathbf{r}) \cdot \tilde{\mathbf{A}}^{(1)}(\mathbf{r})^{T} \right) \right] \, d\mathbf{r} \nonumber \\
    &= -(1-\beta)\frac{3\pi b^2}{175} \mathbf{e}_z,
\end{align}
\begin{align}
   \mathbf{C}_{3} &= -\int_{\Omega} \mathbf{R}(\mathbf{r}): \left[-\tilde{\mathbf{v}}^{(1)}(\mathbf{r}) \cdot \nabla \tilde{\mathbf{g}}^{(1)}(\mathbf{r}) + \frac{1}{2} \left( \tilde{\mathbf{A}}^{(1)}(\mathbf{r}) \cdot \tilde{\mathbf{g}}^{(1)}(\mathbf{r})  + \tilde{\mathbf{g}}^{(1)}(\mathbf{r}) \cdot \tilde{\mathbf{A}}^{(1)}(\mathbf{r})^{T} \right) \right]\, d\mathbf{r} \nonumber \\
   &= \frac{-18\pi(1813-1727b)}{25025} \mathbf{e}_z,
\end{align}
\end{subequations}
and $\tilde{\mathbf{g}}^{(1)}(\mathbf{r}) = \tilde{\mathbf{v}}^{(1)}(\mathbf{r}) \cdot \nabla \tilde{\bm{\tau}}_{p}^{(1)}(\mathbf{r}) - 1/2 \left( \tilde{\mathbf{A}}^{(1)}(\mathbf{r}) \cdot \tilde{\bm{\tau}}_{p}^{(1)}(\mathbf{r}) + \tilde{\bm{\tau}}_{p}^{(1)}(\mathbf{r}) \cdot \tilde{\mathbf{A}}^{(1)}(\mathbf{r})^{T} \right) $.  The convolutions in Equation \ref{eq:forceform} can be rewritten in a more convenient form that will ultimately make the expression for the force more compact by applying the identity: 
\begin{align}
&\hat{a}(\omega) \left[ \hat b( \omega ) * \left( \hat{c}(\omega) \left[ \hat{d} (\omega) * \hat e( \omega ) \right] \right) \right] \\
&\qquad = \frac{1}{(2\pi)^2} \iiint_{-\infty}^\infty  \hat{a}(\omega_{1} + \omega_2 + \omega_3 ) \hat{b}(\omega_{1})\hat{c}(\omega_2 + \omega_{3}) \hat d( \omega_2 ) \hat e( \omega_3 ) \delta(\omega-\omega_{1} - \omega_{2} - \omega_{3}) \, d\omega_{1}d\omega_{2}d\omega_{3}, \nonumber
\end{align}
to yield:
\begin{align}
    \mathbf{\hat F}^{(3)}( \omega ) = \\ -\mathbf{e}_z \frac{1}{(2\pi)^2} & \iiint_{-\infty}^\infty \frac{1}{ \Bar \lambda^2} \Big[ (1-\beta)^2  \frac{3\pi b^2}{175} \frac{1}{\eta^*( \omega_2 + \omega_3 ) } \Big< \lambda \chi( \omega_1 + \omega_2 + \omega_3, \lambda ) \chi( \omega_1, \lambda ) \Big>_\lambda \Big< \lambda \chi( \omega_2 + \omega_3, \lambda ) \chi( \omega_3, \lambda ) \Big>_\lambda \nonumber \\
    & + (1-\beta)^2 \frac{3\pi b^2}{175} \frac{1}{\eta^*( \omega_2 + \omega_3 ) } \Big< \lambda \chi( \omega_1 + \omega_2 + \omega_3, \lambda ) \chi( \omega_2 + \omega_3, \lambda ) \Big>_\lambda \Big< \lambda \chi( \omega_2 + \omega_3, \lambda ) \chi( \omega_3, \lambda ) \Big>_\lambda \nonumber \\
    & + (1-\beta) \frac{18\pi(1813-1727b)}{25025} \Big< \lambda^2 \chi( \omega_1 + \omega_2 + \omega_3, \lambda ) \chi( \omega_2 + \omega_3, \lambda ) \chi( \omega_3, \lambda ) \Big>_\lambda \Big]  \nonumber \\
    & \times \hat V( \omega_1 ) \hat V( \omega_2 ) \hat V( \omega_3 ) \delta(\omega-\omega_{1} - \omega_{2} - \omega_{3}) \, d\omega_{1}d\omega_{2}d\omega_{3}.
\end{align}
Equivalently, the third-order contribution to the force can be expressed in terms of a nonlinear transfer function: 
\begin{equation}
  \mathbf{\hat F}^{(3)}(\omega) = -\mathbf{e}_z \frac{1}{(2\pi)^2} \iiint_{-\infty}^{\infty} R^*_3 (\omega_{1}, \omega_{2}, \omega_{3})  \hat V(\omega_{1}) \hat V(\omega_{2}) \hat V(\omega_{3})  \delta(\omega - \omega_{1} - \omega_{2} - \omega_{3}) \, d\omega_{1} d\omega_{2} d\omega_{3},  \label{eq:TOF}
\end{equation}
where the transfer function is like a dimensionless, third-order complex resistivity: 
\begin{align}
    R^*_3(\omega_1, \omega_2, \omega_3 ) &=  \frac{1}{\Bar \lambda^2} \Big[ (1-\beta)^2 \frac{3\pi b^2}{175} \frac{1}{\eta^*( \omega_2 + \omega_3 ) } \Big< \lambda \chi( \omega_1 + \omega_2 + \omega_3, \lambda ) \chi( \omega_1, \lambda ) \Big>_\lambda \Big< \lambda \chi( \omega_2 + \omega_3, \lambda ) \chi( \omega_3, \lambda ) \Big>_\lambda \nonumber \\
    & + (1-\beta)^2 \frac{3\pi b^2}{175} \frac{1}{\eta^*( \omega_2 + \omega_3 ) } \Big< \lambda \chi( \omega_1 + \omega_2 + \omega_3, \lambda ) \chi( \omega_2 + \omega_3, \lambda ) \Big>_\lambda \Big< \lambda \chi( \omega_2 + \omega_3, \lambda ) \chi( \omega_3, \lambda ) \Big>_\lambda \nonumber \\
    & + (1-\beta) \frac{18\pi(1813-1727b)}{25025}  \Big< \lambda^2 \chi( \omega_1 + \omega_2 + \omega_3, \lambda ) \chi( \omega_2 + \omega_3, \lambda ) \chi( \omega_3, \lambda ) \Big>_\lambda \Big]. \label{eq:R3}
\end{align}
Section \ref{generalize} is devoted to understanding and interpreting this transfer function as a term in a Volterra series expansion of the force in terms of the velocity. 

\section{Generalizing, interpreting and visualizing the model predictions}\label{generalize}

For asymptotically weak deformations, we have derived expressions for the leading order nonlinearities in the velocity and pressure fields in a Johnson-Segalman fluid flowing around a stationary spherical particle with an arbitrarily unsteady, lineal flow imposed in the far-field having charcteristic magnitude $ v_c $.  In dimensional form, these weak nonlinearities scale with $ v_c^2 $.  We have also derived the leading order nonlinearity in the force exerted on this stationary particle, which in dimensional form scales with $ v_c^3 $.

The asymptotic solutions derived in Section \ref{methods}, and the definition of the third-order force as written in Equation \ref{eq:TOF} are not limited to this specific constitutive model and could be calculated for a variety of models or measured in experiments.  In the Appendix, we repeat the calculation for the Giesekus model and compare the asymptotic behavior of the different models in this section.  For both of the constitutive models studied in this work, or for other models and materials, the force exerted on the particle in an unsteady uniform flow field can be expressed in terms of a Volterra series, which we will show serves as a useful conceptual framework for representing the nonlinear force-displacement relationship for a particle immersed in a complex fluid. 

\subsection{Volterra series representation of the force-velocity relationship}\label{volterra}

The Volterra series is a general polynomial representation of a functional describing a scalar, time-dependent input-output relationship. It is analogous to a Taylor series for a scalar, time invariant function \cite{Volterra1959, Bemporad2010}.  For the remainder of the manuscript, we will work in dimensional terms and describe a particle moving through the fluid. The $ \mathbf{e}_z $ component of the dimensional force exerted on a particle executing a lineal translation through a complex fluid with dimensional velocity, $ V( t ) \mathbf{e}_z $, has the Volterra series in Fourier space:
\begin{equation}
    \mathbf{e}_z \cdot \mathbf{\hat F}(\omega) = -\sum_{\substack{n=1 \\ n \in \mathrm{odds}}}^{\infty} \frac{1}{(2\pi)^{n-1}} \int \overset{n}\cdots \int_{-\infty}^{\infty} \zeta^*_{n} (
    \omega_{1}, ...,\omega_{n}) \delta(\omega - \sum_{m = 1}^{n} \omega_{m}) \prod_{m=1}^{n} \hat V(\omega_{m}) d\omega_{m}.
    \label{eq:gen_VS}
\end{equation}
The quantities, $\zeta^{*}_{n}( \omega_1, \ldots, \omega_n ) $, are called nth-order Volterra kernels, which can be recognized as complex resistivities.  The Volterra series is truncated at third order is:
\begin{align}
    \mathbf{e}_z \cdot \mathbf{\hat F}(\omega) &=  -\int_{-\infty}^{\infty} \zeta^*_{1} (\omega_{1} ) \delta(\omega-\omega_{1} ) \hat V(\omega_{1}) d\omega_{1} \nonumber \\
    &- \frac{1}{(2\pi)^2}\iiint_{-\infty}^{\infty} \zeta^*_{3} (
    \omega_{1}, \omega_{2}, \omega_{3})  \hat V(\omega_{1}) \hat V(\omega_{2}) \hat V(\omega_{3}) \delta(\omega -  \omega_{1} - \omega_2 - \omega_3 ) \, d\omega_{1} d\omega_{2} d\omega_{3}.
    \label{eq:TO_VS}
 \end{align}
We can immediately identify the Volterra series coefficients with the dimensionless responses determined in the asymptotic expansions of the previous section:
\begin{subequations}
\begin{equation}
    \zeta^*_1( \omega )  = 6 \pi \eta_0 a \left[ \beta + (1-\beta) \left< ( 1 + i \omega \lambda )^{-1} \right>_\lambda \right],
\end{equation}
\begin{equation}
    \zeta_3^*( \omega_1, \omega_2, \omega_3 ) = \frac{\eta_0 a \Wi^2}{v_c^2} R_3^*( \omega_1, \omega_2, \omega_3 ) = \frac{\eta_0 \bar \lambda^2}{a} R_3^*( \omega_1, \omega_2, \omega_3 ).
\end{equation}
\end{subequations} 
The minus sign introduced in the Volterra series comes from the fact that we have changed the frame of reference from describing a particle held stationary in a far-field flow to considering a particle executing a lineal translation in a stationary fluid.

Writing the force exerted on the particle as a Volterra series in the lineal particle velocity -- or, as we will see in Section \ref{forcecontrolled}, the lineal velocity as a Volterra series in the force -- has several implications. A generalizable relationship between the time-dependent force and velocity with this prescribed lineal motion makes it far easier to relate experimental measurements of this nonlinear response when different temporal protocols are used to drive particle motion in the same fluid.  Section \ref{microrheology} will discuss this in the context of microrheology experiments in particular.

The first- and third-order complex resistivities each have real and imaginary components. In keeping with the notation that is conventional in rheology, we describe these two parts separately: $ \zeta^{*}_{1}(\omega) = \zeta_{1}'(\omega) - i\zeta_{1}''(\omega) $ and $ \zeta^{*}_{3}(\omega_1,\omega_2, \omega_3) = \zeta_{3}'(\omega_1,\omega_2, \omega_3) - i\zeta_{3}''(\omega_1,\omega_2, \omega_3) $.  For all complex fluids, $\zeta'_{1}(\omega), \zeta''_1(\omega)$ are positive and real.  There is no restriction on the sign of $\zeta_{3}'(\omega_1,\omega_2, \omega_3)$ or $\zeta_{3}''(\omega_1,\omega_2, \omega_3)$, but both are real numbers.  The signs of these quantities when $ \omega_1, \omega_2, \omega_3 \ll 1 $ can be interpreted in terms of shear thinning/thickening (real part) or strain softening/hardening (imaginary part), instead.  For finite values of the frequency, physical rationalizations for the signs of these terms are still somewhat controversial.  At zero frequency, the third order transfer function reflects how the drag coefficient either increases or decreases with faster particle motion.  Thus the relationship to thinning/thickening is plainly evident.

Thus far, much of the derivation has been in dimensionless terms.  The complex resistivities presented from here on are dimensional.  It is useful to reflect on their units and how they might be made into quantities that describe the fluid properties independent of the particle size.  The dimensional, first-order resistivity has dimensions in SI units of N s/m.  It scales linearly with the particle radius, $ a $; thus, $ \zeta_1^*( \omega ) / a $ depends only on the properties of the fluid.  In fact, $ \zeta_1^*( \omega ) / ( 6 \pi a ) $ is the dimensional complex viscosity of the fluid.  The dimensional third-order complex resistitivity has dimensions in SI units of N s$^3$/m$^3$.  It scales linearly with the inverse of the particle radius, $ a $; thus, $ \zeta_3^*( \omega_1, \omega_2, \omega_3 ) a $ depends only on the properties of the fluid as well.

It will prove useful to understand some of the properties of the third-order complex resistivity itself. There are two key symmetries inherent in the third-order complex resistivity $\zeta^{*}_{3}(\omega_{1}, \omega_{2}, \omega_{3})$ that impact the weakly nonlinear force-velocity relationship: Hermitian symmetry, arising from the properties of the Fourier transform; and permutation symmetry, arising from the properties of the Volterra series representation.

Hermitian symmetry of a function, $ f( \omega ): \mathbb{R} \rightarrow \mathbb{C} $, implies that the value of that function for argument $-\omega$ is the complex conjugate of the value for argument at $\omega$.  Hermitian symmetry is guaranteed for $\hat{V}(\omega)$ because it is a Fourier transformation of a real-valued time signal. Thus, $\mbox{Re}[\hat{V}(-\omega)] = \mbox{Re} [\hat{V}(\omega)]$, and $\mbox{Im} [\hat{V}(-\omega)] = -\mbox{Im}[\hat{V}(\omega)]$.  The same is true for $ \mathbf{e}_z \cdot \mathbf{\hat F}( \omega ) $.  Because of the Hermitian symmetry of the force and the velocity, $ \zeta_1^*( \omega ) $ also exhibits Hermitian symmetry, and the third-order resistivity satisfies the symmetry relations: $ \mathrm{Re}[ \zeta_3^*( -\omega_1, -\omega_2, -\omega_3 ) ] = \mathrm{Re}[ \zeta_3^*( \omega_1, \omega_2, \omega_3 ) ] $ and $ \mathrm{Im}[ \zeta_3^*( -\omega_1, -\omega_2, -\omega_3 ) ] = -\mathrm{Im}[ \zeta_3^*( \omega_1, \omega_2, \omega_3 ) ] $.

A unique form of the third-order complex resistivity should also possess a permutation symmetry that arises from the construction of the third order term in the Volterra series.  It is clear that simply swapping the names of the frequencies in the triple integral in equation \ref{eq:TO_VS} (\emph{e.g.} $ \omega_1 \rightarrow \omega_2 \rightarrow \omega_3 \rightarrow \omega_1 $) cannot change the result of the expression.  In fact, all the Volterra series kernels are symmetric with respect to the permutations of the frequencies. As such, for the the third-order complex resistivity: 
\begin{equation*} \zeta^{*}_{3} (\omega_{1}, \omega_{2}, \omega_{3}) = \zeta^{*}_{3} (\omega_{1}, \omega_{3}, \omega_{2}) = \zeta^{*}_{3} (\omega_{2}, \omega_{1}, \omega_{3}).
\end{equation*}
The expression for $ R^*_3( \omega_1, \omega_2, \omega_3 ) $ in equation \ref{eq:R3} does not possess this permutation symmetry, but the symmetric version of the third-order complex resistivity can easily be crafted by computing:
\begin{align}
    \zeta_3^*( \omega_1, \omega_2, \omega_3 ) = -\frac{\eta_0 \Bar \lambda^2}{6 a} \Big[ &R_3^*( \omega_1, \omega_2, \omega_3 ) + R_3^*( \omega_2, \omega_3, \omega_1 ) + R_3^*( \omega_3, \omega_1, \omega_2 ) \\
    & + R_3^*( \omega_3, \omega_2, \omega_1 ) + R_3^*( \omega_2, \omega_1, \omega_3 ) + R_3^*( \omega_1, \omega_3, \omega_2 ) \Big]. \nonumber
\end{align}
We will assume that the third-order transfer functions are always represented in their symmeterized form for the remainder of the work unless otherwise noted.

\subsection{Visualizing the third-order complex resistivity}\label{visualizing}

Here, we explore two methods for visualizing the third-order complex resistivity in the Johnson-Segalman and Giesekus models.  The first and simpler of the two visualization methods depicts the real and imaginary parts of $\zeta^{*}_{3}( n_1 \omega, n_2 \omega, n_3 \omega ) $ as a function of $ \omega $ for different integer triplets.  One of the major benefits of this visualization strategy is that it is very similar to strategies already ubiquitous in visualizing measurements of the complex modulus, $G^{*}(\omega)$, and complex viscosity, $\eta^{*}(\omega)$, in both traditional rheology and microrheology experiments. The familiarity should make it easier to compare the third-order complex resistivity arising from different models or even experimental measurements. 

\begin{figure}[t]
\begin{subfigure}{.495\textwidth}
  \centering
    \includegraphics[trim=3cm 8cm 4.5cm 8cm, clip, width = \linewidth]{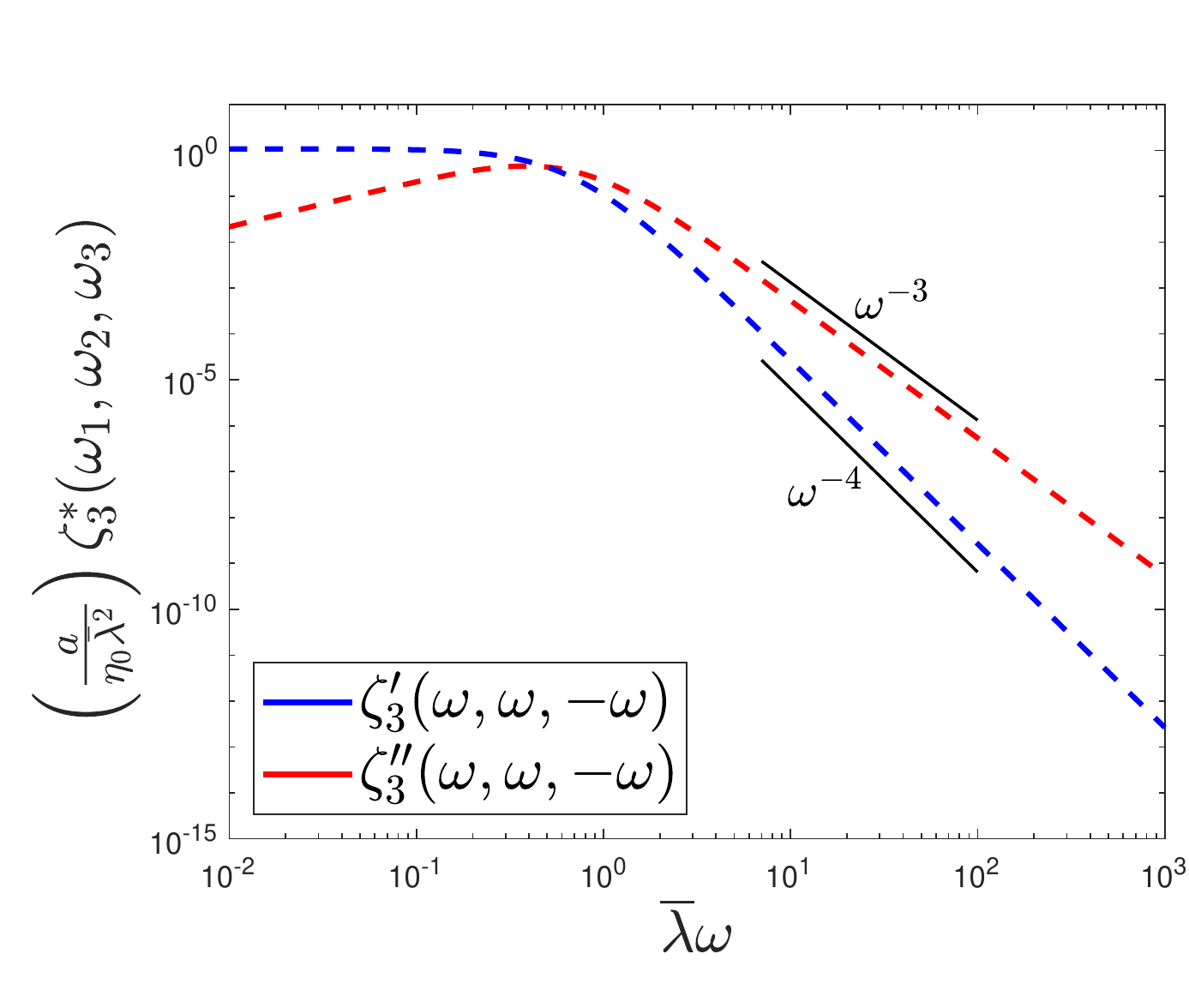} 
\end{subfigure}
\hfill
\begin{subfigure}{.495\textwidth}
  \centering
  \includegraphics[trim=3cm 8cm 4.5cm 8cm, clip, width = \linewidth]{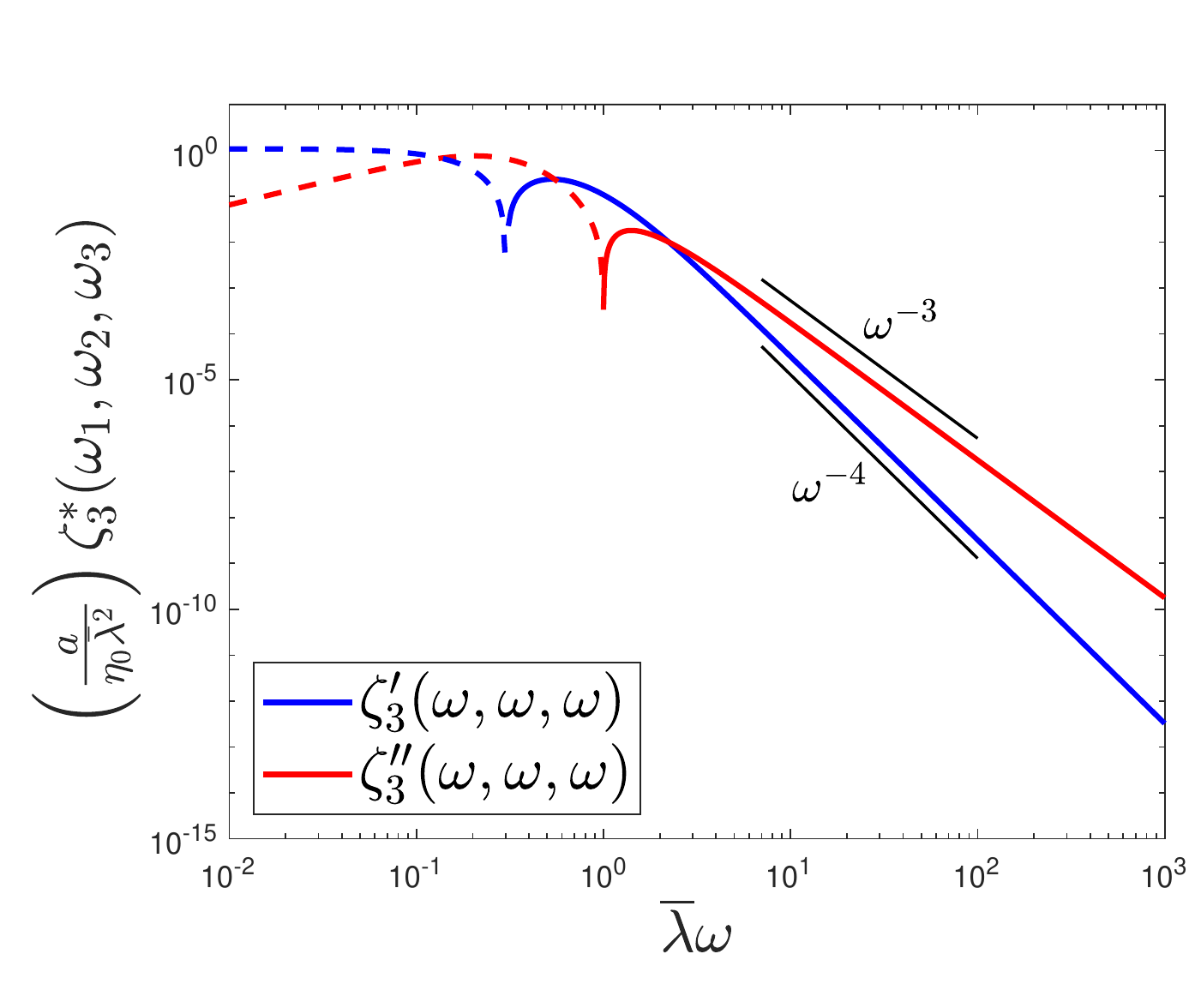} 
\end{subfigure}
\caption{Projections of $\zeta^{*}_{3}(\omega_{1}, \omega_{2}, \omega_{3})$ for a Johnson-Segalman fluid with $b = 0.5$ and $\beta = 0.5$, with the first harmonic response on the left and the third harmonic response on the right. Solid lines indicate positive values, and dashed lines indicate negative values.}
\label{fig:JSZetaProjections2}
\end{figure}

In the projections shown throughout this work, two sets of integer triplets $\{n_{1}, n_{2}, n_{3}\} $ will be shown: $\{1, 1, -1\} $ and $\{1, 1, 1\} $. The third-order resistivity along these coordinates corresponds to the nonlinearities excited by a lineal velocity that oscillates sinusoidally at a single frequency, $ \omega $.  These values could be measured, for example in a displacement controlled microrheology experiment in which the probe particle is made to oscillate at small displacement amplitude and with frequency $ \omega $.  The frequency would be swept across a range of values -- as it is in a typical `frequency sweep', linear response measurement -- to probe the nonlinear viscoelasticity on a variety of different time scales \cite{Gupta2019}.  As in a medium amplitude oscillatory shear (MAOS) experiment, the third-order resistivities, $ \zeta_3^*( \omega, \omega, -\omega ) $ and $ \zeta_3^*( \omega, \omega, \omega ) $, would correlate with the nonlinearities in the Fourier transformation of the force at frequencies $ \omega $ and $ 3 \omega $ or the first and third harmonic, respectively \cite{Ewoldt2013}.  Figure \ref{fig:JSZetaProjections2} depicts these third-order resistivities for a particular Johnson-Segalman fluid with a single relaxation time: $ P( \lambda ) = \delta( \lambda - \Bar \lambda ) $ and a modest value of the slip parameter, $ b = 0.5 $.  

\begin{figure}[p]
\begin{subfigure}[t]{.495\textwidth}
  \centering
  \includegraphics[trim=3cm 10cm 4.5cm 8cm, clip, width = \linewidth]{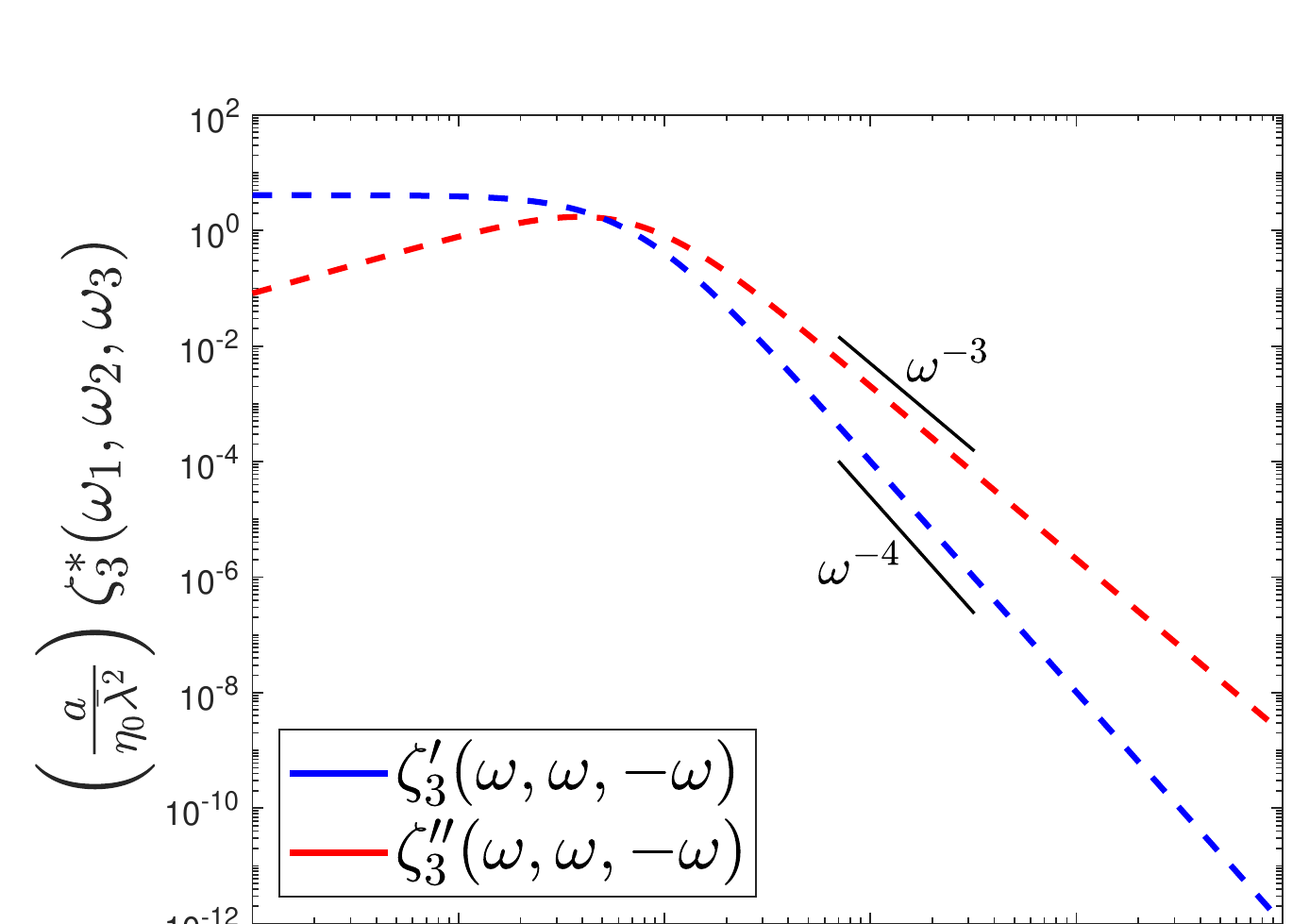}
\end{subfigure}
\hfill
\begin{subfigure}[t]{.495\textwidth}
  \centering
  \includegraphics[trim=3cm 10cm 4.5cm 8cm, clip, width = \linewidth]{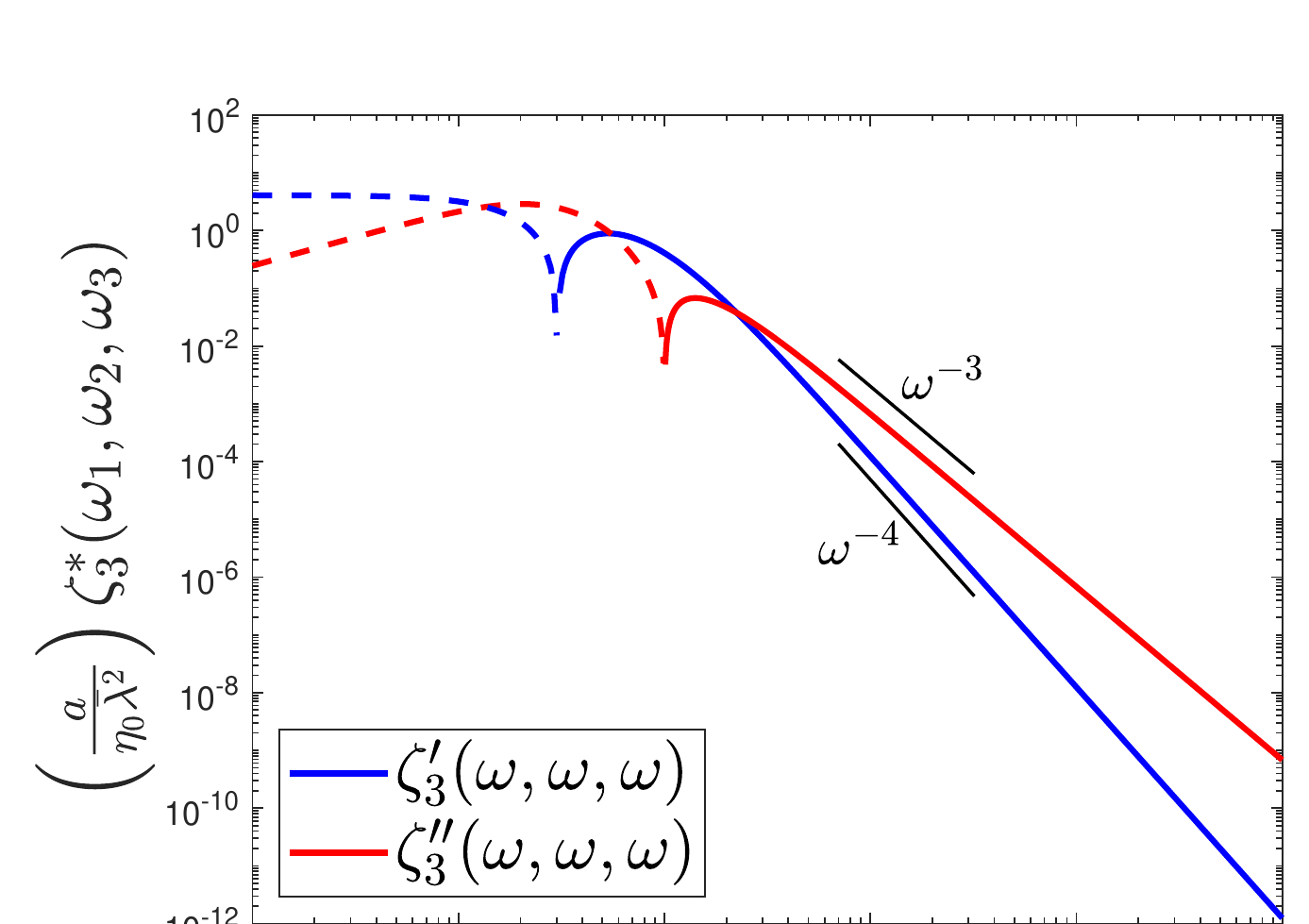}
\end{subfigure}
\begin{subfigure}[t]{.495\textwidth}
  \centering
  \includegraphics[trim=3cm 10cm 4.5cm 8cm, clip, width = \linewidth]{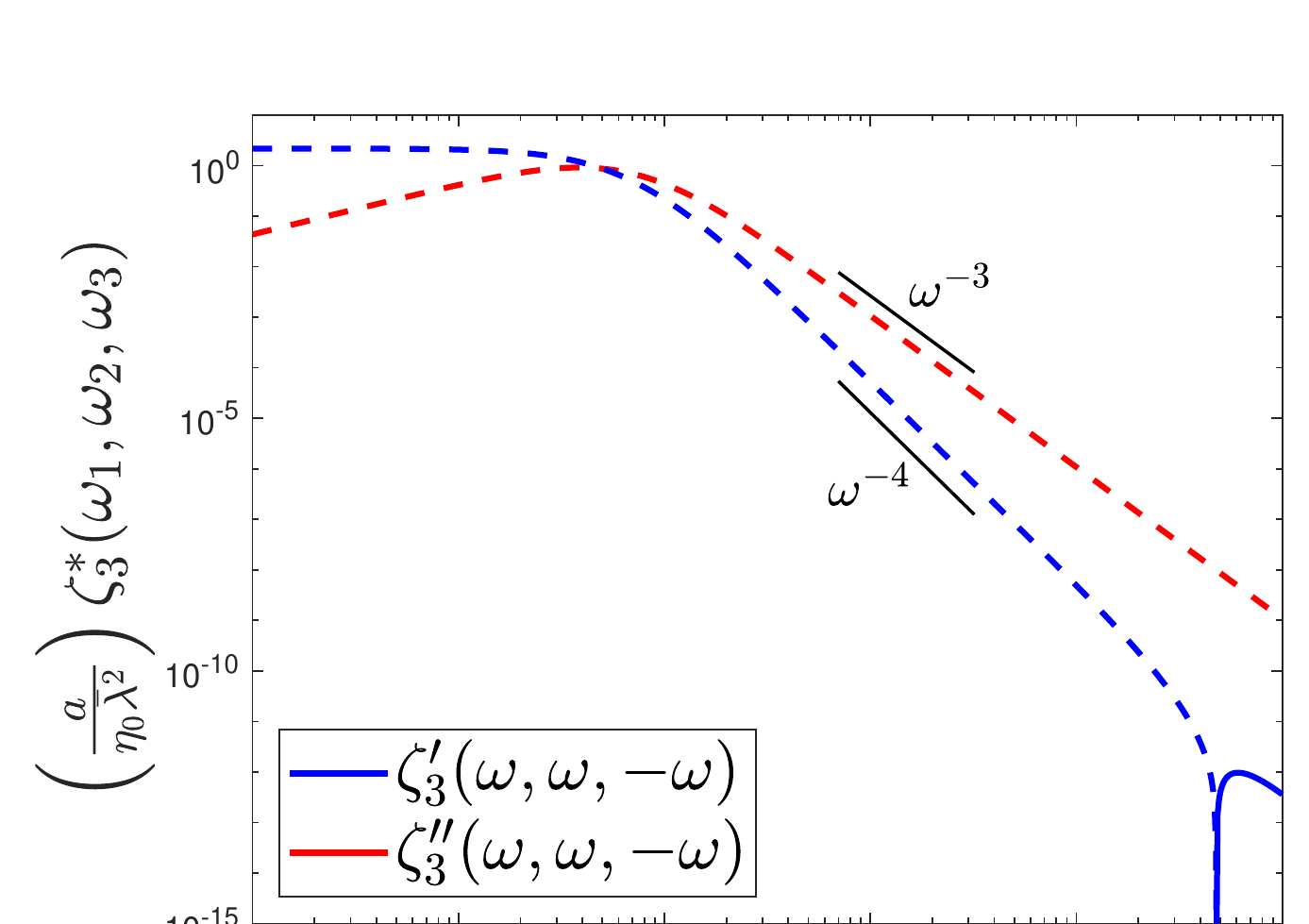}
\end{subfigure}
\begin{subfigure}[t]{.495\textwidth}
  \centering
  \includegraphics[trim=3cm 10cm 4.5cm 8cm, clip, width = \linewidth]{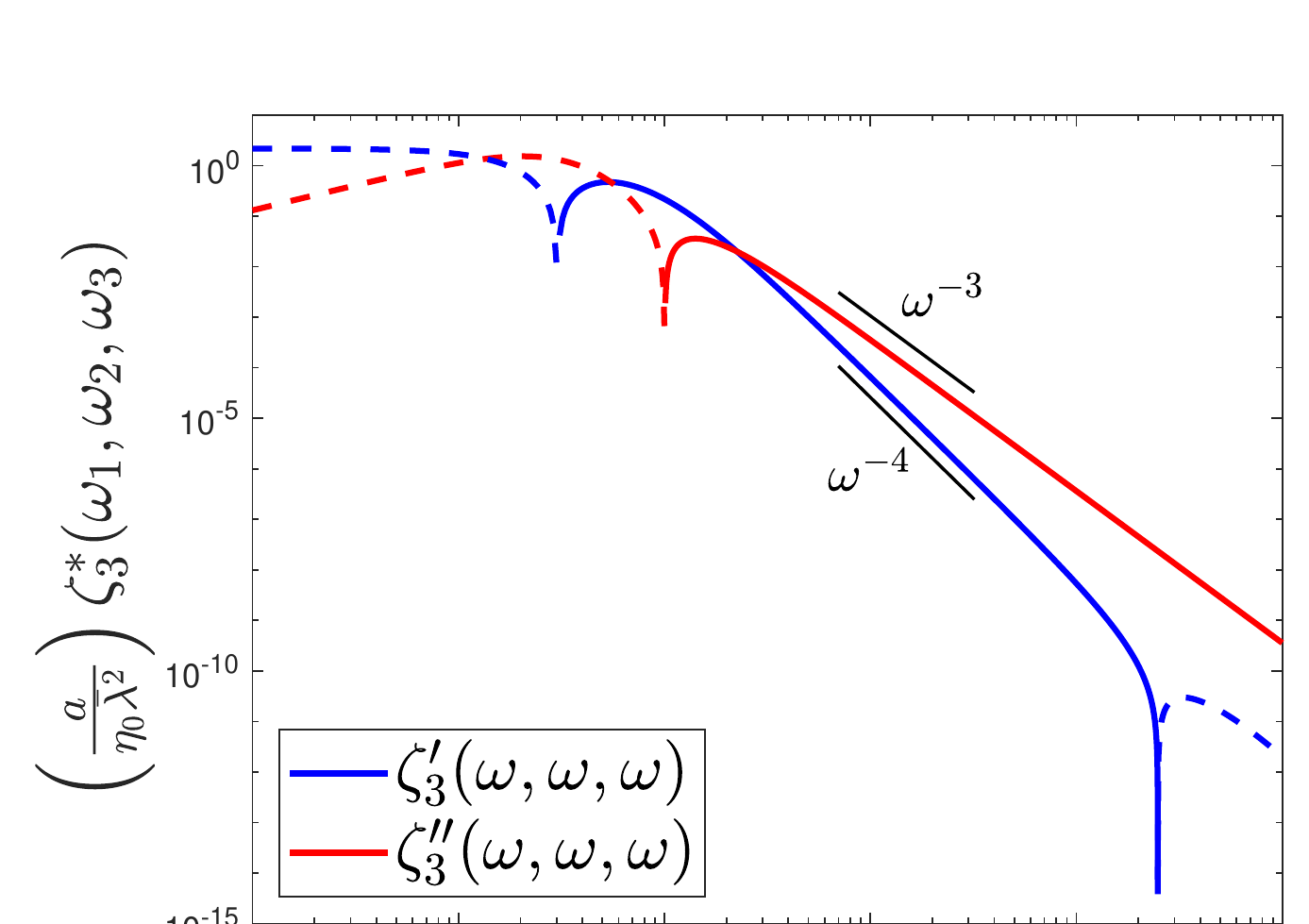}
\end{subfigure}
\begin{subfigure}[t]{.495\textwidth}
  \centering
  \includegraphics[trim=3cm 8cm 4.5cm 8cm, clip, width = \linewidth]{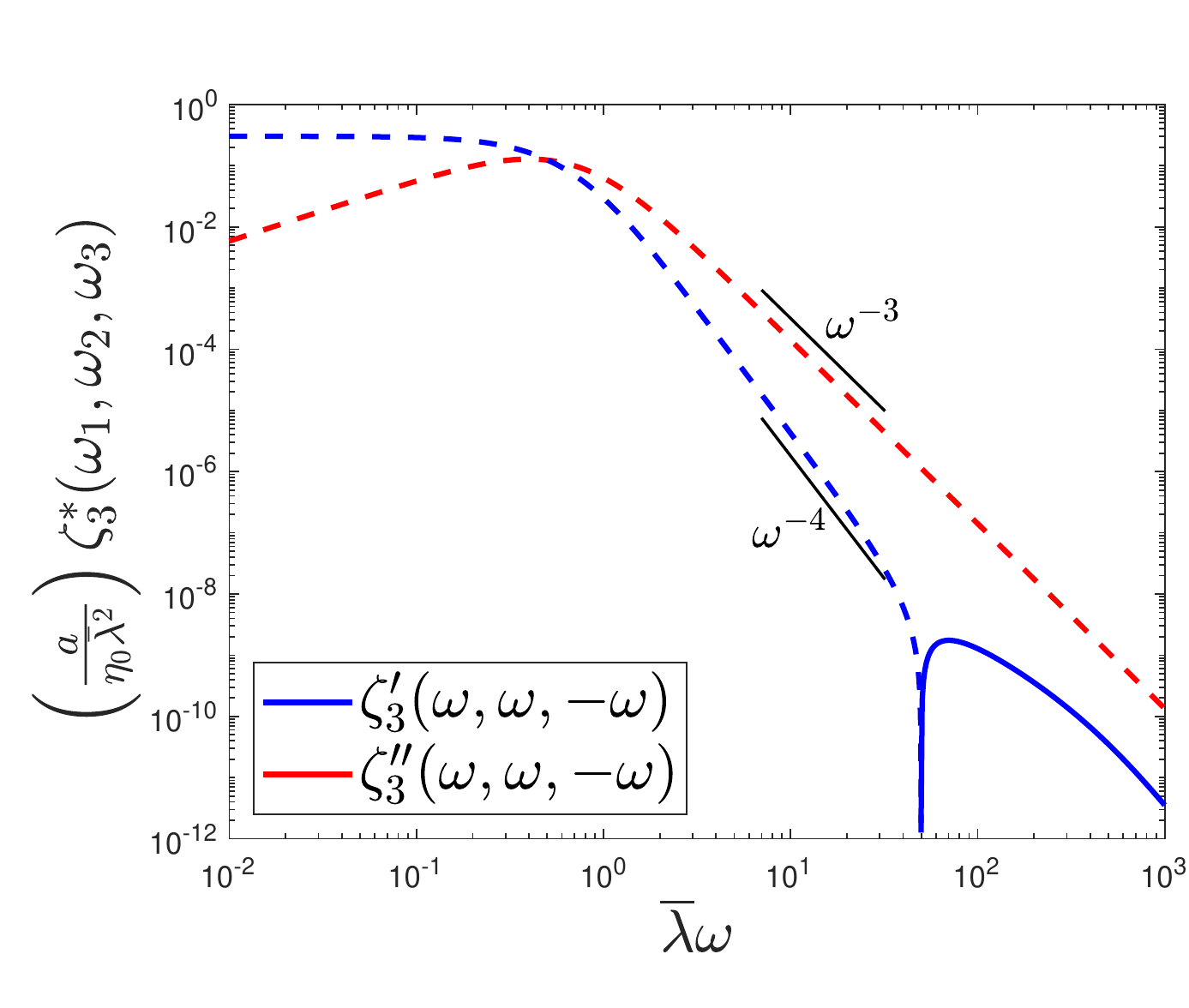} 
\end{subfigure}
\begin{subfigure}[t]{.495\textwidth}
  \centering
  \includegraphics[trim=3cm 8cm 4.5cm 8cm, clip, width = \linewidth]{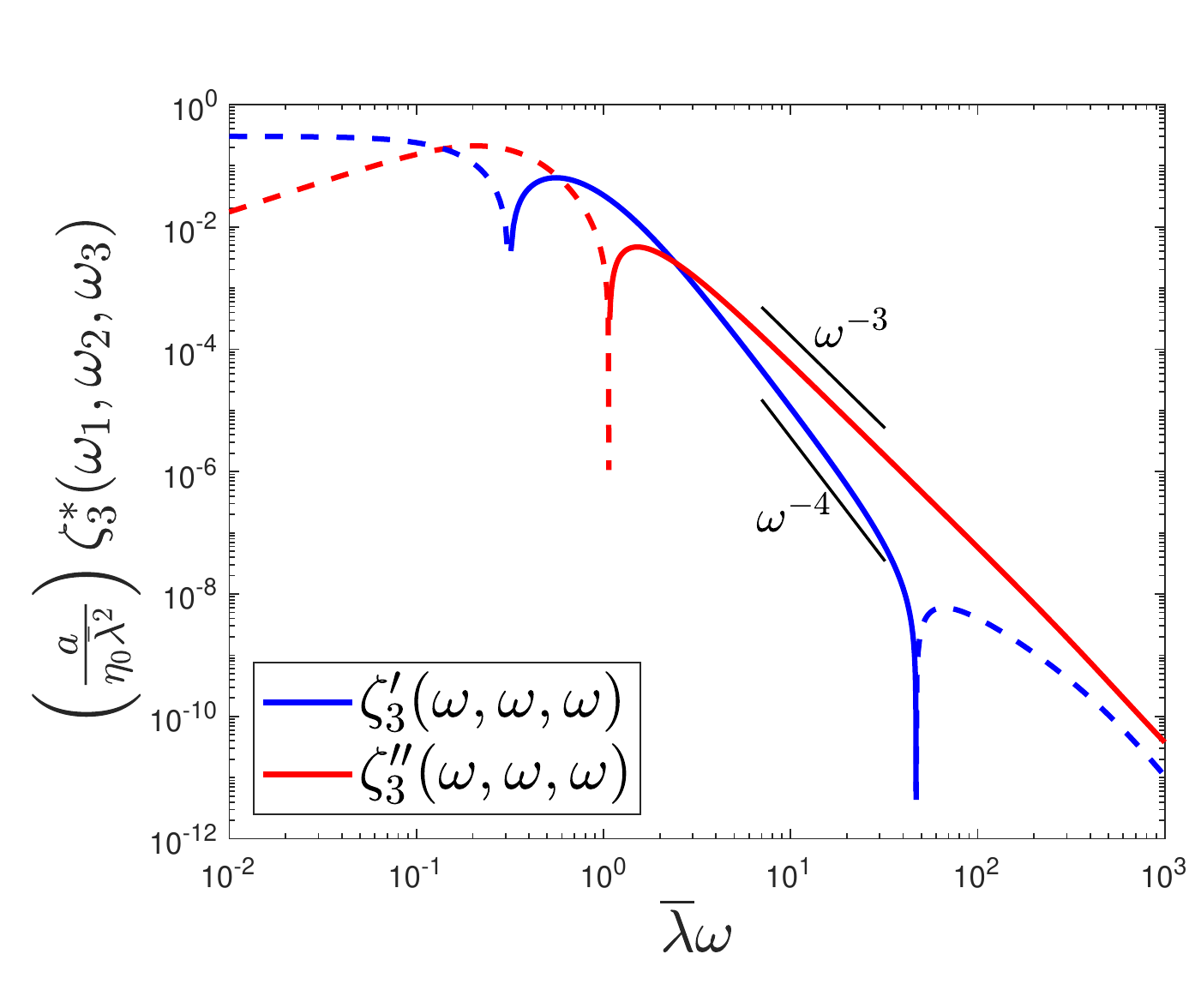}
\end{subfigure}

\caption{Projections of $\zeta^{*}_{3}(\omega_{1}, \omega_{2}, \omega_{3})$ for a Johnson-Segalman fluid. In all cases, $\beta = 10^{-3}$, while the slip parameter $b$, for the top, middle, and bottom rows respectively is: 0, 0.5, 1. The first harmonic response is on the left, and the third harmonic on the right; solid lines indicate positive values, and dashed lines indicate negative values.}
\label{fig:JSZetaProjections}
\end{figure}

The resulting curves are feature-rich just like those of intrinsic nonlinearities probed via simple shear rheometry \cite{Lennon2020a, Lennon2020}.  For instance, the third-order complex resistivity changes sign.  At low frequencies, the leading nonlinearities act to diminish the force exterted on the particle relative to the linear response. At these low frequencies, the response function is dominated by its real part, and thus describes a primarily viscous response and thinning of the fluid upon being deformed.  At higher frequencies, the sign of the response measured on the third harmonic of the force, $ \zeta_3^*( \omega, \omega, \omega ) $, changes, and the nonlinear response functions measured on both the first and third harmonic of the force are dominated by their imaginary parts.  This suggests the nonlinear response is more elastic on these time scales.  The nature of these features depends strongly on the adjustable parameters -- and even the structure -- of the model. For instance, Figure \ref{fig:JSZetaProjections} depicts the same projections of the third-order complex resistivity for a strongly coupled Johnson-Segalman fluid, $ \beta \ll 1 $, with a single relaxation time for different values of the slip parameter.

For the Johnson-Segalman model, the third-order complex resistivity depends in simple, but revealing ways on the slip parameter, $ b $, and on the ratio of the solvent to the zero shear viscosity, $ \beta $.  When $ b = 0 $, as for a single mode, corotational Maxwell fluid, $ \mathbf{C}_1 = \mathbf{C}_2 = 0 $, and the nonlinear response to a single tone oscillation is characterized by:
\begin{subequations}
\begin{equation}
    \zeta_3^*( \omega, \omega, \omega ) = \left( \frac{4662 \pi}{3575} \right)\frac{\eta_0 \Bar \lambda^2 ( 1 - \beta )}{a} \left( \frac{1}{1 + i \Bar \lambda \omega } \right) \left( \frac{1}{1 + 2 i \Bar \lambda \omega } \right) \left( \frac{1}{1 + 3 i \Bar \lambda \omega } \right), \label{eq:thirdharmcrm}
\end{equation}
\begin{equation}
    \zeta_3^*( \omega, \omega, -\omega ) = \left( \frac{4662 \pi}{3575} \right)\frac{\eta_0 \Bar \lambda^2 ( 1 - \beta )}{a} \left( \frac{1}{1 + 2 i \Bar \lambda \omega + ( \Bar \lambda \omega )^2 + 2 i ( \Bar \lambda \omega )^3 } \right). \label{eq:firstharmcrm}
\end{equation}
\end{subequations}
At high frequency, $ \omega \Bar \lambda \gg 1 $, these response functions are dominated by their imaginary parts, which have opposite signs and each decay as $ \omega ^{-3} $.  For most values of $ b $ and in the weak coupling limit for which $ \beta $ approaches $ 1 $ asymptotically, these transfer functions take on an almost identical form to that given by the corotational Maxwell model too.   In particular, for values of $ b $ smaller than $ 1727( 1 - \beta )/1813  $, and $ \beta \rightarrow 1 $, equations \ref{eq:thirdharmcrm} and \ref{eq:firstharmcrm} are good asymptotic approximations when the rational multiple of $ \pi $ in the pre-factor is replaced by $ 18 ( 1813 - 1727 b ) / 25025 $.  The error incurred in this approximation is of order $ ( 1 - \beta )^2 $.

In the strong coupling limit, $ \beta \rightarrow 0 $, there are no convenient simplification of third-order complex resistivity.  However, we can observe that because $ \eta^*( \omega ), \chi( \omega, \Bar \lambda ) \sim \omega^{-1} $ at high frequencies, we should expect that $ \zeta_3^*( \omega, \omega, \pm \omega ) $ will decay as $ \omega^{-3} $.  Figure \ref{fig:JSZetaProjections} shows just how feature rich the nonlinear transfer functions are in the strong coupling limit.  They can exhibit multiple sign changes whose existence and position depends sensitively on the value of the slip parameter.  This is also known from analysis of intrinsic nonlinearities of Johnson-Segalman fluids in simple shear flows \cite{Lennon2020, Ramlawi2020a}.

Similar analyses can be done for the Giesekus model, for which the full form of the third-order complex resistivity is shown in Appendix \ref{giesekus}. The single mode Giesekus constitutive model defines the polymeric contribution to the stress as:
\begin{multline}
    \bm{\tau}_{p} (\textbf{r}, t) + \De \frac{\partial \bm{\tau}_{p}}{\partial t} (\textbf{r}, t) + \\ \Wi\left( \alpha [\bm{\tau}_{p} (\textbf{r}, t) \cdot \bm{\tau}_{p} (\textbf{r}, t)] + \mathbf{v} (\textbf{r}, t) \cdot \nabla \bm{\tau}_{p}(\textbf{r}, t) - [ \bm{\tau}_{p} (\textbf{r}, t) \cdot \nabla \mathbf{v}(\textbf{r}, t) + \mathbf{v}^{T} (\textbf{r}, t) \cdot \bm{\tau}_{p} (\textbf{r}, t) ] \right) = 2\mathbf{e}(\textbf{r}, t).
\end{multline}

\begin{figure}[t!]
\begin{subfigure}{.495\textwidth}
  \centering
  \includegraphics[trim=3cm 9.85cm 4.5cm 8cm, clip, width = \linewidth]{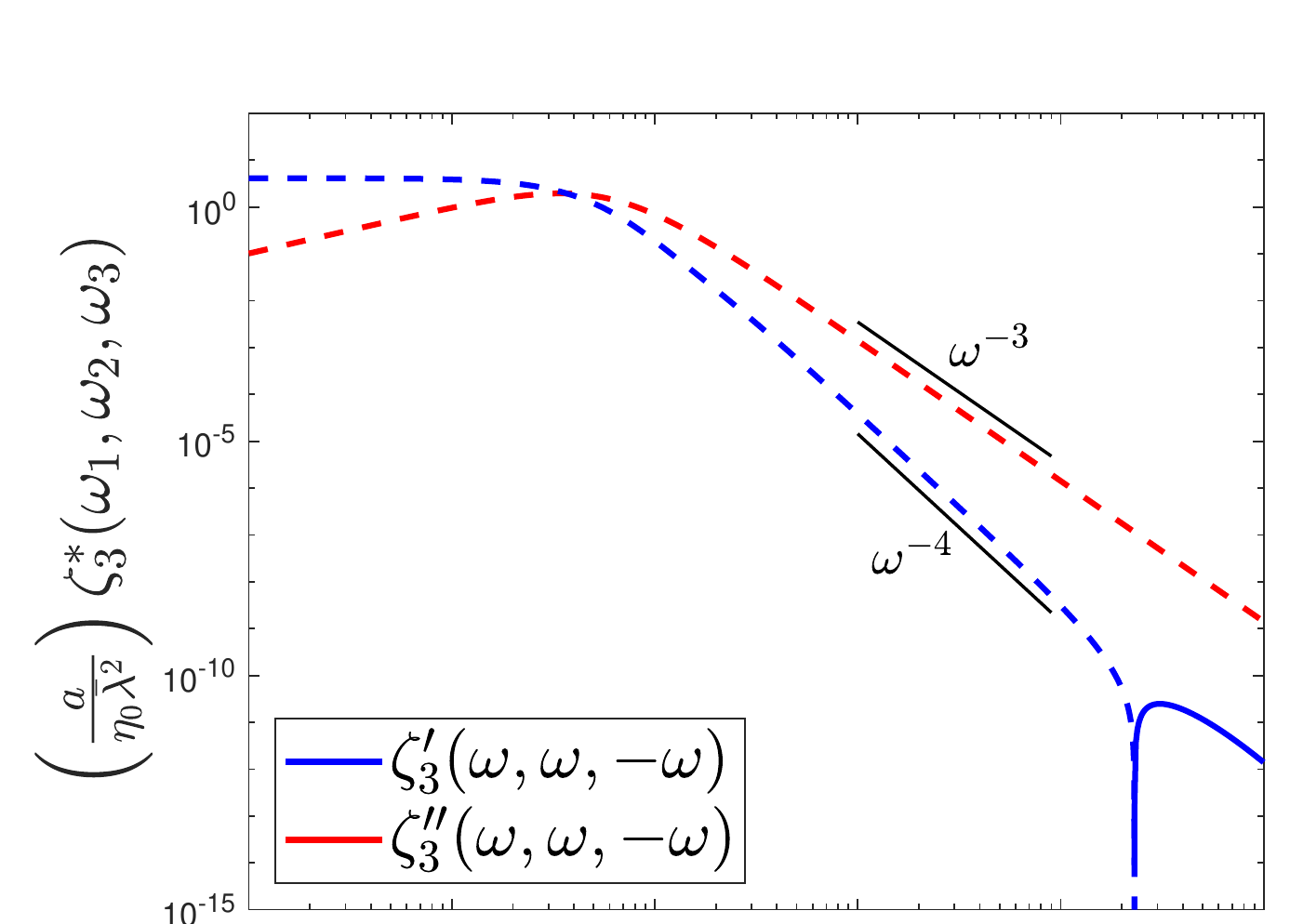}
\end{subfigure}
\hfill
\begin{subfigure}{.495\textwidth}
  \centering
 \includegraphics[trim=3cm 9.85cm 4.5cm 8cm, clip, width = \linewidth]{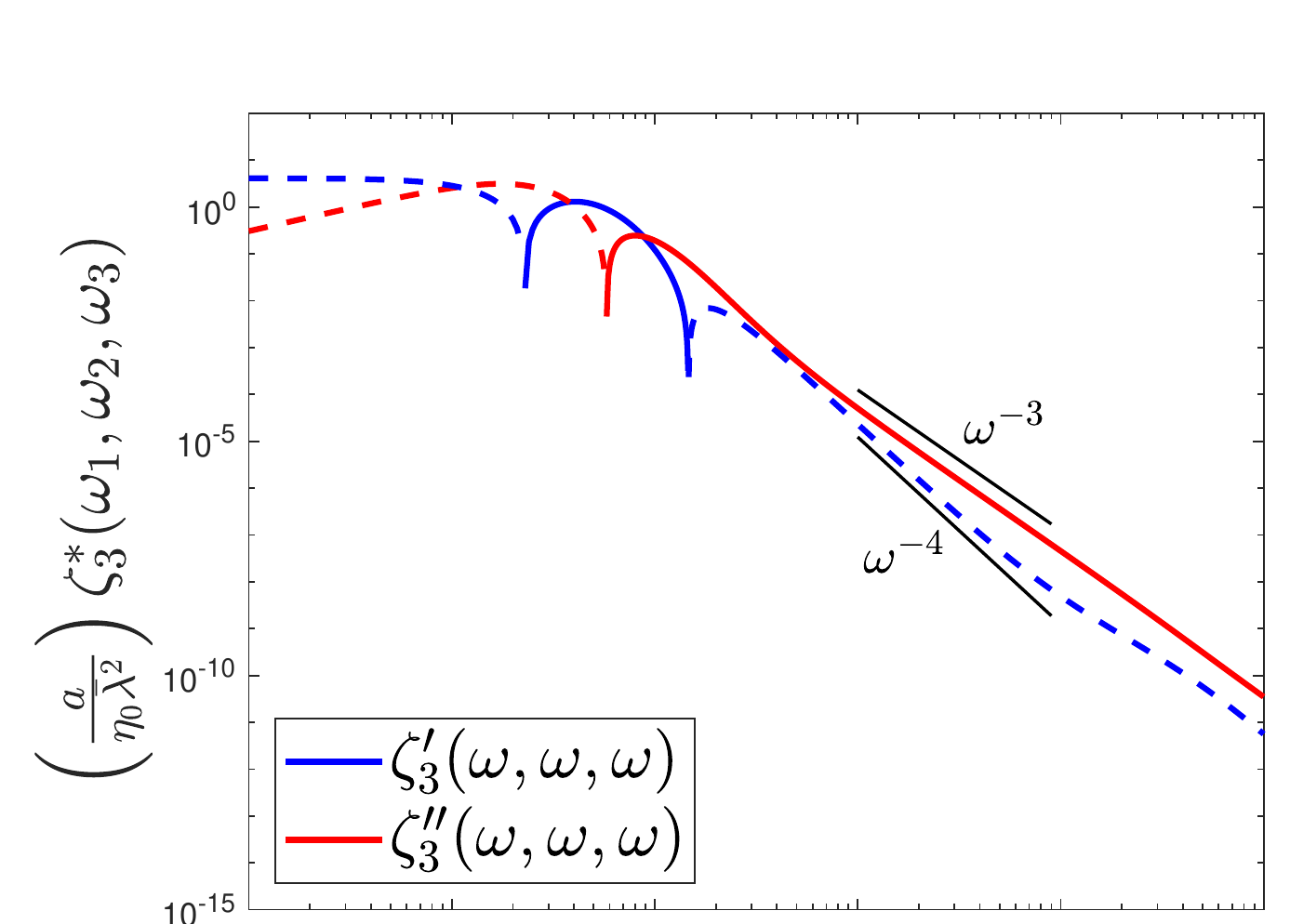}
\end{subfigure}
\begin{subfigure}{.495\textwidth}
  \centering
  \includegraphics[trim=3cm 8cm 4.5cm 8cm, clip, width = \linewidth]{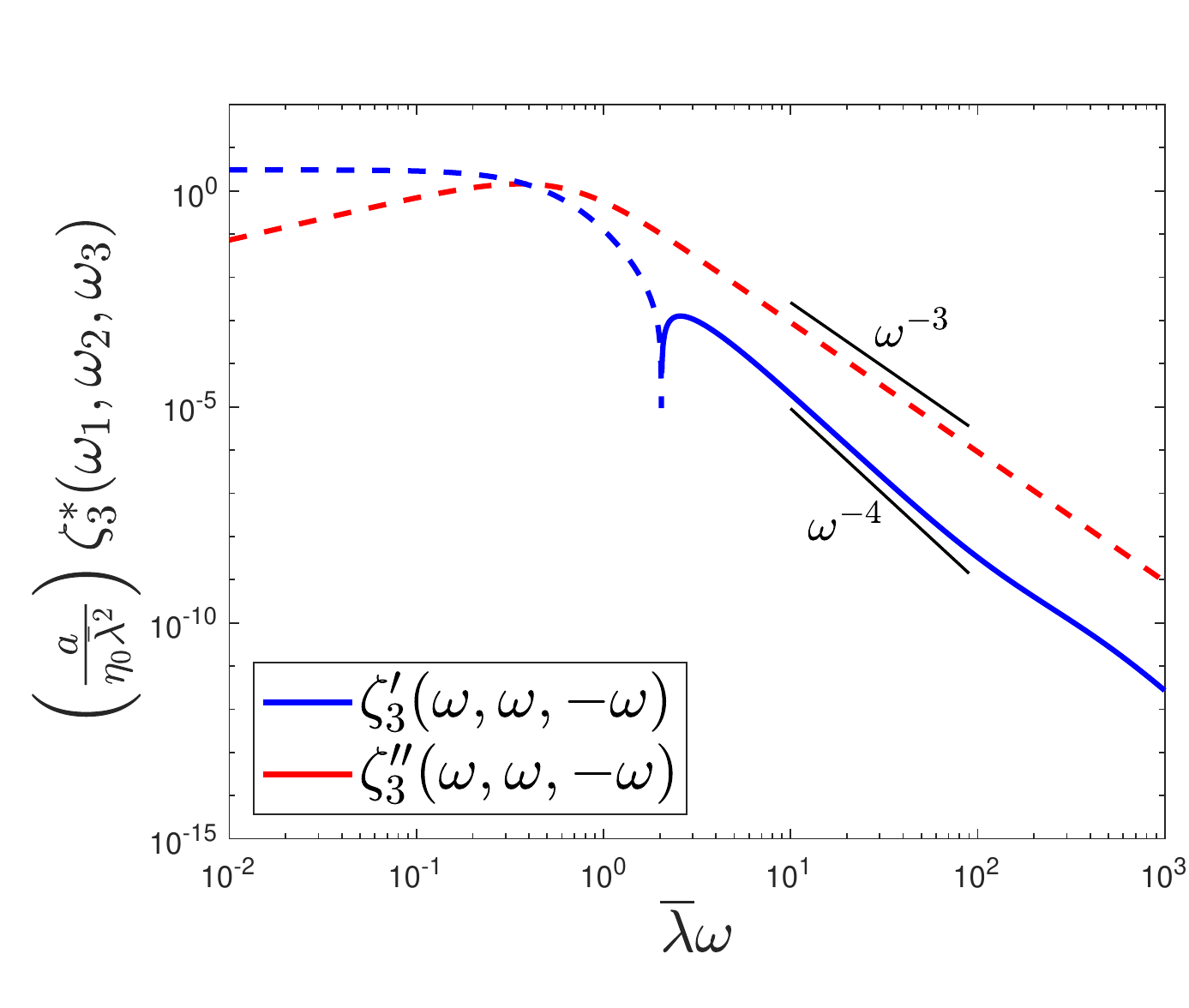}
\end{subfigure}
\hfill
\begin{subfigure}{.495\textwidth}
  \centering
  \includegraphics[trim=3cm 8cm 4.5cm 8cm, clip, width = \linewidth]{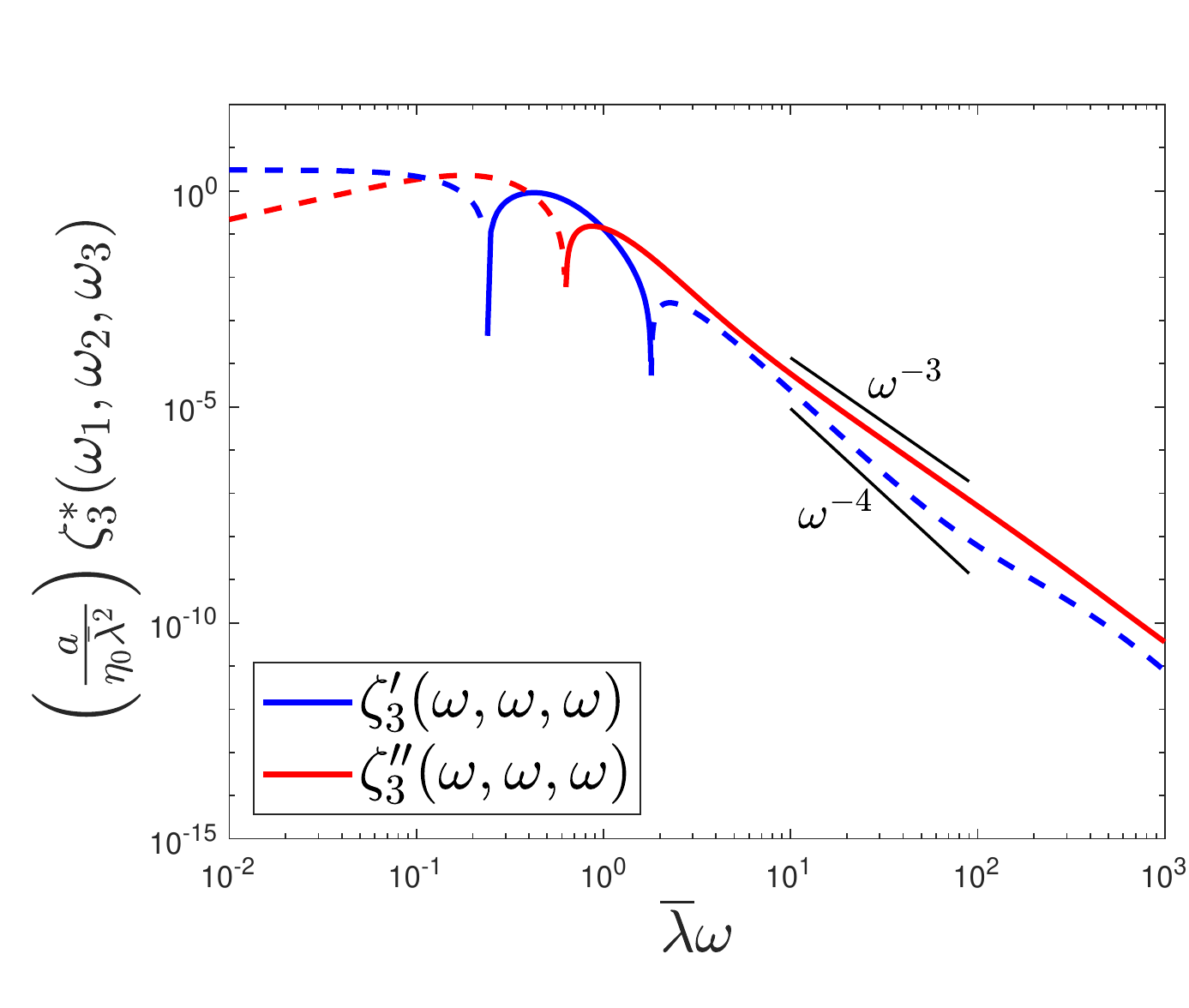}
\end{subfigure}
\caption{Projections of $\zeta^{*}_{3}(\omega_{1}, \omega_{2}, \omega_{3})$ for a single mode Giesekus fluid with $\beta = 10^{-3}$ and  $\alpha = 0.5$ (top row), $\alpha = 0.3$ (bottom row). The first harmonic response is on the left, and the third harmonic on the right; solid lines indicate positive values, and dashed lines indicate negative values.}
\label{fig:GiesekusZetaProjections}
\end{figure}
It contains the adjustable ``mobility factor'' $\alpha$, which scales the quadratic nonlinearity in the stress and is associated with anisotropic hydrodynamic drag on polymer molecules \cite{Bird1987}. When $\alpha = 0$, the Oldroyd-B model is recovered.

By the same logic used in examining the results of calculations with the Johnson-Segalman model, it is expected that the third-order resistivity of a single-mode Giesekus fluid would decay at high frequencies, $ \omega \Bar \lambda \gg 1 $, as $\omega^{-3}$ due to the presence of a dominant $\alpha$-independent term that scales as $\omega^{-3}$. Projections of the third-order complex resistivity for a Giesekus fluid, like those in Figure \ref{fig:GiesekusZetaProjections}, show a variety of other features that can be used to make inferences about the fluid behavior. For example, at $\alpha = 0.3$ and $\alpha = 0.5$, the third harmonic measurement, $\zeta^{*}_3(\omega,\omega,\omega)$, features two sign changes in its real part. Over that distinct range of moderate frequencies, $\zeta_3'(\omega,\omega,\omega)$ is positive, while it is negative at both low frequencies and higher frequencies. The real part of the first harmonic measurement, $\zeta_3'(\omega, \omega, -\omega)$, also features a sign change for both values of $\alpha$, though it occurs at significantly higher frequency for $\alpha = 0.5$. Both the first and third harmonics are dominated by the real part at low frequency and the imaginary part at high frequency.  This is an indication of a transition from viscous to elastic nonlinearities when the variation in the deformation rate occurs on shorter time scales. These changes in response are remarkably sensitive to the value of $\alpha$.

While it is clear that these projections of the third-order complex resistivity can provide insights into distinguishing features of different fluid responses, they do not represent the full wealth of information captured in the entire transfer function. We introduce a second and somewhat more complicated visualization strategy for understanding $\zeta^{*}_{3}(\omega_{1}, \omega_{2}, \omega_{3})$, which is intended to more fully capture the data-rich nature of this function by eliminating the constraint of only showing $\zeta^{*}_3(n_1 \omega, n_2 \omega, n_3 \omega)$ over a range of characteristic frequencies $\omega$.

\begin{figure}[b]
\centering
\includegraphics[width=0.75\linewidth]{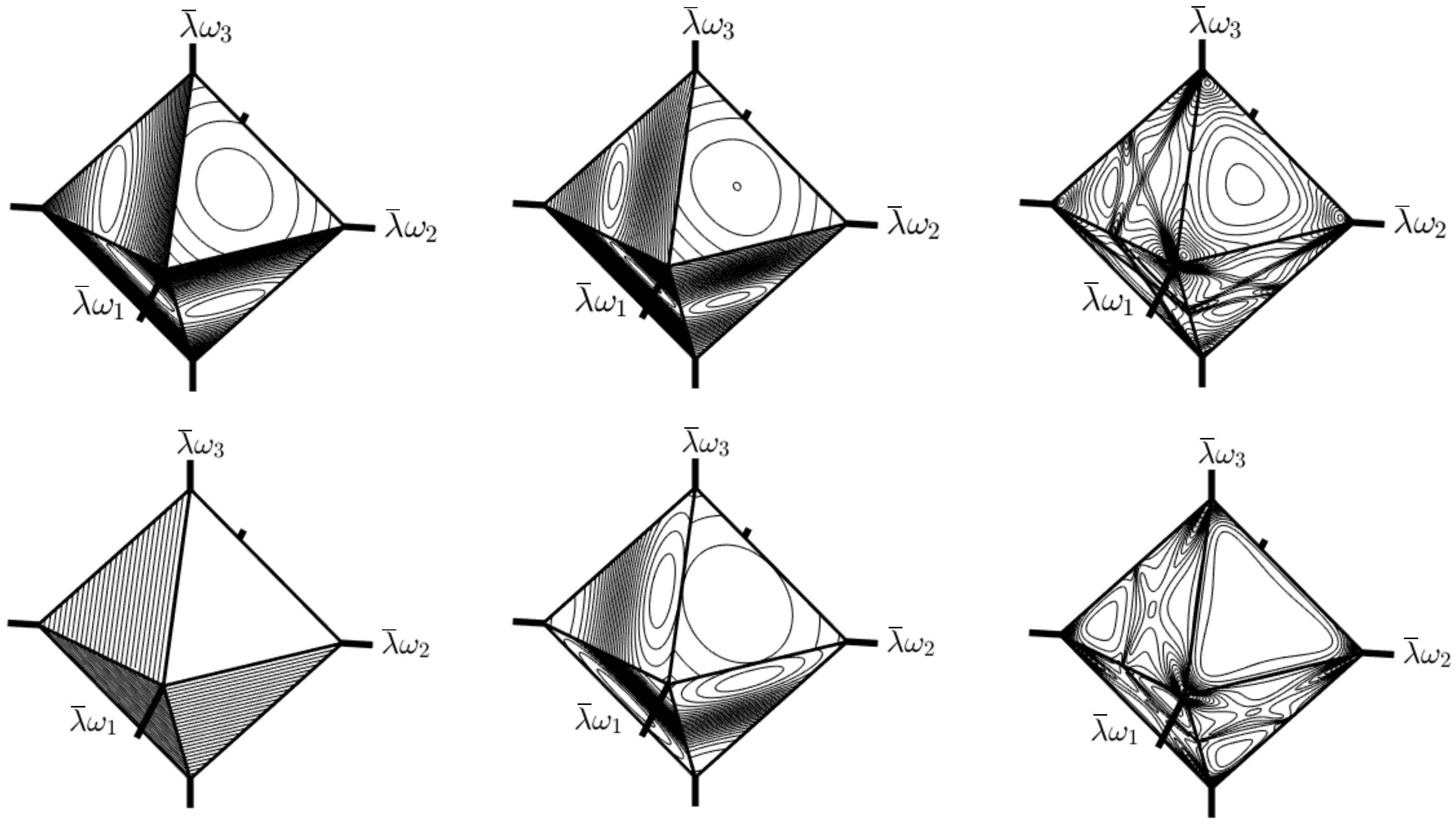}
\caption{Contours of $\zeta^{*}_{3}(\omega_{1}, \omega_{2}, \omega_{3})$ for one particular Johnson-Segalman fluid shown in three dimensions on different constant 1-norm surfaces. The top row shows  $\zeta'_{3}(\omega_{1}, \omega_{2}, \omega_{3})$, and the bottom row shows  $\zeta''_{3}(\omega_{1}, \omega_{2}, \omega_{3})$. In both rows, from left to right $\Bar{\lambda} \left\Vert\bm{\omega} \right\Vert _{1} = 0.1,1,10$.}
\label{fig:full_octos}
\end{figure}

The third-order complex resistivity exists in a three-dimensional frequency space characterized by $(\omega_{1}, \omega_{2}, \omega_{3})$. We will examine a constant $L^{1}$-norm subspace of the three-dimensional frequency space, with the $ L^1$-norm given by:
\[ \left\Vert  \bm{\omega} \right\Vert_{1} = |\omega_{1}| + |\omega_{2}| + |\omega_{3}|\]
 In the three frequency space, a constant $ L^1 $-norm surface is a regular octahedron, as shown in Figure \ref{fig:full_octos}. The lines on the surface of the different octahedra are the contours of the real (top row) and imaginary (bottom row) parts of $\zeta^{*}_{3}(\omega_{1}, \omega_{2}, \omega_{3})$ on the surfaces given by a particular value of $\Bar{\lambda} \left\Vert\bm{\omega} \right\Vert _{1}$.

These complete representations of the third-order complex resistivity on the constant $L^{1}$-norm surface show how this function behaves and changes throughout the entire frequency space. However, due to the symmetries outlined in Section \ref{volterra}, Figure \ref{fig:full_octos} also contains redundant information. We can take advantage of these symmetries to simplify visualization of the contours of $\zeta^{*}_{3}(\omega_{1}, \omega_{2}, \omega_{3})$. Large portions of the following method for doing so were initially developed by Lennon, et al. \cite{Lennon2020} for visualization of the third-order complex modulus used in Medium-Amplitude Parallel Superposition (MAPS) rheology. For the purpose of brevity, only the essential parts of this visualization strategy will be described here.

\begin{figure}[t]
  \centering
  \includegraphics[trim=4cm 6cm 4cm 4cm, clip, width=\linewidth]{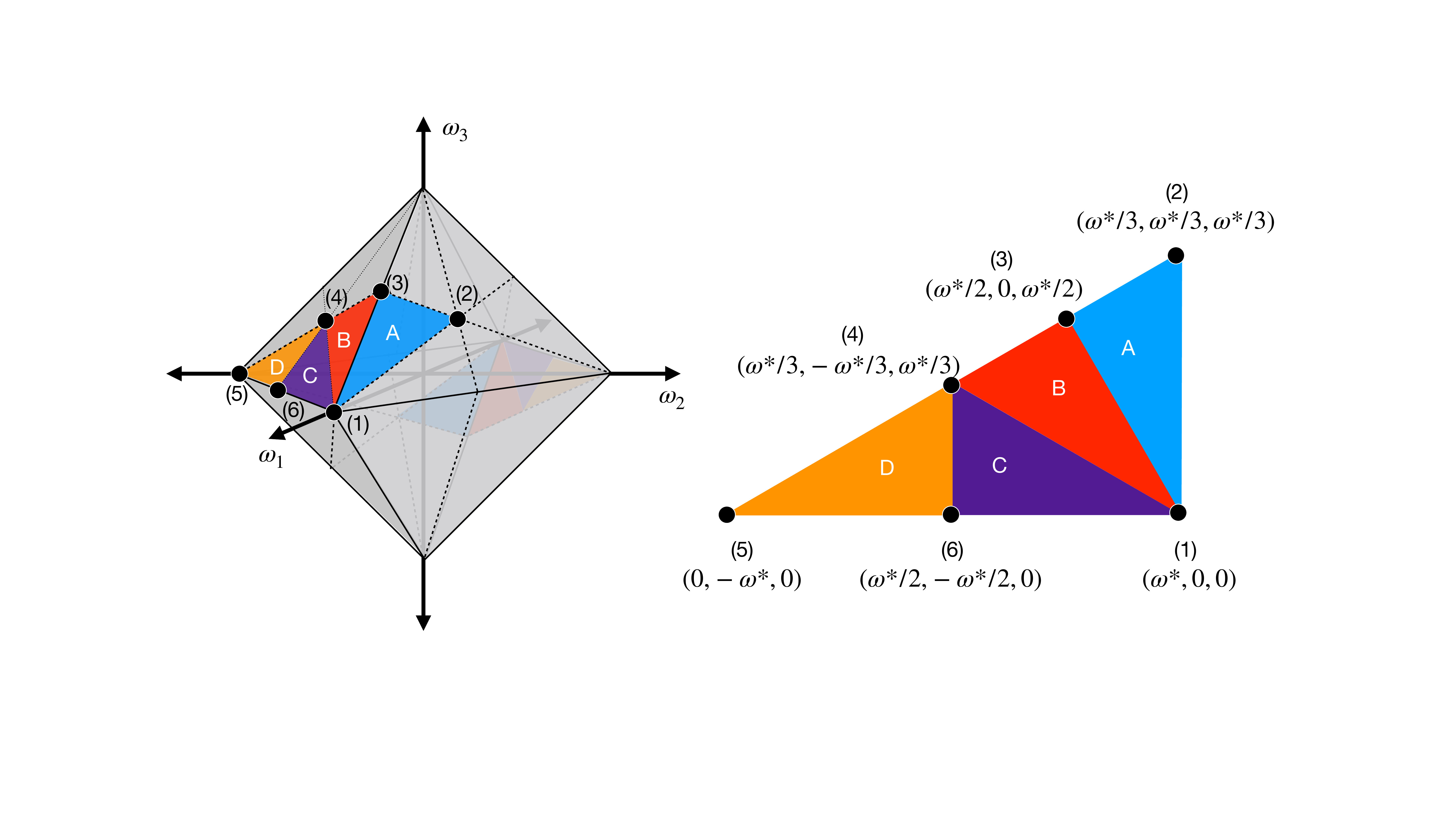}
\caption{Subregions of the constant $L^{1}$-norm space ($ |\omega_1| + |\omega_2| + |\omega_3| = \omega^* $) used to fully define  $\zeta^{*}_{3}(\omega_{1}, \omega_{2}, \omega_{3})$. Reproduced from \cite{Lennon2020} and modified with permission.}
\label{fig:MAPS_spaces}
\end{figure}

Permutation symmetry divides the octahedral surface into 12 subregions, indicated by the dashed lines in Figure \ref{fig:MAPS_spaces}. Of these 12, only two can be specified uniquely due to permutation symmetry; these are colored at the front and back of the octahedron. However, these two regions are directly related by Hermitian symmetry. So, the behavior of the third-order complex resistivity can be fully understood from its behavior in the region highlighted in the front of the octahedron in Figure \ref{fig:MAPS_spaces}. 

In Figure \ref{fig:MAPS_spaces}, this triangular subregion that fully defines the behavior of the third-order complex resistivity is lifted from the surface of the octahedron and projected flat into a two dimensional space. This region can be further divided into four subspaces, geometrically defined by the boundaries: 
\begin{subequations}
\begin{align}
    &A: \quad \omega_{1} \geq \omega_{2} \geq \omega_{3}, \quad \,\,\,\,\, \omega_{1}, \omega_{2}, \omega_{3} \geq 0, \\
    &B: \quad \omega_{1} \geq \omega_{3} \geq -\omega_{2}, \quad \omega_{1}, \omega_{3} \geq 0 \geq \omega_{2} ,\\
    &C: \quad \omega_{1} \geq -\omega_{2} \geq \omega_{3}, \quad \omega_{1}, \omega_{3} \geq 0 \geq \omega_{2},  \\
    &D: \quad -\omega_{2} \geq \omega_{1} \geq \omega_{3}, \quad \omega_{1}, \omega_{3} \geq 0 \geq \omega_{2} .
\end{align}
\label{eq: subspaces}
\end{subequations}

Through proper application of the symmetries previously detailed, any point in 3D frequency space can be associated with a point that satisfies one of 4 inequalities in Equation \ref{eq: subspaces}, thus allowing the behavior of $\zeta^{*}_3 (\omega_1, \omega_2, \omega_3)$ to be depicted and described in one of these four subregions.

In MAPS rheology, the specific vertices marked in Figure \ref{fig:MAPS_spaces} are related to common flow protocols used to make weakly nonlinear rheological measurements with traditional rheology equipment (medium-amplitude oscillatory shear (MAOS) and parallel superposition rheology (PS)). In microrheology experiments, the typical imposed flow protocol is a single-tone oscillation. If extended into the nonlinear regime by imposing medium-amplitude single-tone oscillation, this would analogous to the MAOS single-tone flow protocol. However, microrheology experiments do not have to be limited to measurement of the third-order complex resistivity at vertices on the periphery of this triangle. Exploration of the interior of this domain could carry a wealth of information about material behavior, which might be used to improve model or parameter identification or predict behavior of such fluids in different flow protocols.

The highlighted area shown in Figure \ref{fig:MAPS_spaces} can be ``lifted'' from the surfaces shown in figure \ref{fig:full_octos}, to depict the contours of the third-order complex resistivity in two dimensions without loss of non-redundant information, due to the previously stated symmetries. This can be done for different values of the frequency $L^{1}$-norm and for a variety of constitutive models if the analytical expression for $\zeta^{*}_{3}(\omega_{1}, \omega_{2}, \omega_{3})$ is known. 

\begin{figure}[t]
\begin{subfigure}{.495\textwidth}
  \centering
  \includegraphics[trim=4cm 10.75cm 4cm 9.6cm, clip, width = \linewidth]{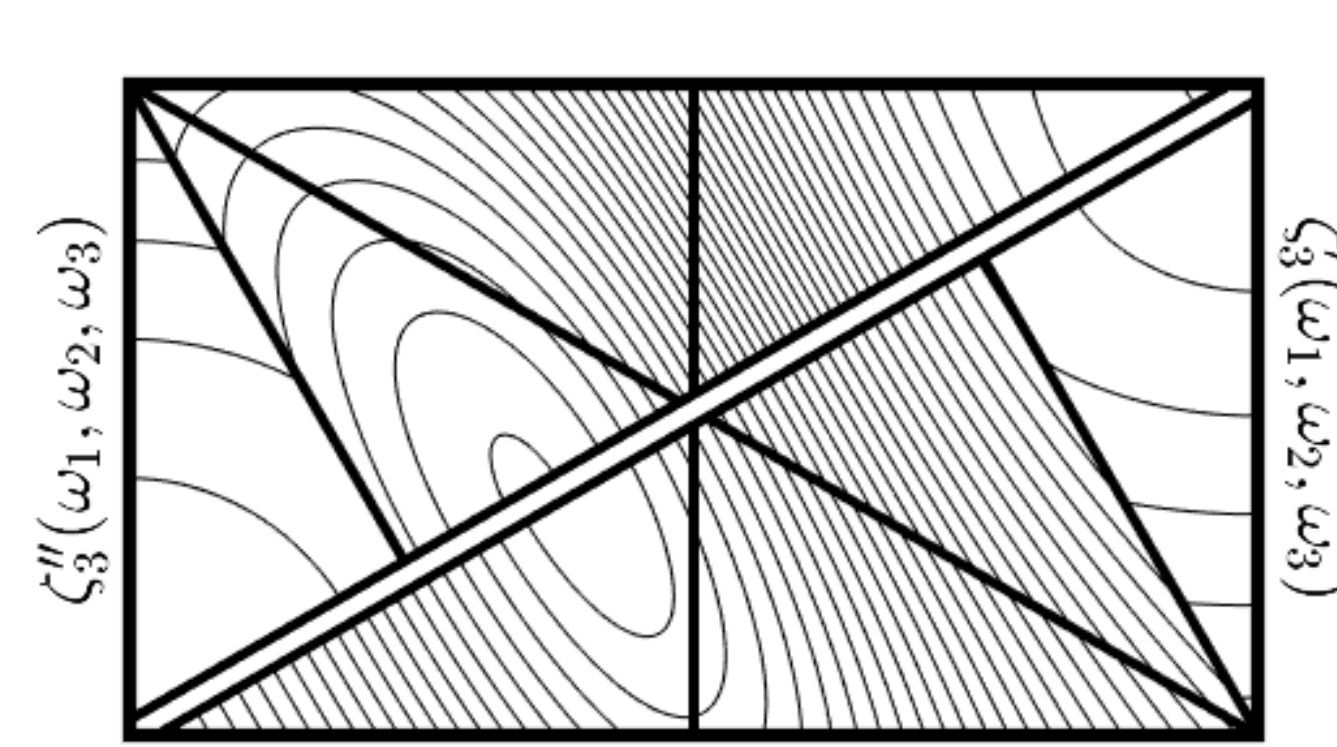}
\end{subfigure}
\hfill
\begin{subfigure}{.495\textwidth}
  \centering
  \includegraphics[trim=4cm 10.75cm 4cm 9.6cm, clip, width = \linewidth]{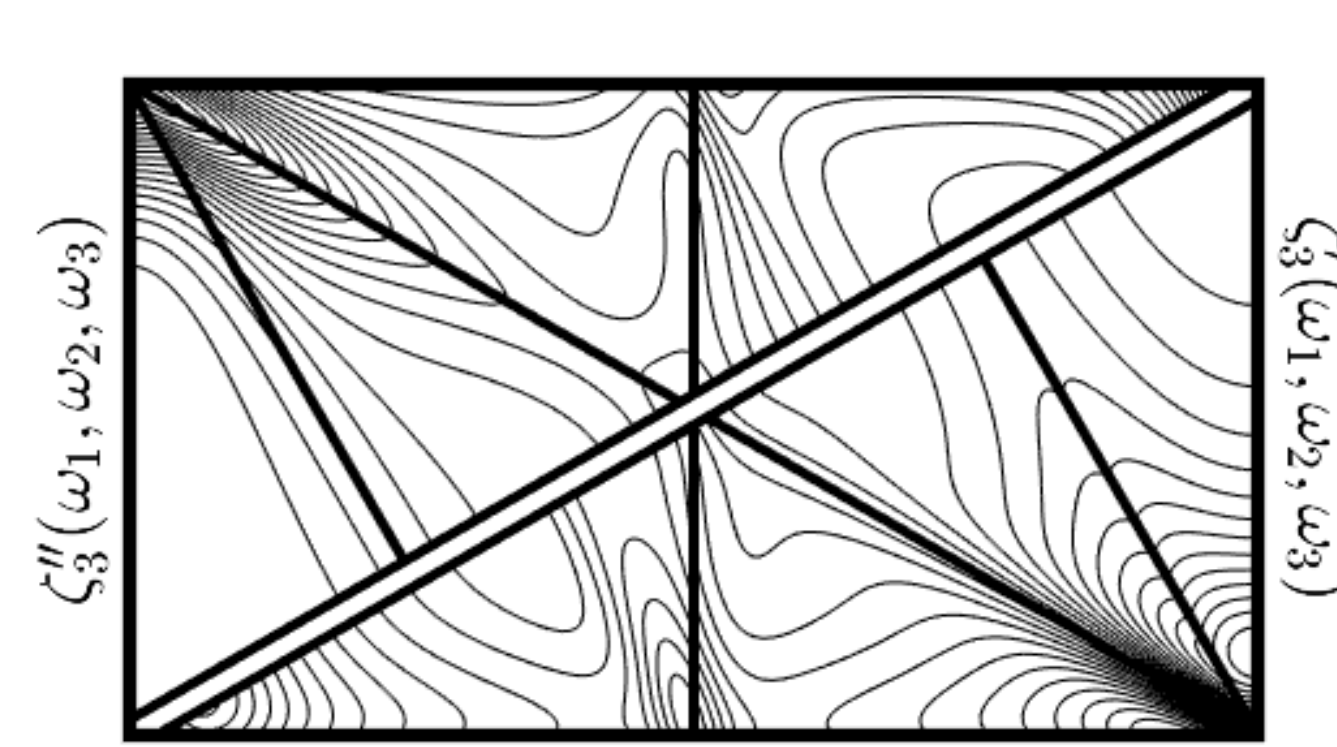} 
\end{subfigure}
\begin{subfigure}{.495\textwidth}
  \centering
  \includegraphics[trim=4cm 10.75cm 4cm 9.6cm, clip, width = \linewidth]{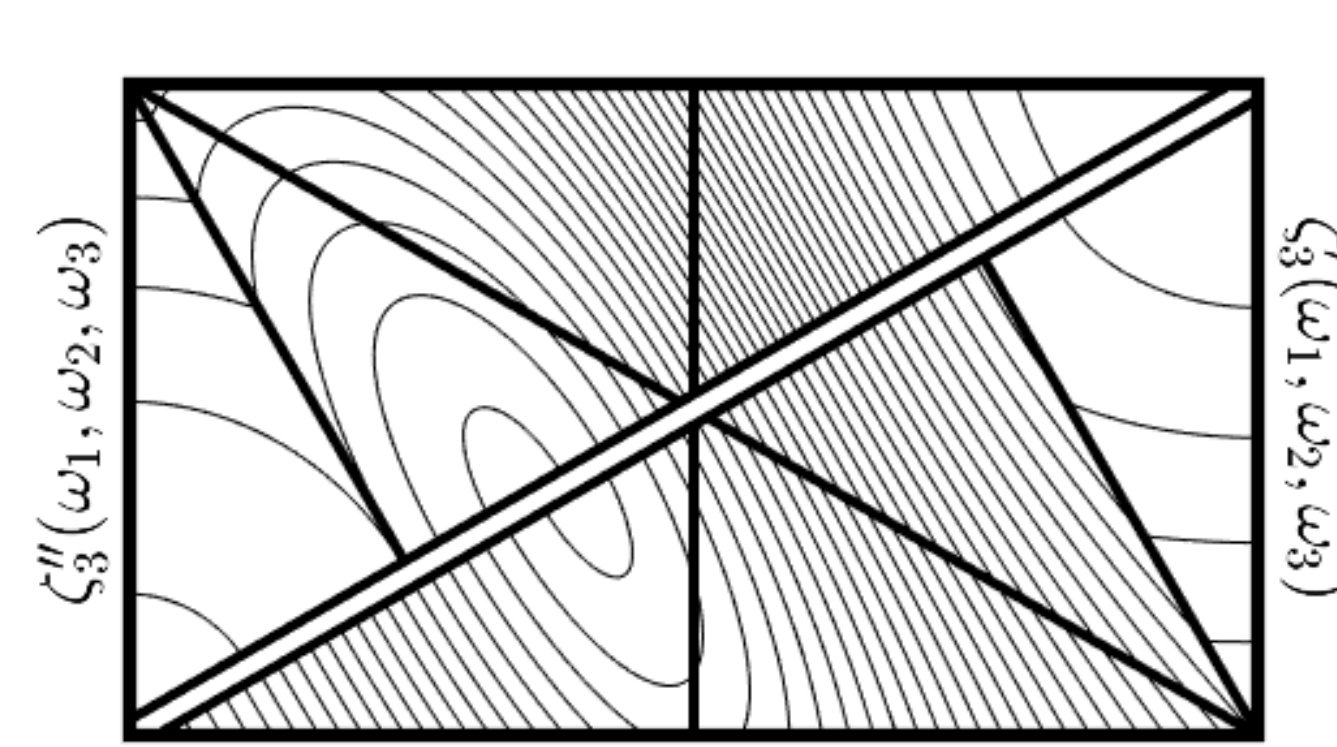}
  \caption{$\Bar{\lambda} \left\Vert\bm{\omega} \right\Vert _{1} = 1$}
\end{subfigure}
\hfill
\begin{subfigure}{.495\textwidth}
  \centering
  \includegraphics[trim=4cm 10.75cm 4cm 9.6cm, clip, width = \linewidth]{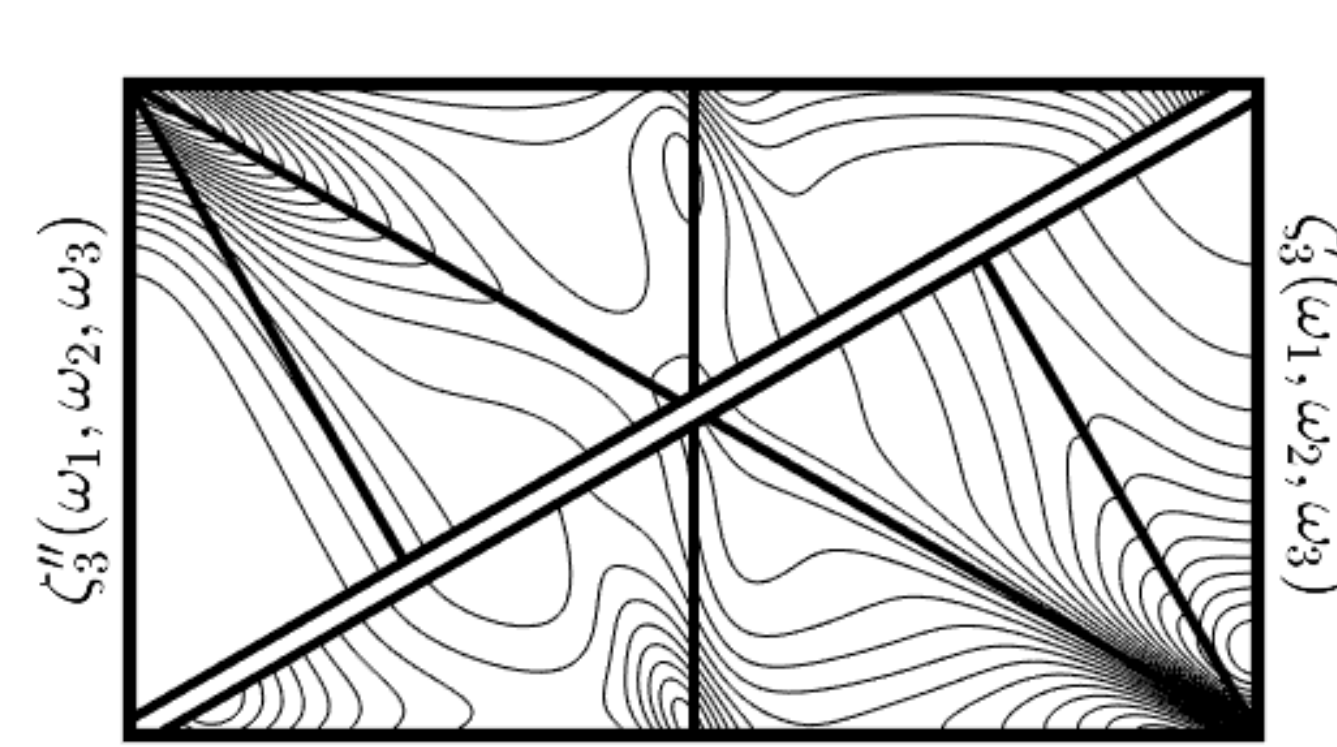}
  \caption{$\Bar{\lambda} \left\Vert\bm{\omega} \right\Vert _{1} = 10$}
\end{subfigure}
\caption{Linearly spaced contours of $\zeta^{*}_{3}(\omega_{1}, \omega_{2}, \omega_{3})$ at different values of $\Bar{\lambda} \left\Vert\bm{\omega} \right\Vert _{1}$ for a Johnson-Segalman fluid with $\beta = 10^{-3}$, and $b = 1$ (top row), or $ b = 0.5 $ (bottom row).}
\label{fig:JSZetaContours}
\end{figure}

Figure \ref{fig:JSZetaContours} shows examples of these contour plots for the Johnson-Segalman model in the strong coupling limit $\beta \rightarrow 0$ at both moderate ($b = 0.5$) and high ($b = 1$) values of the slip parameter. These plots can be generated at different values of $\Bar{\lambda} \left\Vert\bm{\omega} \right\Vert _{1}$, with higher values reflecting features of the nonlinear response excited by variations in deformation rate on shorter time scales.

\begin{figure}[h]
\begin{subfigure}{.495\textwidth}
  \centering
  \includegraphics[trim=4cm 10.75cm 4cm 9.6cm, clip, width = \linewidth]{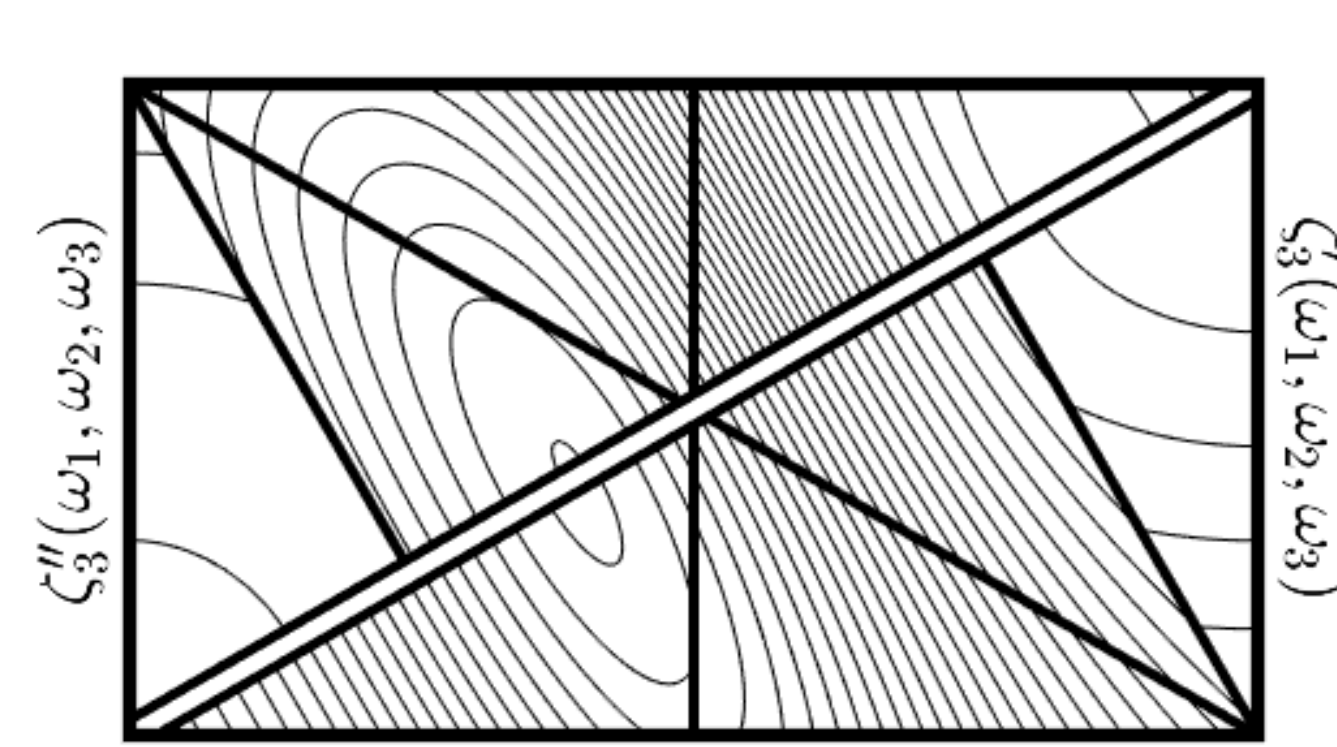}
    \caption{$\Bar{\lambda} \left\Vert\bm{\omega} \right\Vert _{1} = 1$}
\end{subfigure}
\hfill
\begin{subfigure}{.495\textwidth}
  \centering
  \includegraphics[trim=4cm 10.75cm 4cm 9.6cm, clip, width = \linewidth]{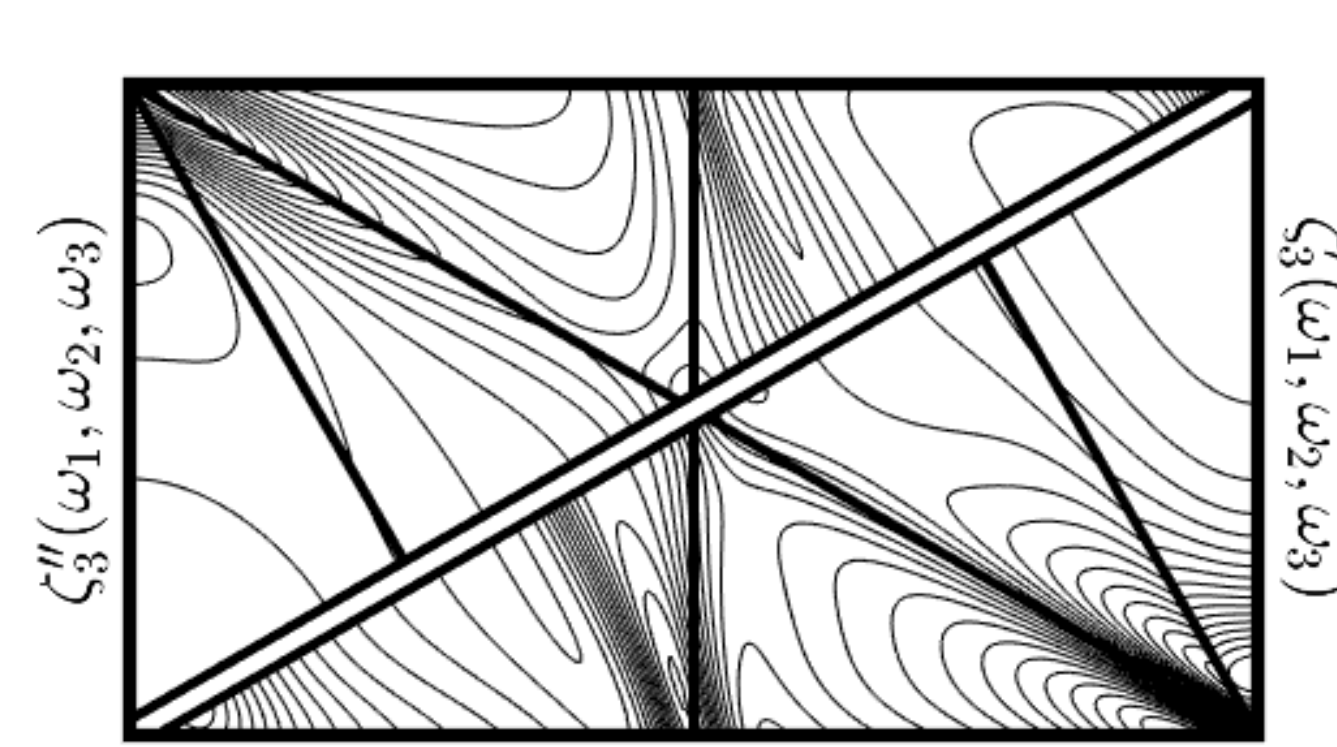}
    \caption{$\Bar{\lambda} \left\Vert\bm{\omega} \right\Vert _{1} = 10$}
\end{subfigure}
\caption{Linearly spaced contours of $\zeta^{*}_{3} (\omega_{1}, \omega_{2}, \omega_{3})$ at different values of $\Bar{\lambda} \left\Vert\bm{\omega} \right\Vert _{1}$ for a Giesekus fluid with $\alpha = 0.3$ and $\beta =10^{-3}$. \\}
\label{fig:GiesekusZetaContours}
\end{figure}

The same can be done for the Giesekus model, as shown in Figure \ref{fig:GiesekusZetaContours} in the strong coupling limit $\beta \rightarrow 0$ with $\alpha = 0.3$. It is worth noting that the qualitative nonlinear response of the Johnson-Segalman and Giesekus models is fairly similar at low frequencies $\Bar{\lambda} \left\Vert\bm{\omega} \right\Vert _{1}$, or for velocities that change slowly in time. However, at higher $\Bar{\lambda} \left\Vert\bm{\omega} \right\Vert _{1}$, the nonlinear responses have starkly different qualitative features that are easily recognized by these contours.

The weakly nonlinear behavior of viscoelastic materials described by these constitutive models is highly sensitive to the time scale on which it is probed, the model structure, and the model parameters.  In all of the contour plots shown, the behavior of the third-order complex resistivity changes drastically with even small changes in the $\bm{\omega}$ coordinate.  Measurements taken at one set of $\bm{\omega}$ coordinates are not necessarily informative about the behavior of $\zeta_3^{*}(\omega_1, \omega_2, \omega_3)$ elsewehere in three-dimensional $\bm{\omega}$ space.  However, a frequency sweep with a single-tone oscillation in the velocity or, better yet, a method of probing broadly the response across the surface of the constant $ L^1 $-norm surface might be used to learn which fluid is generating the response and the parameters of that fluid model with great sensitivity \cite{Lennon2021}.

\newpage
\section{Example flows calculated from the weakly nonlinear response} \label{examples}

To clearly illustrate the flexibility and utility of the solutions we have derived for the weakly nonlinear response of a particle to lineal motion, we will present some specific examples. First, we examine two different `start-up' problems for motion in a Johnson-Segalman fluid.  One of these is the calculation of the force on a particle suddenly set into motion.  The other is the calculation of the velocity of a particle in response to a suddenly applied force.  Then, we discuss the motion of a particle in a harmonic trap whose basin of attraction moves in a line as in an active microrheology experiment.   

\subsection{The force on a particle suddenly set into motion at constant velocity} \label{startup}

Here, we apply $\zeta^{*}_{3}(\omega_{1}, \omega_{2}, \omega_{3})$ derived previously for the single mode Johnson-Segalman model to predict the weakly nonlinear force exerted on a particle on start-up of a steady lineal motion of the particle.  The total force exerted on the particle by the fluid, $ \mathbf{e}_z \cdot \mathbf{F}( t ) $ resulting from a lineal motion with $ V( t ) = -V H( t ) $, where $ H(t )$ is the Heaviside step function, can be written as the the sum of a linear contribution denoted: $F^{(1)}(t)$, and a weakly nonlinear contribution denoted: $F^{(3)} (t)$ such that $ \mathbf{e}_z \cdot \mathbf{F}(t) =F^{(1)} (t) + F^{(3)} (t) $.

\begin{figure}[H]
    \centering
   \includegraphics[scale = 0.6]{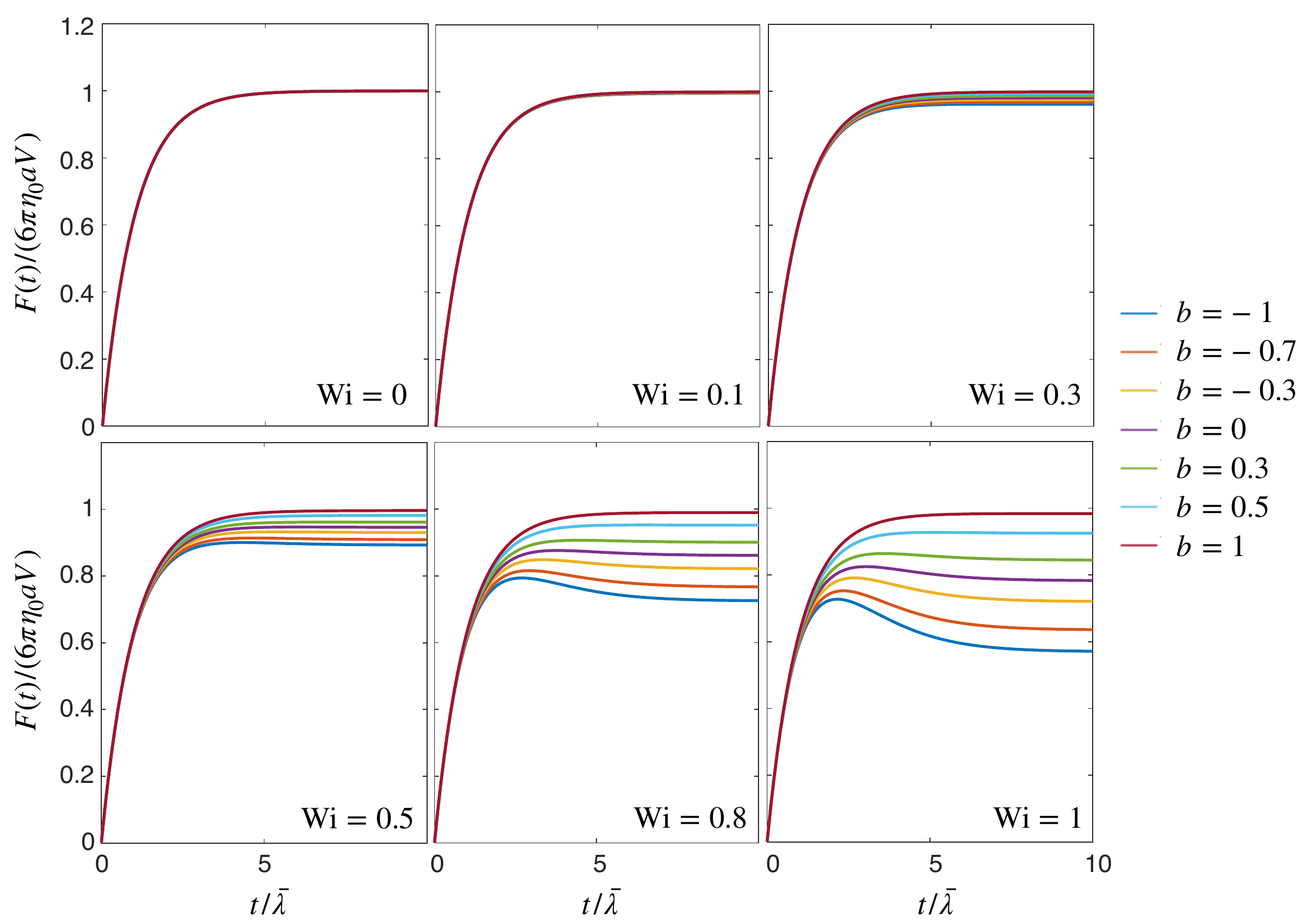}
    \caption{Total dimensionless force over time following the sudden lineal motion of a particle with both linear and third-order contributions at a variety of Weissenberg numbers and slip parameters with $ \beta = 10^{-3} $.}
    \label{fig:startuptotal}
\end{figure}

In Fourier space, the lineal speed is:
\begin{equation} 
    \hat{V}(\omega) = -V \left( \pi \delta(\omega) + \frac{1}{i\omega} \right),
\end{equation}
from which we find that the linear contribution to the force in Fourier space is:
\begin{equation}
    \hat F^{(1)}(\omega) = V \zeta^{*}_{1} (\omega) \left( \pi \delta(\omega) + \frac{1}{i\omega} \right),
\end{equation}
or in real space, for $ t > 0 $:
\begin{equation}
   F^{(1)}(t) =  6 \pi \eta_0 a V \left( \beta + (1-\beta) (1-e^{-\frac{t}{\Bar \lambda}}) \right).
\end{equation}
As shown in Figure \ref{fig:startuptotal}, the force at low Weissenberg numbers grows towards a plateau on long time scales in an exponential fashion.  At third order, the contribution to the force in Fourier space takes the form: 
\begin{equation} \hat{F}^{(3)} (\omega) = -\frac{1}{(2 \pi)^2} \iiint_{-\infty}^{\infty} \zeta^{*}_{3} (\omega_{1}, \omega_{2}, \omega_{3})  \hat{V}(\omega_{1}) \hat{V}(\omega_{2}) \hat{V}(\omega_{3}) \delta (\omega - \omega_{1} - \omega_{2} - \omega_{3}) d\omega_{1} d\omega_{2} d\omega_{3},
\end{equation}
In the time domain, this contribution to the force is: 
\begin{equation} 
F^{(3)} (t) = V^3 \iiint_{-\infty}^{t} \mathcal{F}^{-1}_{\omega_{1}, \omega_{2}, \omega_{3}} \left[ \zeta^{*}_{3} (\omega_{1}, \omega_{2}, \omega_{3}) \right] (\tau_{1}, \tau_{2}, \tau_{3}) d\tau_{1} d\tau_{2} d\tau_{3}  ,
\label{eq:TOF_invFT}
\end{equation}
For the single relaxation time Johnson-Segalman fluid, this contribution can be written as:
\begin{equation}
    F^{(3)}( t ) = \frac{6 \pi \eta_0 \Bar \lambda^2 V^3}{a} ( A^{(3)}( t ) + B^{(3)}( t ) + C^{(3)}( t ) ) = 6 \pi \eta_0 a V \mathrm{Wi}^2 ( A^{(3)}( t ) + B^{(3)}( t ) + C^{(3)}( t ) ) ,
\end{equation}
with $ \Wi = V \Bar \lambda / a $, and:
\begin{subequations}
\begin{align}
A^{(3)}(t) = -\frac{b^2 (1-\beta)^2}{350} & \left( 1 + \frac{1}{\beta ^ {2} \Bar \lambda} \left[ \beta (1-\beta) ^{2}  \Bar \lambda e^{-\frac{t}{\Bar \lambda} \left( \frac{2-\beta}{1-\beta} \right) } - (1-\beta)^{2} \Bar \lambda e^{-\frac{t}{\Bar \lambda} \left( \frac{1}{1-\beta} \right) } \right. \right. \\ & \left. \left. -\beta \Bar \lambda e^{-\frac{2t}{\Bar \lambda}} + \left( \Bar \lambda( 1-2\beta+2\beta^2 - \beta^3 ) -t \beta ( 1 + \beta ) \right) e^{-\frac{t}{\Bar \lambda}} \right] \right), \nonumber
\end{align}
\begin{align}
B^{(3)}(t) =  -\frac{b^2 (1-\beta)^2 }{350} & \left[ 1 + \frac{\beta^{3} }{(1-\beta)^3} e^{-\frac{t}{\Bar \lambda} \left(\frac{1}{\beta} \right)}  \right. \\ & \left. -\frac{1}{2(1-\beta)^{3} \Bar \lambda^{2}} \left( 2\beta^2 \Bar \lambda^2 - 2 \beta (1-\beta) \Bar \lambda(t+\Bar \lambda) + (1-\beta)^2 (t^2 + 2 \Bar \lambda t +2 \Bar \lambda^2) \right) e^{-\frac{t}{\Bar \lambda}} \right], \nonumber
\end{align}
\begin{equation}
C^{(3)}(t) = -\frac{3 (1813-1727b)(1-\beta) }{25025} \left( 1- \frac{1}{2 \Bar  \lambda^2} \left( t^2+ 2 \lambda t + 2 \Bar \lambda^2 \right) e^{-\frac{t}{\Bar \lambda}} \right).
\end{equation}
\end{subequations}

\begin{figure}[H]
  \centering
  \includegraphics[width = 0.75\linewidth]{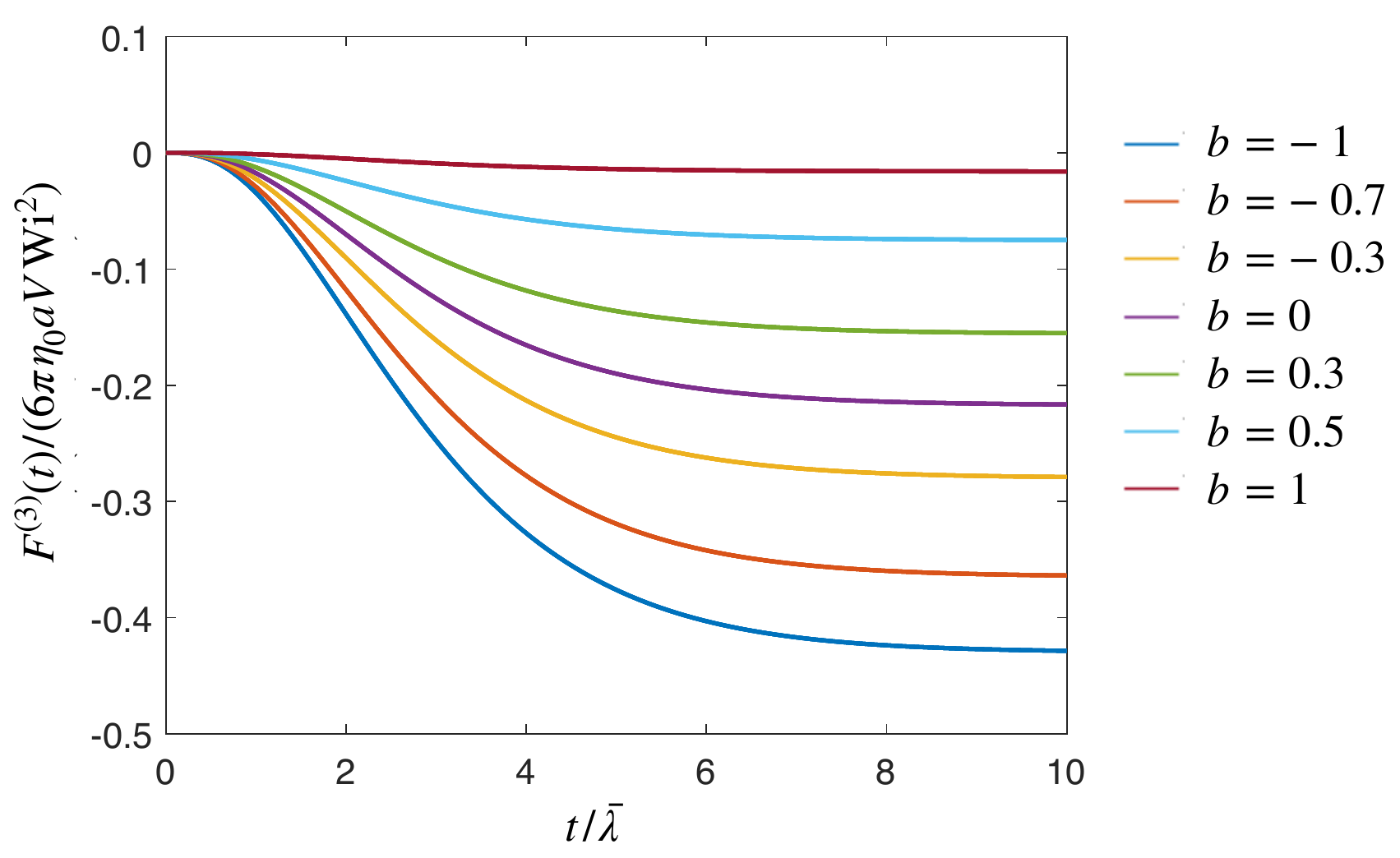}
\caption{Dimensionless third-order force for $ \beta = 10^{-3} $ over time with a variety of slip parameters as indicated by the legend at the right of the figure.}
\label{fig:startupf3}
\end{figure}

Though some of these expressions are long and appear complicated, they simply reflect the product of different exponential decays with polynomial functions of time.  When $ b = 0 $ as in the corotational Maxwell model, the exponential decay occurs only on the time scale $ \Bar \lambda $.  However, for other values of $ b $ additional exponential decay modes with time scales proportional to $ \Bar \lambda $ but weighted on either the relative solvent or polymer viscosity govern the start-up response. This relatively simple function is easy to visualize, as in Figures \ref{fig:startupf3} and \ref{fig:startuptotal}.  This asymptotic model for the force during start-up of lineal motion predicts many of the features and behaviors one would expect from other start-up problems in rheology.

In Figure \ref{fig:startuptotal}, the $\Wi = 0$, or linear, case predicts growth of the force from 0 to a steady plateau.  With increasing $\Wi$, the terminal force normalized by the expectation for steady linear response is decreasing. For larger Weissenberg numbers, this reduction becomes increasingly significant, and the predicted total force begins to have features like a small overshoot before reaching the steady state. These kinds of features are qualitatively similar to some experimental observations of start-up flows around spheres \cite{Becker1994}. 

Variation of the slip parameter $b$, which has a range of $-1$ to $1$, affects the total predicted force in a predictable manner. The greatest reduction in the total predicted force occurs at $-1$, which corresponds to the lower-convected Maxwell model with a Newtonian solvent, and that reduction weakens for higher values of $b$. The least reduction in force occurs for $b = 1$, which reduces the constitutive model to the Oldroyd-B model. This type of relationship would be intuitively expected, as it is the non-affine fluid motion that contributes to drag reduction via thinning \cite{Lee2012}.

\subsection{Expressing the velocity in terms of the force in the weakly nonlinear limit} \label{forcecontrolled}

To this point, we have determined the unsteady force exerted on a sphere in a viscoelastic fluid as a function of an imposed, unsteady, lineal velocity of the particle. However, it may be desirable to express the lineal, unsteady velocity, $ V( t ) $, in terms of the force exerted on the particle by the fluid, $ F( t ) $.  This expression in the weakly nonlinear limit can be written as an analogous set of terms from a truncated Volterra series:
\begin{equation}
    \hat{V}(\omega) = -\xi^{*}_{1}(\omega)\hat{F}(\omega)  -\frac{1}{(2\pi)^2} \iiint_{\infty}^{\infty} \xi^{*}_{3}(\omega_{1}, \omega_{2}, \omega_{3})   \hat{F}(\omega_{1})\hat{F}(\omega_{2})\hat{F}(\omega_{3}) \delta(\omega - \omega_{1} - \omega_{2} - \omega_{3}) \, d\omega_{1} d\omega_{2} d\omega_{3}, \label{eq:velocityforcerel}
\end{equation}
where $\xi^{*}_1(\omega)$ is called the first-order complex mobility and $ \xi^*_3( \omega_1, \omega_2, \omega ) $ is the third-order complex mobility.  The mobility and resistivity kernels are directly related to one another: $ \xi_1^*( \omega ) = 1 / \zeta_1^*( \omega ) $, and
\begin{equation}
    \xi^{*}_{3}(\omega) = -\frac{\zeta^{*}_{3} (\omega_{1}, \omega_{2}, \omega_{3})}{\zeta^{*}_{1}(\omega_{1} + \omega_{2} + \omega_{3})\zeta^{*}_{1} (\omega_{1})\zeta^{*}_{1} (\omega_{2})\zeta^{*}_{1} (\omega_{3}) }.
\end{equation}
just as was shown for the complex moduli and complex compliance in previous work on MAPS rheology \cite{Lennon2020}.  Because $ \zeta_1^*( \omega ) \sim a $ and $ \zeta_3^*( \omega_1, \omega_2, \omega_3 ) \sim a^{-1} $, $ \xi_3^*( \omega_1, \omega_2, \omega_3 ) \sim a^{-5} $.  From this, we can see that the third-order complex mobility is highly sensitive to the particle size.

\begin{figure}[h]
    \centering
   \includegraphics[trim = 4cm 8cm 4cm 8cm, clip, width = 0.8\linewidth]{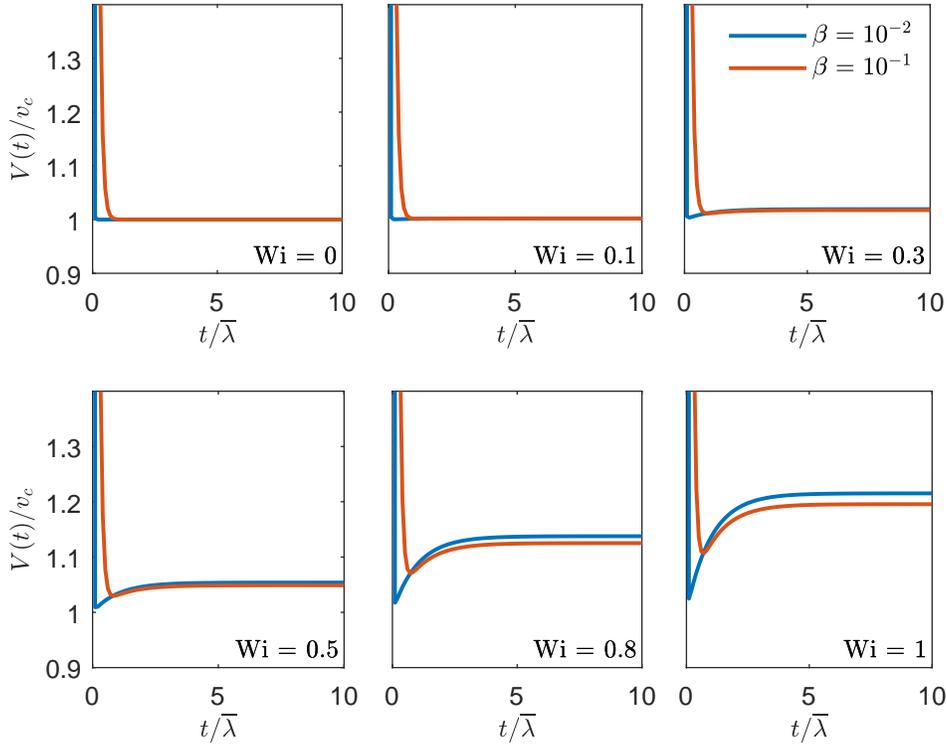}
    \caption{Dimensionless velocity of a particle over time following a suddenly applied force including both the linear and third-order contributions at a variety of Weissenberg numbers with $b = 0$.}
    \label{fig:vfstartup}
\end{figure}

This mobility relation, for instance, makes it possible to determine the time dependent velocity of a particle suddenly set into motion by a force that is constant in time, so long as the magnitude of that force is sufficiently small.  In the limit that the particle inertia is negligible, one would simply replace the force the fluid exerts on the particle, $ F( t ) $, with the negative of the impulsive force, $ -F H(t) $, where $ H(t) $ is the Heaviside step function. The predicted time dependent velocity for a particle suddenly impelled by a constant force is shown in Figure \ref{fig:vfstartup} for a fluid described by the corotational Maxwell model.  The characteristic velocity on which the particle velocity is made dimensionless is simply: $ v_c = F / ( 6 \pi \eta_0 a ) $, and the Weissenberg number is defined with respect to this velocity multiplied by $ 6 \pi $.

An interesting observation from experiments with sedimenting particles in viscoelastic fluids is that in some fluids the magnitude of the force is so large that the rapid particle motion induces a sort of elastic recoil.  When this occurs, the particle velocity oscillates rather than reaching a steady state.   At some finite point in time after the particle is set into motion, $ V^\prime(t) = \mathcal{F}^{-1}[ i \omega \hat V( \omega ) ] = 0 $, rather than the particle decelerating monotonically towards its terminal velocity.  The magnitude of the force required to induce this reversal in acceleration is worth considering and has been investigated experimentally \cite{Zhang2018,Mohammadigoushki2016}.  We might approximate it by considering the Volterra series for the velocity truncated at third order and identifying when the time rate of change of the velocity is zero.  This same equation can be interpreted as a direct relationship between an appropriately scaled force -- a Weissenberg number for the problem of a particle impelled by a sudden force -- and a function of the time coordinate:

\begin{align}
     \mathrm{Wi}^2 &= \left( \frac{F \Bar \lambda}{\eta_0 a^2} \right)^2 \\
     & = -\frac{(2 \pi)^2 \Bar \lambda^2}{\eta_0 a^2} \frac{\mathcal{F}^{-1}\left[ i \omega \xi_1^*( \omega ) \hat H( \omega ) \right]}{\mathcal{F}^{-1}\left[ i \omega \iiint_{-\infty}^\infty \xi_3^*( \omega_1, \omega_2, \omega_3 ) \delta( \omega - \omega_1 - \omega_2 - \omega_3 ) \hat H( \omega_1 ) \hat H( \omega_2 ) \hat H( \omega_3 ) \, d \omega_1 d \omega_2 d \omega_3 \right]}. \nonumber
\end{align}

\begin{figure}[b]
\begin{subfigure}{.495\textwidth}
  \centering
  \includegraphics[trim=3cm 8cm 4.5cm 8cm, clip, width = \linewidth]{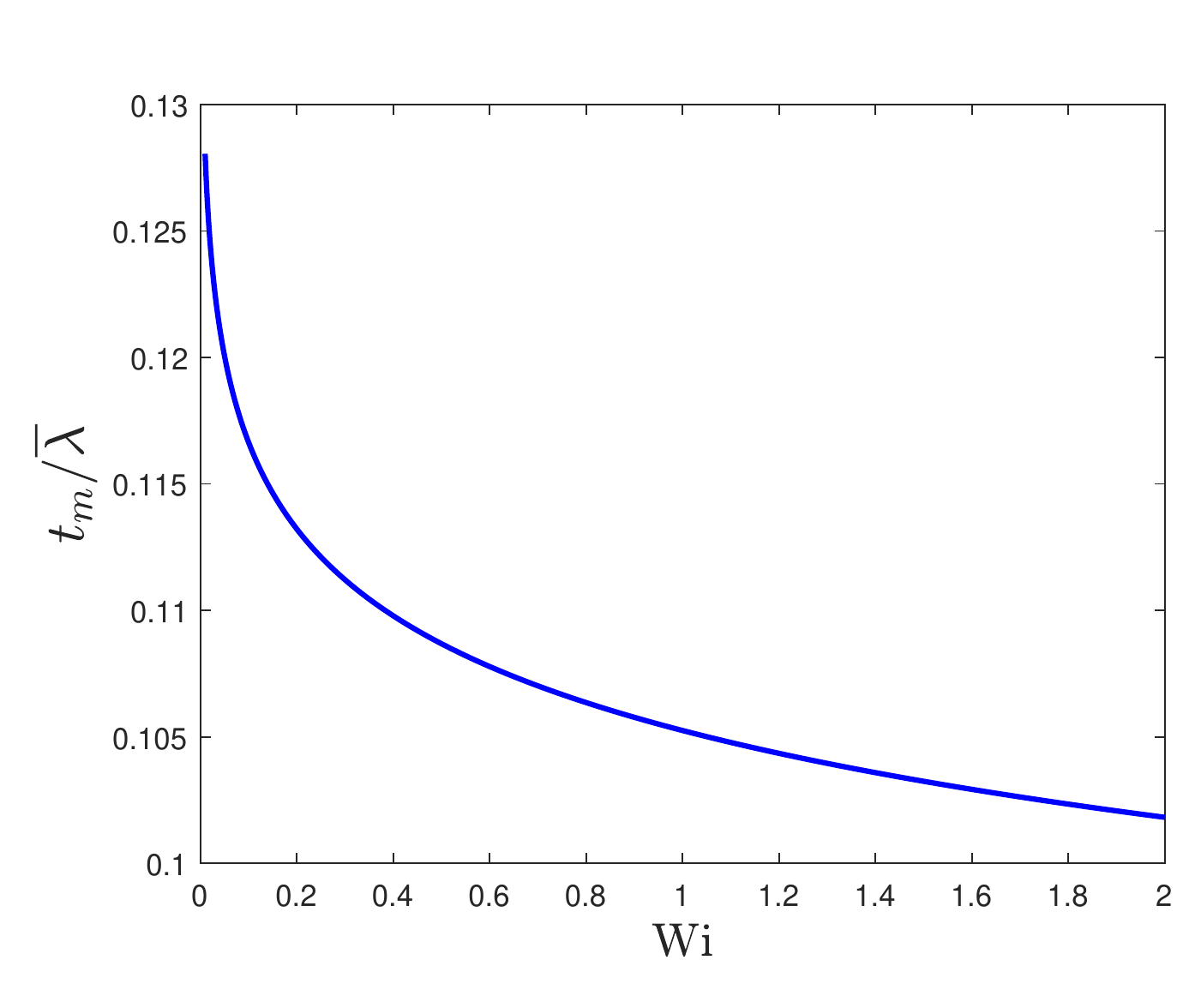}
  \caption{Strong-coupling limit ($\beta = 0.01$)}
\end{subfigure}
\hfill
\begin{subfigure}{.495\textwidth}
  \centering
  \includegraphics[trim=3cm 8cm 4.5cm 8cm, clip, width = \linewidth]{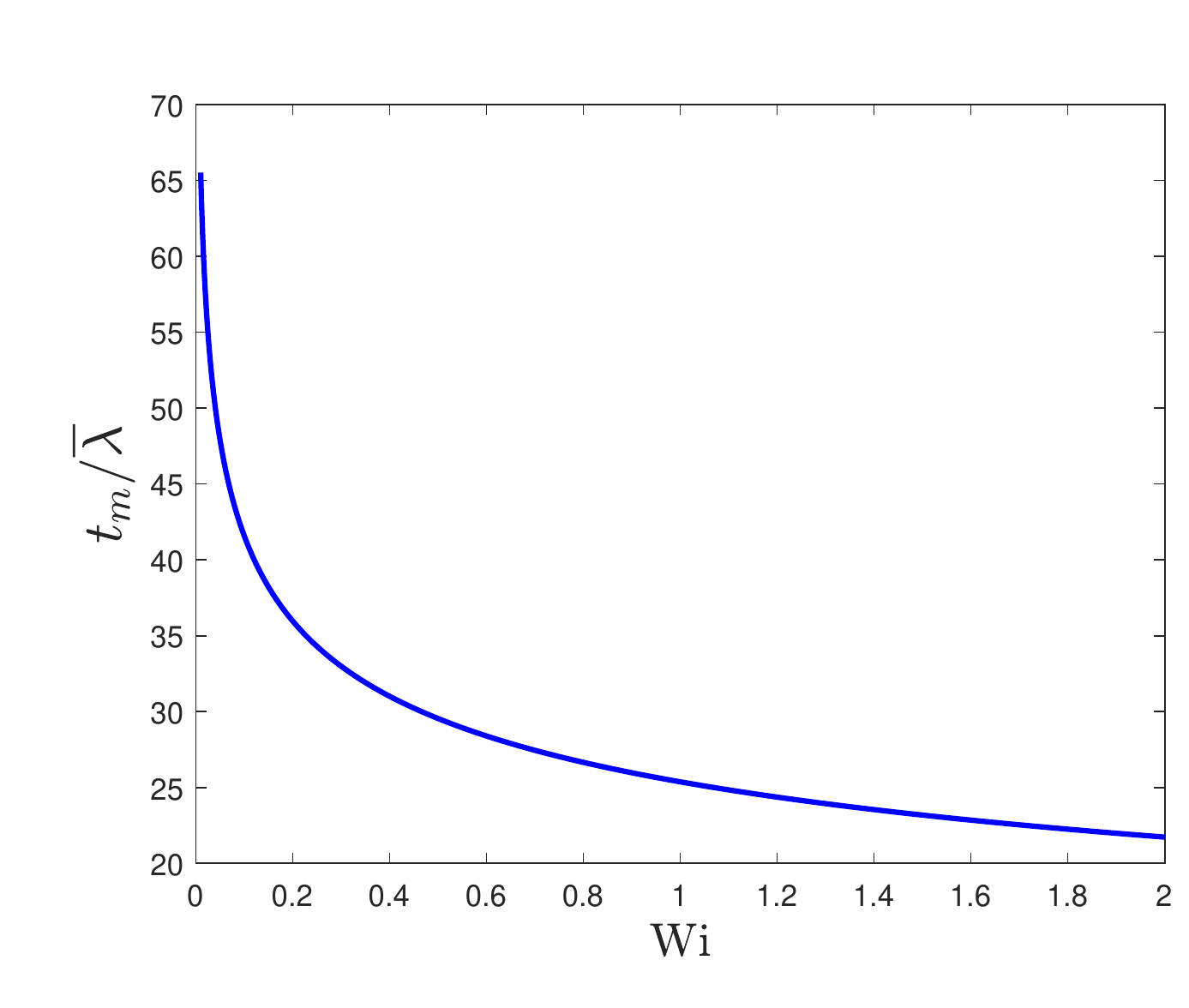}
  \caption{Weak-coupling limit ($\beta = 0.99$)}
\end{subfigure}
\caption{Time of undershoot, $t_m / \Bar{\lambda}$, as a function of $\mathrm{Wi}$. Note that while the overall trend is similar for both the strong-coupling and weak-coupling limits, the magnitudes are vastly different due to the dependence on $\beta$, with the undershoot occurring at much shorter times in the strong-coupling limit.}
\label{fig:onset}
\end{figure}

For a given value of the Weissenberg number, this relationship gives the corresponding points in time at which the time rate of change in the velocity might become zero.  We plot this relationship in Figure \ref{fig:onset} for the corotational Maxwell model with a single relaxation time in both the strong- and weak-coupling limits.  Because the Weissenberg number must be real valued, we consider only positive values of $ \Wi^2 $.

The relationships we have derived here do not predict oscillations in the velocity, but rather an undershoot in velocity that subsequently levels out to the terminal velocity as seen in Figure \ref{fig:startuptotal}. Therefore, these results are not directly analogous to the experimental observations of oscillations in the settling velocity, which tend to occur at much higher Weissenberg numbers than those for which our calculations are applicable \cite{Zhang2018, Lee2012}. It is thought that these oscillations may be partially caused by inertial effects for which our calculations do not account.  However the model appears to describe a related effect in the weakly nonlinear limit. 

As clearly seen in Figure \ref{fig:onset}, in both strong- and weak-coupling cases for the corotational Maxwell model, an undershoot in the velocity occurs for all Weissenberg numbers, though at significantly longer times for low $ \mathrm{Wi} $. Though the dependence of the time to reach the undershoot on $ \mathrm{Wi} $ is very similar in both the weak- and strong-couplings limits, the undershoot in the velocity occurs much sooner in the strong-coupling limit than the weak-coupling limit.  Additionally, the relative scale of the undershoot is larger in the strong-coupling limit.

\subsection{The weakly nonlinear response of a particle in a microrheology experiment} \label{microrheology}
 
In one active microrheology experiment, a spherical probe particle is driven to move along with a trap produced by a laser or magnetic tweezer.  If the trap is relatively stiff or the displacement between the particle and the focus of the trap is small, the force exerted by the trap on the particle is well approximated as linear in the displacement: $ \mathbf{F}_\mathrm{trap}(t) = -k( \mathbf{X}(t) - \mathbf{X}_\mathrm{trap}(t)) $, where $ k $ is the stiffness.  We will assume for these purposes that there is no nonlinearity in the trapping force, though this assumption is easily relaxed.  In such microrheology experiments, the position of the trap $ \mathbf{X}_\mathrm{trap}( t ) = X_\mathrm{trap}(t) \mathbf{e}_z $ can be controlled to achieve a lineal motion.  The position of the particle $ \mathbf{X}( t ) = X( t ) \mathbf{e}_z $ can be measured and for small, lineal trap motions, the particle will execute a lineal motion too.  We aim then to determine a Volterra series relationship between the particle position and the trap location truncated in the weakly nonlinear limit:
\begin{align}
    \hat X( \omega ) &= \psi_1^*( \omega ) \hat X_\mathrm{trap}( \omega ) \\
    &+ \frac{1}{(2 \pi)^2} \iiint_{-\infty}^\infty \psi_3^*( \omega_1, \omega_2, \omega_3 ) \hat X_\mathrm{trap}( \omega_1 ) \hat X_\mathrm{trap}( \omega_2 ) \hat X_\mathrm{trap}( \omega_3 ) \delta( \omega - \omega_1 - \omega_2 - \omega_3 ) \, d \omega_1 d \omega_2 d \omega_3. \nonumber
\end{align}
The newly specified Volterra series coefficients, $ \psi_1^*( \omega ) $ and $ \psi_3^*( \omega_1, \omega_2, \omega_3 ) $, will be functions of the complex resistivities or mobilities determined previously.  In the following, we use the spectral derivative: $ \hat V( \omega ) = i \omega \hat X( \omega ) $ to re-express the lineal velocity of the particle in terms of its position.

At first order in the trap position, the relationship between the the particle velocity and the force exerted on the particle by the trap is simply: $ i \omega \hat X( \omega ) = -k \xi_1^*( \omega ) ( \hat X( \omega ) - \hat X_\mathrm{trap}( \omega ) ) $, from which we find:

\begin{equation}
    \hat X( \omega ) = \frac{k \xi^{*}_{1} (\omega) }{i\omega + k \xi^{*}_{1} (\omega)} \hat X_\mathrm{trap}( \omega ).
\end{equation}

Therefore, $ \psi^{*}_{1} (\omega) = k \xi^{*}_{1} (\omega) / ( i\omega + k \xi^{*}_{1} (\omega) ) $, which is a well-known expression for the linear response function in an active microrheology experiment.  In the weak trap limit, $ k \ll \eta_0 a \omega $, $ \psi_1^*( \omega ) \approx -(ik/\omega) \xi_1^*( \omega ) $, so that the linear response function in the microrheology experiment is directly proportional to the first-order complex mobility.  In the stiff trap limit, $ k \gg \eta_s a \omega $, $ \psi_1( \omega ) \approx 1 - i \omega \zeta_1^*( \omega ) / k $ , so that the linear microrheology response function is $ O( 1 ) $ with a small perturbation proportional to the first-order complex resistivity.

\begin{figure}[t]
\begin{subfigure}{.495\textwidth}
  \centering
  \includegraphics[trim=3cm 8cm 4.5cm 8cm, clip, width = \linewidth]{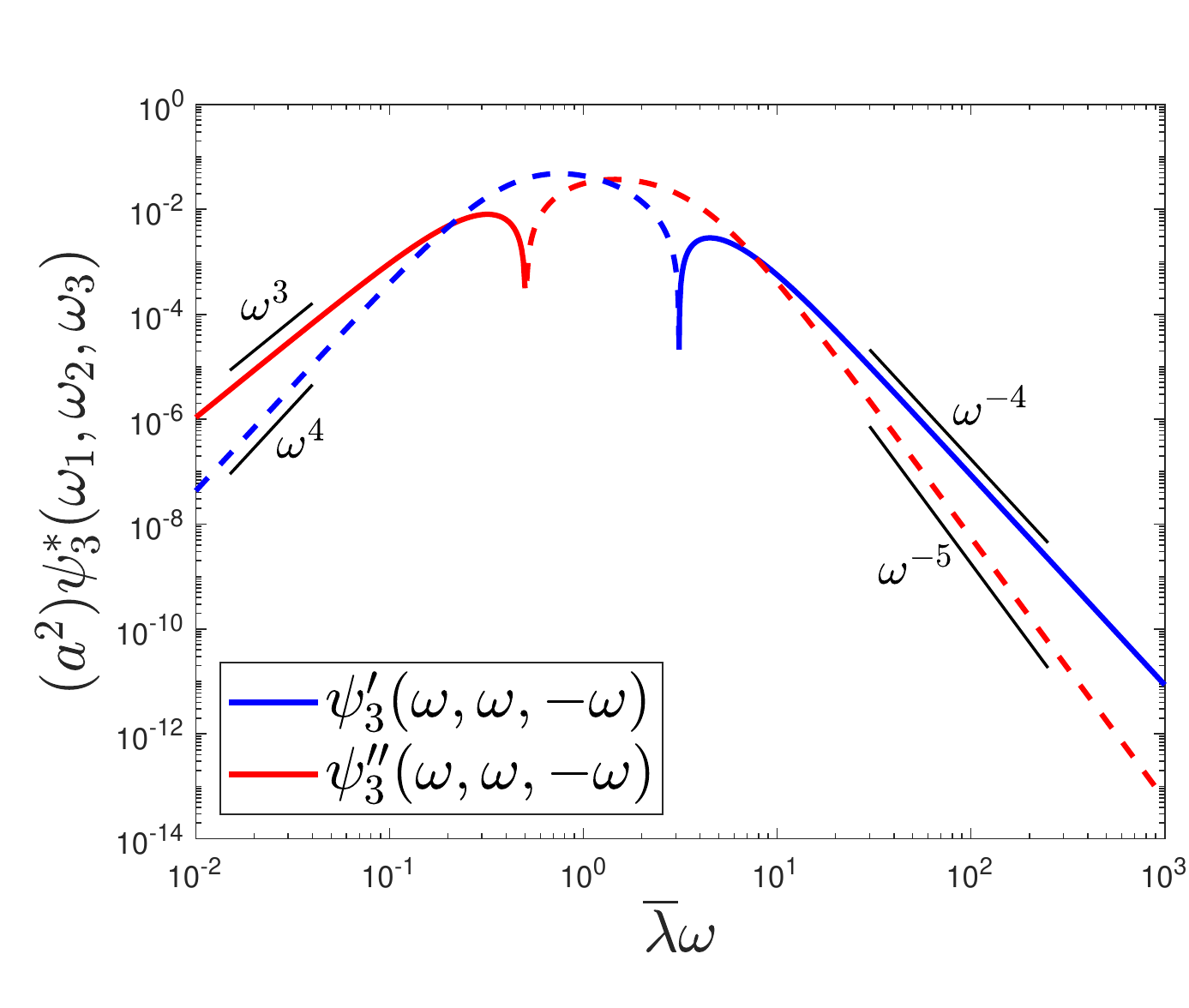}
\end{subfigure}
\hfill
\begin{subfigure}{.495\textwidth}
  \centering
  \includegraphics[trim=3cm 8cm 4.5cm 8cm, clip, width = \linewidth]{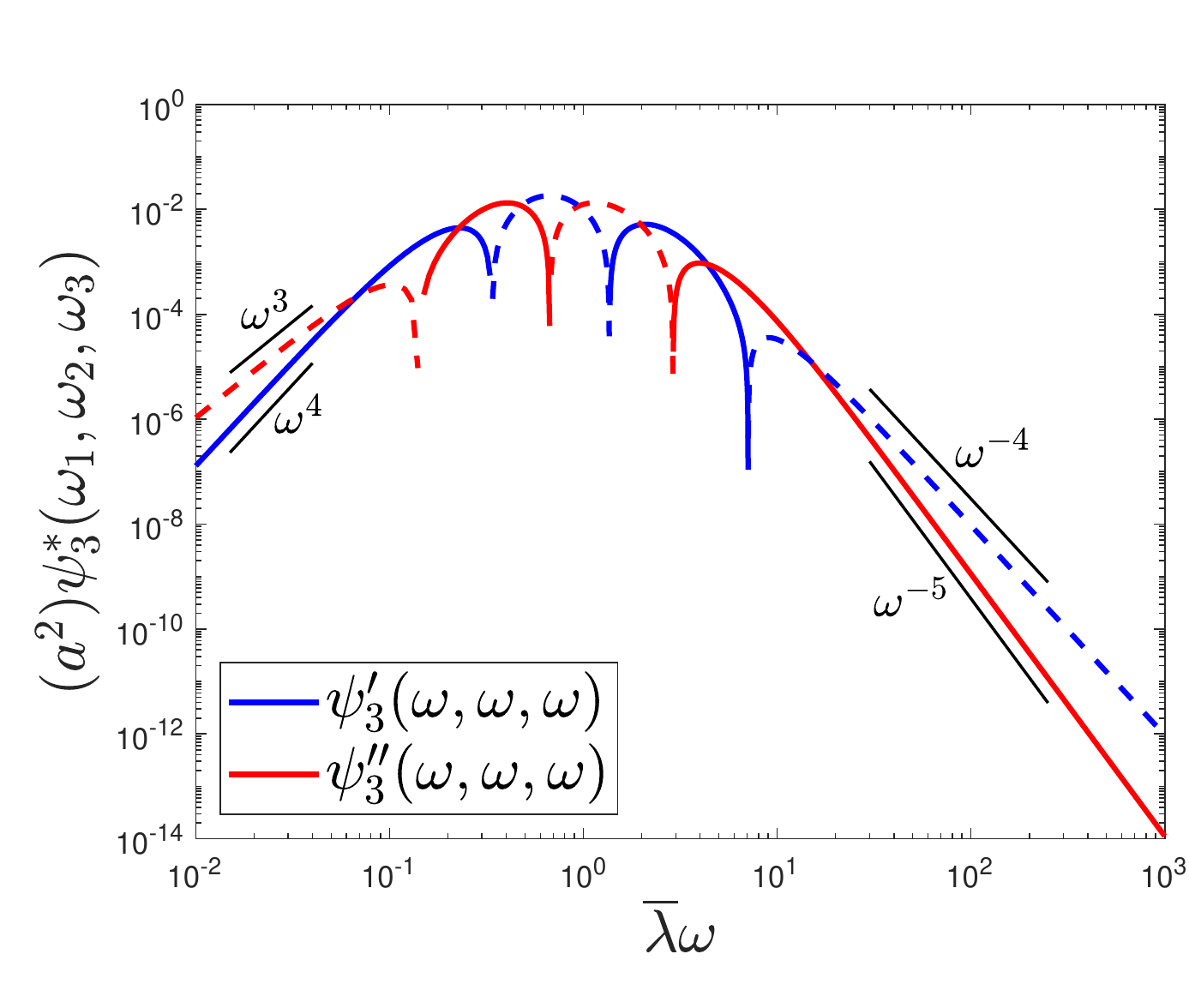}
\end{subfigure}
\begin{subfigure}{.495\textwidth}
  \centering
  \includegraphics[trim=3cm 8cm 4.5cm 8cm, clip, width = \linewidth]{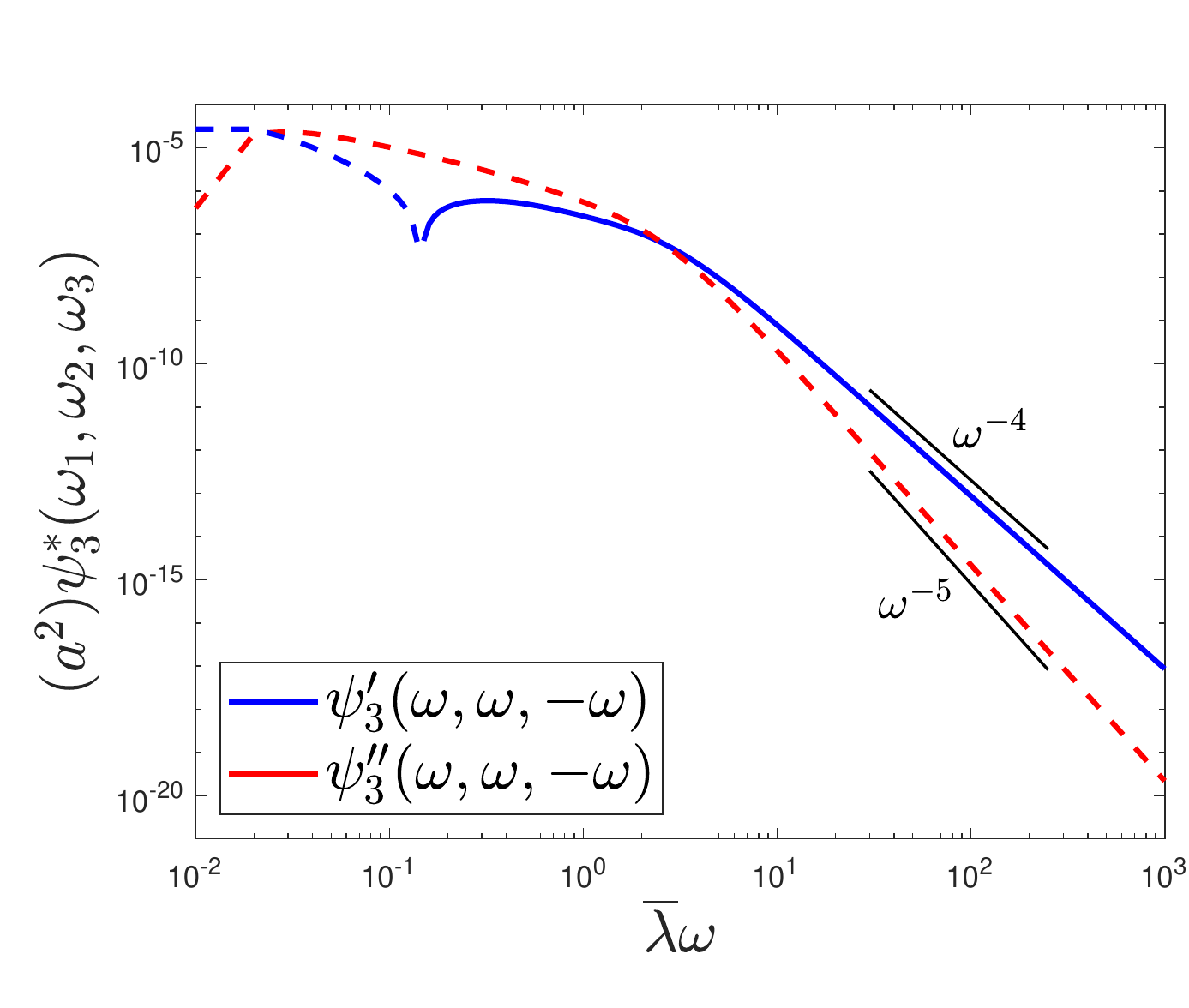}
\end{subfigure}
\hfill
\begin{subfigure}{.495\textwidth}
  \centering
  \includegraphics[trim=3cm 8cm 4.5cm 8cm, clip, width = \linewidth]{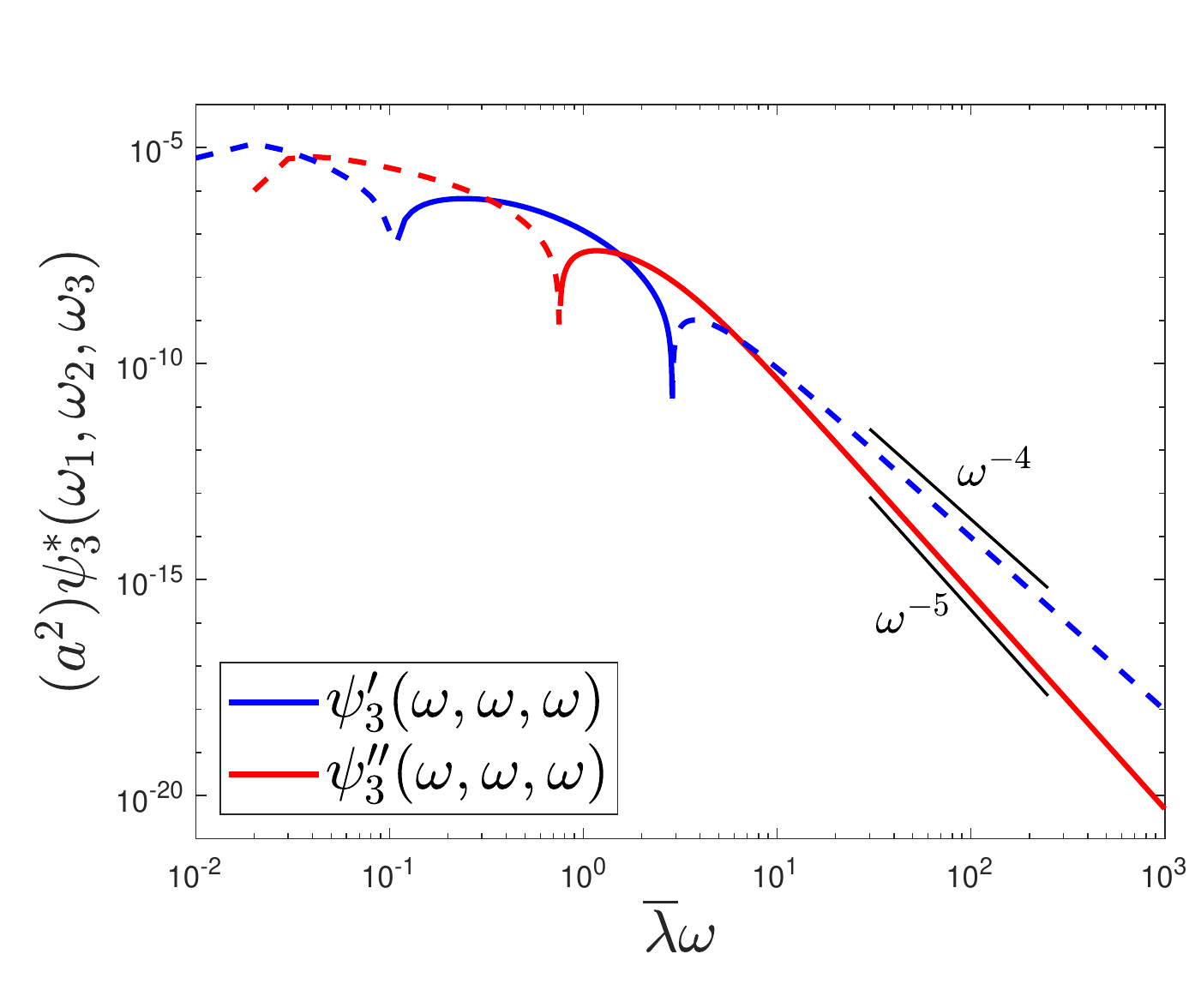}
\end{subfigure}
\caption{Projections of $\psi^{*}_{3} (\omega_{1}, \omega_{2}, \omega_{3})$ for a fluid with $b = 0.5$, $\beta = 0.5$, subjected to a single-tone oscillation in a system with trap stiffness $ k \bar \lambda / (\eta_0 a ) = 1 $ (top row), $ k \bar \lambda / (\eta_0 a ) = 10^{-2} $ (bottom row). }
\label{fig:JSBetaProjs}
\end{figure}

There are, of course, additional higher order contributions to the particle position.  We can find the third order contribution easily by writing particle position as the linear contribution plus a perturbation denoted $ \delta \hat X( \omega ) $.  The trapping force in frequency space is: $ \mathbf{e}_z \cdot \mathbf{\hat F}_\mathrm{trap}(\omega) = -k ( \psi_1^*( \omega ) - 1 ) \hat X_\mathrm{trap} -k \delta \hat X( \omega ) $.  This can be substituted into equation \ref{eq:velocityforcerel} for the Volterra series of velocity in terms of applied force, which can then be truncated at third order in the trap position.  This gives an expression for perturbation to the linear response of the particle valid at third order:
\begin{align}
    i\omega \delta \hat{X}(\omega) &= -\xi^{*}_{1}(\omega) k \delta \hat{X}( \omega ) \\
    & - \iiint_{-\infty}^{\infty} k^3 \xi^{*}_{3} (\omega_{1}, \omega_{2}, \omega_{3}) ( \psi_1^*( \omega_1 ) - 1 )  ( \psi_1^*( \omega_2 ) - 1 ) ( \psi_1^*( \omega_3 ) - 1 ) \nonumber \\
    & \quad \quad \quad  \hat{X}_\mathrm{trap} (\omega_1) \hat{X}_\mathrm{trap} (\omega_{2}) \hat{X}_\mathrm{trap}(\omega_{3}) \delta(\omega - \omega_1 - \omega_2 - \omega_3 ) \, d\omega_{1} d\omega_{2} d\omega_{3}. \nonumber
\end{align}
Solving for the perturbation to the linear response at third order in the trap position, we find that the third-order response function in a microrheology experiment is:
\begin{align}
    &\psi_3^*( \omega_1, \omega_2, \omega_3 ) = \label{eq:beta3} \\
    & \quad -\left( \frac{k^3 \xi_3^*( \omega_1, \omega_2, \omega_3 ) }{ i ( \omega_1 + \omega_2 + \omega_3 ) + k\xi_1^*( \omega_1 + \omega_2 + \omega_3 ) } \right) \left(  \frac{i \omega_1 }{ i\omega_1 + k \xi^{*}_{1} (\omega_1) } \right) \left(  \frac{ i \omega_2 }{ i\omega_2 + k \xi^{*}_{1} (\omega_2) } \right) \left(  \frac{ i \omega_3 }{ i\omega_3 + k \xi^{*}_{1} (\omega_3) } \right). \nonumber
\end{align}

As with the linear response function for the microrheology experiment, the trap stiffness has a significant impact on the value of $\psi^{*}_{3}(\omega_{1}, \omega_{2}, \omega_{3})$. While $\psi^{*}_{3}(\omega_{1}, \omega_{2}, \omega_{3})$ approaches 0 in the limits of both large and small values of $k$, the ways this Volterra kernel decays in these limits are distinct. For a weak trap, $ \psi_3^*( \omega_1, \omega_2, \omega_3 ) \approx ik^3 \xi_3^*( \omega_1, \omega_2, \omega_3 ) / ( \omega_1 + \omega_2 + \omega_3 ) $, so that the response function is directly proportional to the third-order complex mobility and the cube of the trap stiffness.  Whereas for stiff traps, $ \psi_3^*( \omega_1, \omega_2, \omega_3 ) \approx -i \omega_1 \omega_2 \omega_3 \zeta_3^*( \omega_1, \omega_2, \omega_3 ) / k $.  In this case the transfer function is directly proportional to the third-order complex resistivity and inversely proportional to the trap stiffness.  Inference of the material properties: $ \xi_3^*( \omega_1, \omega_2, \omega_3 ) a^5 $ or $ \zeta_3^*( \omega_1, \omega_2, \omega_3 ) a $, from a microrheology experiment will require inverting the expression in equation \ref{eq:beta3}.  Doing so requires either careful consideration of whether the experiment is conducted in the weak and strong trapping limits so that limiting expressions can be applied or precise measurement of the trap stiffness and first-order complex mobility.

Projections and contours of $\psi^{*}_{3}(\omega_{1}, \omega_{2}, \omega_{3})$ can be visualized the same way that those of $\zeta^{*}_{3} (\omega_{1}, \omega_{2}, \omega_{3})$ are in Section \ref{visualizing}. The first and third harmonic projections of $\psi^{*}_{3}(\omega_{1}, \omega_{2}, \omega_{3})$, made dimensionless on $a^2$, are shown in Figure \ref{fig:JSBetaProjs}. The trap stiffness $ k $ is made dimensionless on $\eta_0 a / \Bar{\lambda} $, and the ratio of these two groups can be varied freely as we have in Figure \ref{fig:JSBetaProjs}. It can be clearly seen that for a Johnson-Segalman fluid with a trap having moderate stiffness $k \bar \lambda / (\eta_0 a ) = 1 $, $\psi^{*}_3 (\omega_1, \omega_2, \omega_3)$  trends toward zero at both high and low frequencies, though at low frequencies this scales like $\psi^{*}_3 (\omega_1, \omega_2, \omega_3) \sim \omega^{3}$, and at high frequencies like $\psi^{*}_3 (\omega_1, \omega_2, \omega_3)\sim \omega^{-4}$. For both the first and third harmonic measurements, there is a sign change observed in both components of $\psi^{*}_3(\omega_1, \omega_2, \omega_3)$, as well as the change in scaling behavior, within the moderate frequency range of approximately ($10^{-1} < \bar \lambda \omega < 10$). However, in the case of a weak trap $k \bar \lambda / (\eta_0 a ) = 10^{-2} $, $\psi^{*}_3 (\omega_1, \omega_2, \omega_3)$ exhibits the type of scaling expected, bearing a strong resemblance -- particularly at high frequency -- to the expected behavior of $\xi^{*}_3 (\omega_1, \omega_2, \omega_3)$, at an overall lower magnitude due to the cubic scaling with $ k $.

\section{Conclusions} \label{conclusions}
In this work, we have shown a general method for describing the unsteady creeping flow of an incompressible, isothermal viscoelastic fluid around a sphere executing a lineal translation. The asymptotic solutions for the pressure and flow fields in the fluid accompanying such translations were calculated up to second order in the amplitude of the lineal translation for the Johnson-Segalman and Giesekus fluids.  It was demonstrated that these velocity and pressure profiles can be used to calculate the force at third order in the magnitude of the translational velocity. These analytical solutions are valid for low Weissenberg numbers, or in the weakly nonlinear limit.  We believe that this same method of solution, which solves the momentum balance, conservation of mass and constitutive equations in Fourier space using a kind of separation of variables, can be applied straightforwardly to other constitutive models. 

The method and calculations shown here have two key implications: they provide a more general fluid mechanical understanding of and ability to model unsteady phenomena in flows of viscoelastic fluids around a sphere without resorting to numerical approximation, and they provide a framework for understanding the first effects of nonlinearity in unsteady particle motions.  The analytical solution for the weakly nonlinear force was shown in this work to be able to describe flow phenomena in a ``start-up'' flow that were previously observed only in experiment and numerical simulation, like a force overshoot at high Weissenberg numbers \cite{Becker1994}, and a pronounced enhancement of overall drag reduction with increasing non-affine deformation \cite{Lee2012}.  

While the solution method described a particle executing a prescribed translation, due to the generality and simplicity of the asymptotic solution, a Volterra series representation could be constructed and inverted to describe the weakly nonlinear response to an imposed force instead. This reconstitution of the solution may expand the usefulness of these types of asymptotic solutions for validation of numerical and computational techniques to unsteady flows. Additionally, it could be used to provide insight into some of the less well-understood dynamic phenomena observed in experiments such as velocity oscillation during sedimentation.  Beyond the improved ability to model and understand the fluid mechanics of these unsteady flows, the technique presented here also provides a framework for weakly nonlinear microrheology. While researchers have both made measurements in the nonlinear regime in physical systems using microrheology and modeled the nonlinear response in idealized systems via material specific theory and numerical simulation, there is no general, unified method for understanding such measurements, comparing them to constitutive models, or comparing them between different materials. 

We have proposed that the third-order complex resistivity and mobility, when scaled appropriately on the particle size, are the key quantities of interest in weakly nonlinear microrheology.  These are functions that can be both measured via active microrheology and modeled via the methods shown in this work. The first-order complex resistivity is already a well-understood, measurable quantity in microrheology experiments. The extension of this quantity to the weakly nonlinear regime via the Volterra series expansion shown in this work arises naturally and provides a generalized way of representing nonlinear microrheology measurements.

Ultimately, taking advantage of this new representation will require new experimental protocols for efficiently probing the third-order complex resistivity or mobility.  Some techniques developed for weakly nonlinear rheology at the macro-scale like MAOS or MAPS might be applied to microrheology as well. However, this requires mastering the coordination of trap and particle motion beyond the constant velocity and single-tone oscillation protocols most commonly used in active microrheology.  This more careful design of complex temporal protocols to probe a larger subset of the total space characterized by the third-order complex resistivity and mobility is left for future work. We think that this is a worthwhile effort as it could provide a method of deriving a more complete characterization of rheological responses for efficient model identification and parameter estimation in fluids that are precious or otherwise difficult to study in bulk.

\bibliography{article}

\newpage
\appendix
\section{Derivation of the third-order complex resistivity for a Giesekus fluid}\label{giesekus}

The third-order complex resistivity is derived here for a fluid described by the Giesekus constitutive model.  For a single relaxation time Giesekus fluid, the dimensional polymer stress is: 

\begin{multline} 
    \bm{\tau}_{p} (\textbf{r}, t) +  \frac{\alpha \lambda}{\eta_{p}}  (\bm{\tau}_{p} (\textbf{r}, t) \cdot \bm{\tau}_{p}  (\textbf{r}, t)) +  \\ \lambda \left( \frac{\partial \bm{\tau}_{p} (\textbf{r}, t)}{\partial t} + \mathbf{v} (\textbf{r}, t) \cdot \nabla\bm{\tau}_{p} (\textbf{r}, t) - \bm{\tau}_{p}(\textbf{r}, t) \cdot \nabla v(\textbf{r}, t) - \nabla v^{T}(\textbf{r}, t) \cdot \bm{\tau}_{p}(\textbf{r}, t)\right) = 2 \eta_{p} \mathbf{e} (\textbf{r}, t).
\end{multline}
The parameter $ \alpha $ governs the strength of a nonlinearity proportional to the square of the polymer stress.  The momentum balance and continuity equation are the same as in the derivation described for the Johnson-Segalman model.  When the same scaling relationships are used as in the main text, a dimensionless expression for the polymer stress can be written as:
\begin{multline}
    \bm{\tau}_{p} (\textbf{r}, t) + \De \frac{\partial \bm{\tau}_{p}}{\partial t} (\textbf{r}, t) + \\ \Wi\left[ \alpha \bm{\tau}_{p} (\textbf{r}, t) \cdot \bm{\tau}_{p} (\textbf{r}, t) + \mathbf{v} (\textbf{r}, t) \cdot \nabla \bm{\tau}_{p}(\textbf{r}, t) - \left( \bm{\tau}_{p} (\textbf{r}, t) \cdot \nabla \mathbf{v}(\textbf{r}, t) + \mathbf{v}^{T} (\textbf{r}, t) \cdot \bm{\tau}_{p} (\textbf{r}, t) \right) \right] = 2\mathbf{e}(\textbf{r}, t).
\end{multline}

The same ordered expansion of variables in $ \Wi $ and matched order equations can be formulated as in the main text.  In frequency space, the $ n $th order contribution to a polymer stress mode with relaxation time, $ \lambda $, satisfies the equation: 
\begin{multline}
   \hat{\bm{\tau}}_{p}^{(n)} (\textbf{r}, \omega,\lambda) =  \chi(\omega, \lambda) \Bigg( 2 \hat{\mathbf{e}}^{(n)} (\textbf{r}, \omega) - \frac{\lambda}{\Bar \lambda} \sum_{m =1}^{n-1} \left[ \alpha \hat{\bm{\tau}}_{p}^{ (n-m)} (\textbf{r}, \omega, \lambda) * \hat{\bm{\tau}}_{p}^{(m)} (\textbf{r}, \omega, \lambda) +  \right. \\ \left. \hat{\mathbf{v}}^{(n-m)} (\textbf{r}, \omega) * \nabla\hat{\bm{\tau}}_{p}^{(m)} (\textbf{r}, \omega, \lambda) - \left( \hat{\bm{\tau}}_{p}^{(n)} (\textbf{r}, \omega, \lambda) * \nabla \hat{\mathbf{v}}^{(n-m)} (\textbf{r}, \omega) + \hat{\mathbf{v}}^{(n-m) T} (\textbf{r}, \omega) * \hat{\bm{\tau}}_{p}^{(m)} (\textbf{r}, \omega, \lambda) \right) \right] \Bigg).
\end{multline}

The first order solution for the pressure and velocity fields and the force is the same as for the Johnson-Segalman model. At second order, the equations and solutions differ. The second order equations to be solved are:
\begin{subequations}
\begin{equation}
     -\nabla \hat{p}^{(2)} (\textbf{r}, \omega)  + \beta \nabla \cdot \hat{\bm{\tau}}_{s}^{(2)} (\textbf{r}, \omega) + (1-\beta) \nabla \cdot \left< \bm{\tau}_{p}^{(2)} (\textbf{r}, \omega, \lambda) \right>_\lambda = 0, \qquad \nabla \cdot \hat{\mathbf{v}}^{(2)} (\textbf{r}, \omega) = 0 ,
\end{equation}
\begin{equation}
    \hat{\bm{\tau}}_{s}^{(2)} (\textbf{r}, \omega) = 2\hat{\mathbf{e}}^{(2)} (\textbf{r}, \omega),
\end{equation}
\begin{multline}
   \hat{\bm{\tau}}_{p}^{(2)} (\textbf{r}, \omega,\lambda) =  \chi(\omega, \lambda)\left( 2 \hat{\mathbf{e}}^{(2)} (\textbf{r}, \omega) - \frac{\lambda}{\Bar \lambda} \left[ \alpha \hat{\bm{\tau}}_{p}^{(1)} (\textbf{r}, \omega, \lambda) * \hat{\bm{\tau}}_{p}^{(1)} (\textbf{r}, \omega, \lambda) + \hat{\mathbf{v}}^{(1)} (\textbf{r}, \omega) * \nabla\hat{\bm{\tau}}_{p}^{(1)} (\textbf{r}, \omega,\lambda) \right.\right. \\ \left.\left. - \left( \hat{\bm{\tau}}_{p}^{(1)} (\textbf{r}, \omega,\lambda) * \nabla \hat{\mathbf{v}}^{(1)} (\textbf{r}, \omega) + \hat{\mathbf{v}}^{(1)T}  (\textbf{r}, \omega) * \hat{\bm{\tau}}_{p}^{(1)} (\textbf{r}, \omega,\lambda) \right) \right] \right).
\end{multline}
\end{subequations}

Using the established relationships between the frequency-domain and steady-state solutions at first order: $\hat{\bm{\tau}}_{p}^{(1)} (\textbf{r}, \omega) = \hat V(\omega)\chi(\omega, \lambda) \tilde{\bm{\tau}}_{p}^{(1)} (\textbf{r}, \lambda) $, $ \hat{\mathbf{v}}^{(1)} (\textbf{r}, \omega) = \hat V(\omega) \tilde{\mathbf{v}}^{(1)} (\textbf{r}) $, we can re-express the second-order polymeric stress as:
\begin{multline}
   \hat{\bm{\tau}}_{p}^{(2)} (\textbf{r},\omega,\lambda) =  \chi(\omega, \lambda)\left( 2 \hat{\mathbf{e}}^{(2)}(\textbf{r}) - \alpha [ \hat V(\omega)\chi(\omega,\lambda)] * [\hat V(\omega)\chi(\omega,\lambda)] \tilde{\bm{\tau}}_{p}^{(1)}(\textbf{r}) \cdot \tilde{\bm{\tau}}_{p}^{(1)}(\textbf{r}) \right. \\ \left.+  (\hat V(\omega)* [ \hat V(\omega)\chi(\omega,\lambda)] ) \left[ \tilde{\mathbf{v}}^{(1)} (\textbf{r}) \cdot \nabla\tilde{\bm{\tau}}_{p}^{(1)} (\textbf{r}, t) - (\tilde{\bm{\tau}}_{p}^{(1)} (\textbf{r}) \cdot \nabla \tilde{\mathbf{v}}^{(1)} (\textbf{r}) + \tilde{\mathbf{v}}^{(1)T}(\textbf{r}) \cdot \tilde{\bm{\tau}}_{p}^{(1)}(\textbf{r}) ) \right] \right).
   \label{eq:giesk_tau2}
\end{multline}
We define a rescaled Giesekus parameter denoted:
\begin{equation}
    \alpha^{*} = \alpha \frac{ \left< \lambda \chi(\omega, \lambda) \big( [ \hat V(\omega)\chi(\omega,\lambda) ] * [ \hat V(\omega) \chi(\omega,\lambda) ] \big) \right>_\lambda }{ \left< \lambda \chi(\omega, \lambda) \big( \hat V(\omega)* [ \hat V(\omega)\chi(\omega,\lambda) ] \big) \right>_\lambda },
\end{equation}
which, when applied to Equation \ref{eq:giesk_tau2}, yields the following definition of the second-order polymeric stress:
\begin{multline}
   \hat{\bm{\tau}}_{p}^{(2)} (\textbf{r},\omega,\lambda) =  \chi(\omega, \lambda)\left( 2 \hat{\mathbf{e}}^{(2)}(\textbf{r}) -   (\hat V(\omega)* [ \hat V(\omega)\chi(\omega,\lambda)] )  \left[\alpha^{*} \tilde{\bm{\tau}}_{p}^{(1)}(\textbf{r}) \cdot \tilde{\bm{\tau}}_{p}^{(1)}(\textbf{r}) \right. \right. \\ \left. \left.+  \tilde{\mathbf{v}}^{(1)} (\textbf{r}) \cdot \nabla\tilde{\bm{\tau}}_{p}^{(1)} (\textbf{r}, t) - (\tilde{\bm{\tau}}_{p}^{(1)} (\textbf{r}) \cdot \nabla \tilde{\mathbf{v}}^{(1)} (\textbf{r}) + \tilde{\mathbf{v}}^{(1)T}(\textbf{r}) \cdot \tilde{\bm{\tau}}_{p}^{(1)}(\textbf{r}) ) \right] \right).
   \label{eq:giesk_tau2_r}
\end{multline}
We can then rescale the second order velocity and pressure as done previously for the Johnson-Segalman model:
\begin{subequations}
 \begin{equation}
   \hat{\mathbf{v}}_{2} (\textbf{r}, \omega )= \frac{1}{\Bar{\lambda}\eta^{*}(\omega)}  \left< \lambda \chi(\omega, \lambda) \left( \hat V(\omega)* [\hat V(\omega,\lambda)  \chi(\omega,\lambda) ] \right) \right>_{\lambda} \tilde{\mathbf{v}}^{(2)} (\textbf{r}) ,
   \end{equation}
   \begin{equation}
     \hat{p}_{2} (\textbf{r})= \frac{1}{\Bar \lambda} \left< \lambda \chi(\omega, \lambda) \left( \hat V(\omega)* [\hat V(\omega,\lambda)  \chi(\omega,\lambda) ] \right) \right>_{\lambda} \tilde{p}_{2}(\textbf{r}).
     \end{equation}
\end{subequations}
On substitution of these scaling relations into the constitutive model, solving for the polymeric stress and then substituting that into the the momentum balance and continuity equation, we recover equations for the steady-state velocity and pressure fields in a Giesekus fluid.  Solving these equations as before, we find that the steady state velocity and pressure profiles are:
\begin{subequations}
\begin{multline}
    \tilde{v}_i^{(2)}(\textbf{r}) = (1-\beta) (1-\alpha) \left[  \frac{3}{8r^3}\left( 1 - \frac{3}{r} + \frac{3}{r^2} - \frac{1}{r^3}  \right) r_i  - \frac{9}{8r^4}\left( 1 - \frac{2}{r} + \frac{1}{r^2} \right)\delta_{i3} r_3   \right. \\ - \frac{9}{8r^5} \left(1 - \frac{4}{r} + \frac{5}{r^2} - \frac{2}{r^3} \right) r_i r_3 r_3  \left.  \right],
\end{multline}
\begin{equation}
    \tilde{p}^{(2)}(\textbf{r}) = \frac{1}{4r^3} \left( 9 - \frac{9}{r}-\frac{27}{r^2} + \frac{87}{2r^3} - \frac{27}{r^5} + \frac{9}{r^{7}} \right) + \alpha \frac{3 }{4r^3}\left( -1 + \frac{3}{2r} + \frac{9}{2r^2} - \frac{12}{r^3} + \frac{5}{r^5} -\frac{2}{r^{7}}  \right).
\end{equation}
\end{subequations}

The first and second order velocity fields provide the necessary elements to calculate the third-order contribution to the force via the reciprocal theorem. Following the same procedure as with the Johnson-Segalman fluid, we need to compute the quantity: $\left< \hat{\bm{\tau}}^{(3)}_p( \mathbf{r}, \omega, \lambda ) \right>_\lambda - 2 \left< \chi(\omega,\lambda)\right>_\lambda \hat{\mathbf{e}}^{(3)}(\mathbf{r}, \omega ) $, in order to evaluate the third order contribution to the force as in equation \ref{eq:thirdorderforce}.  An average over the constitutive model at third-order in the deformation amplitude yields: 
\begin{multline}
    \left<\hat{\bm{\tau}}_{p}^{(3)} (\textbf{r}, \omega, \lambda) \right>_\lambda - 2 \left< \chi(\omega, \lambda)\right>_\lambda \hat{\mathbf{e}}^{(3)}(\textbf{r}, \omega) =  \\ -  \left< \left(\frac{\lambda}{\Bar{\lambda}}\right) \chi(\omega, \lambda)\bigg[  \alpha^* \left( \hat{\bm{\tau}}_{p}^{(1)} (\textbf{r}, \omega, \lambda) * \hat{\bm{\tau}}_{p}^{(2)} (\textbf{r}, \omega, \lambda) + \hat{\bm{\tau}}_{p}^{(1)} (\textbf{r}, \omega, \lambda) * \hat{\bm{\tau}}_{p}^{(2)} (\textbf{r}, \omega, \lambda)  \right) \right.  \\ \left. + \hat{\mathbf{v}}^{(1)} (\textbf{r}, \omega) * \nabla \hat{\bm{\tau}}_{p}^{(2)}(\textbf{r}, \omega, \lambda) +  \hat{\mathbf{v}}^{(2)}(\textbf{r}, \omega)*\nabla \hat{\bm{\tau}}_{p}^{(1)}(\textbf{r}, \omega, \lambda)   -\left( \hat{\bm{\tau}}_{p}^{(2)}(\textbf{r}, \omega, \lambda) * \nabla \hat{\mathbf{v}}^{(1)} (\textbf{r}, \omega) \right. \right. \\ \left. \left.+ \hat{\bm{\tau}}_{p}^{(1)}(\textbf{r}, \omega, \lambda)*\hat{\mathbf{v}}^{(2)} (\textbf{r}, \omega) + \hat{\mathbf{v}}^{(1)T} (\textbf{r}, \omega)* \hat{\bm{\tau}}_{p}^{(2)}(\textbf{r}, \omega, \lambda) + \hat{\mathbf{v}}^{(2)T}(\textbf{r}, \omega) * \hat{\bm{\tau}}_{p}^{(1)} (\textbf{r}, \omega, \lambda)     \right)  \bigg] \right>_\lambda.
\end{multline}

Ultimately, substituting the known values for the first and second order fields and computing the volume integral in equation \ref{eq:thirdorderforce} gives the unsteady third-order force expressed as the combination of seven distinct terms: 

\begingroup
\allowdisplaybreaks
\begin{align}
    \mathbf{\hat F}^{(3)}( \omega ) &= \mathbf{C}_{1} \left< \lambda \chi(\omega, \lambda) \left(\left[\hat{V} (\omega)\chi (\omega, \lambda)\right] * \left[ \chi(\omega, \lambda) \left(\frac{ \left< \lambda \chi(\omega, \lambda)\left[\hat{V} (\omega) * \hat{V} (\omega) \chi (\omega, \lambda)\right]\right>_{\lambda} }{\eta(\omega)} \right)\right] \right) \right>_{\lambda} 
    \\ 
    &+\mathbf{C}_{2}  \left< \lambda \chi (\omega, \lambda) \Big( \left[ \hat{V} (\omega)\chi(\omega, \lambda) \right] *\left[\hat{V} (\omega)\chi(\omega, \lambda) * \hat{V} (\omega)\chi(\omega, \lambda) \right]\Big) \right>_{\lambda}   \nonumber
    \\
    &+\mathbf{C}_{3} \left< \lambda  \chi(\omega, \lambda)\left(  \left[ \hat{V} (\omega)\chi(\omega, \lambda) \right]* \left[\chi(\omega, \lambda) (\hat{V} (\omega)*\hat{V} (\omega)\chi(\omega, \lambda)) \right] \right) \right>_{\lambda}  \nonumber
    \\
    &+\mathbf{C}_{4} \left< \lambda  \chi(\omega, \lambda) \left( \left[ \frac{ \left< \chi(\omega, \lambda) (\hat{V} (\omega) *\left[\hat{V} (\omega) \chi (\omega, \lambda)\right]) \right>_{\lambda}  }{\eta^{*}(\omega)}\right] *[ \hat{V} (\omega)\chi(\omega, \lambda) ] \right)\right>_{\lambda}   \nonumber
    \\
    &+\mathbf{C}_{5} \left< \lambda  \chi(\omega, \lambda) \left( \hat{V} (\omega) *\left[ \chi(\omega, \lambda)(\hat{V} (\omega)\chi(\omega, \lambda)) \right] *\left[\hat{V} (\omega)\chi(\omega, \lambda)\right] \right) \right>_{\lambda}   \nonumber
    \\
    &+\mathbf{C}_{6} \left< \lambda  \chi(\omega, \lambda)\left ( \hat{V} (\omega) * \left[ \frac{\chi(\omega, \lambda) \left< \chi(\omega, \lambda)\left[\hat{V} (\omega) * \hat{V} (\omega) \chi (\omega, \lambda)\right] \right>_{\lambda}   }{\eta^{*}(\omega)} \right] \right) \right>_{\lambda}   \nonumber
    \\
    &+\mathbf{C}_{7} \left< \lambda  \chi(\omega, \lambda) \left( \hat{V} (\omega) * \left[ \chi(\omega, \lambda)[\hat{V} (\omega)*\hat{V} (\omega)\chi(\omega, \lambda)] \right] \right) \right>_{\lambda}, \nonumber
\end{align}

where
\begin{subequations}
\begin{align}
    \mathbf{C}_{1} &= -2\alpha (1-\beta) \int_{\Omega}  \mathbf{R}(\textbf{r})  \Big[ \tilde{\bm{\tau}}_{p}^{(1)} (\textbf{r}) \cdot \tilde{\mathbf{e}}^{(2)} (\textbf{r}) + \tilde{\mathbf{e}}^{(2)}(\textbf{r}) \cdot \tilde{\bm{\tau}}_{p}^{(1)}(\textbf{r}) \Big] d\textbf{r} 
    \\
    & = \frac{6\pi(1-\alpha)(1-\beta)^{2} \alpha}{175} \textbf{e}_{z}, \nonumber
    \\
    \mathbf{C}_{2} &= 2\alpha^{2} (1-\beta) \int_{\Omega} \mathbf{R}(\textbf{r}) : \Big[ \tilde{\bm{\tau}}_{p}^{(1)}  (\textbf{r}) \cdot \tilde{\bm{\tau}}_{p}^{(1)}  (\textbf{r}) \cdot \tilde{\bm{\tau}}_{p}^{(1)}  (\textbf{r}) \Big]   d\textbf{r} 
    \\
    & =  \frac{5652\pi\alpha^{2}(1-\beta)}{2275} \textbf{e}_{z}, \nonumber
    \\
    \mathbf{C}_{3} &=  -\alpha(1-\beta) \int_{\Omega}  \mathbf{R}(\textbf{r}) : \Big[ \tilde{\bm{\tau}}_{p}^{(1)}  (\textbf{r}) \cdot \mathbf{g}_{1} (\textbf{r}) + \mathbf{g}_{1}  (\textbf{r})\cdot \tilde{\bm{\tau}}_{p}^{(1)}  (\textbf{r}) \Big]  d\textbf{r}
    \\
    & = - \frac{5652\pi\alpha(1-\beta)}{2275} \textbf{e}_{z}, \nonumber
    \\
    \mathbf{C}_{4} &=(1-\beta) \int_{\Omega} \mathbf{R}(\textbf{r}) : \Big[ -\tilde{\mathbf{v}}^{(2)}  (\textbf{r}) \cdot \nabla \tilde{\bm{\tau}}_{p}^{(1)}  (\textbf{r}) + \tilde{\bm{\tau}}_{p}^{(1)}  (\textbf{r}) \cdot \nabla \tilde{\mathbf{v}}^{(2)}  (\textbf{r}) + \nabla \tilde{\mathbf{v}}^{(2)} (\textbf{r})^{T} \cdot \tilde{\bm{\tau}}_{p}^{(1)}  (\textbf{r}) \Big] d\textbf{r}
    \\
    & = - \frac{3\pi(1-\alpha)(1-\beta)^{2}}{175} \textbf{e}_{z}, \nonumber
    \\
    \mathbf{C}_{5} &= \alpha (1-\beta)\int_{\Omega} \mathbf{R}(\textbf{r}) : \Big[ -\tilde{\mathbf{v}}^{(1)} \cdot \nabla [ \tilde{\bm{\tau}}_{p}^{(1)} \cdot \tilde{\bm{\tau}}_{p}^{(1)} (\textbf{r}) ] + [\tilde{\bm{\tau}}_{p}^{(1)}  (\textbf{r}) \cdot \tilde{\bm{\tau}}_{p}^{(1)} (\textbf{r})] \cdot \nabla \tilde{\mathbf{v}}^{(1)} (\textbf{r}) \\
    & \qquad + \nabla \tilde{\mathbf{v}}^{(1)} (\textbf{r})^{T} \cdot [\tilde{\bm{\tau}}_{p}^{(1)}  (\textbf{r})\cdot \tilde{\bm{\tau}}_{p}^{(1)}(\textbf{r}) )]   \Big]  d\textbf{r}
    \\
    & = - \frac{2826\pi\alpha(1-\beta)}{2275} \textbf{e}_{z}, \nonumber
    \\
    \mathbf{C}_{6} &= 2 (1-\beta)\int_{\Omega} \mathbf{R}(\textbf{r}) : \Big[ -\tilde{\mathbf{v}}^{(1)} (\textbf{r}) \cdot \nabla \tilde{\mathbf{e}}^{(2)} (\textbf{r}) + \tilde{\mathbf{e}}^{(2)} (\textbf{r}) \cdot \nabla \tilde{\mathbf{v}}^{(1)} (\textbf{r}) + \nabla \tilde{\mathbf{v}}^{(1)} (\textbf{r})^{T} \cdot \tilde{\mathbf{e}}^{(2)}  (\textbf{r}) \Big]   d\textbf{r}
    \\
    & =  - \frac{3\pi(1-\alpha)(1-\beta)^{2}}{175} \textbf{e}_{z}, \nonumber
    \\ \nonumber \\ \nonumber \\ \nonumber
    \mathbf{C}_{7} &= (1-\beta) \int_{\Omega} \mathbf{R}(\textbf{r}) : \Big[ -\tilde{\mathbf{v}}^{(1)}  (\textbf{r}) \cdot \nabla \mathbf{g}^{(1)}  (\textbf{r}) + \mathbf{g}^{(1)} (\textbf{r}) \cdot \nabla \tilde{\mathbf{v}}^{(1)}  (\textbf{r}) + \nabla \tilde{\mathbf{v}}^{(1)} (\textbf{r})^{T} \cdot \mathbf{g}^{(1)} (\textbf{r})  \Big] d\textbf{r} \\
    & = - \frac{1548\pi(1-\beta)}{25025}  \textbf{e}_{z}. \nonumber
\end{align}
\end{subequations}

As with the Johnson-Segalman model, we find that the third-order force in a Giesekus fluid can also be expressed in the form of a nonlinear transfer function: 
\begin{equation}
    \mathbf{\hat F}^{(3)}( \omega ) = \mathbf{e}_z \frac{1}{(2 \pi)^2} \iiint_{-\infty}^{\infty} R^{*}_{3}(\omega_{1}, \omega_{2}, \omega_{3}) \delta(\omega-\omega_{1}-\omega_{2}-\omega_{3})\hat{V}(\omega_{1})\hat{V}(\omega_{2})\hat{V}(\omega_{3})d\omega_{1}d\omega_{2}d\omega_{3},
\end{equation}
where
\begin{align}
    R_3^*( \omega_1, \omega_2, \omega_3 ) &= 
    \\
    \frac{1}{\Bar \lambda^2} \Bigg[ &\frac{6\pi(1-\alpha)(1-\beta)^{2} \alpha}{175\eta^{*}(\omega_2 + \omega_3)} \Big< \lambda \chi(\omega_1 + \omega_2 + \omega_3 , \lambda)\chi(\omega_1, \lambda)\chi(\omega_2 + \omega_3, \lambda)  \Big>_{\lambda} \Big<  \lambda \chi(\omega_2 + \omega_3, \lambda) \chi(\omega_3, \lambda) \Big>_{\lambda}   \nonumber
    \\
    & + \frac{5652\pi\alpha^{2}(1-\beta)}{2275} \Big<\lambda^2 \chi(\omega_{1} + \omega_{2} + \omega_{3}, \lambda)\chi(\omega_{1}, \lambda)\chi(\omega_{2}, \lambda)\chi(\omega_{3}, \lambda)\Big>_{\lambda} \nonumber
    \\
    & - \frac{5652\pi\alpha(1-\beta)}{2275} \Big< \lambda^2 \chi(\omega_{1}+\omega_{2} + \omega_{3}, \lambda)\chi(\omega_{1}, \lambda)\chi(\omega_{3}, \lambda)\chi(\omega_{2} + \omega_{3}, \lambda) \Big>_{\lambda} \nonumber
    \\ 
    & - \frac{3\pi(1-\alpha)(1-\beta)^{2}}{175\eta^{*}(\omega_2 + \omega_3)} \Big< \lambda \chi(\omega_1 + \omega_2 + \omega_3 , \lambda)\chi(\omega_1, \lambda)  \Big>_{\lambda} \Big<  \lambda \chi(\omega_2 + \omega_3, \lambda) \chi(\omega_3, \lambda) \Big>_{\lambda}  \nonumber
    \\
    & - \frac{2826\pi\alpha(1-\beta)}{2275} \Big< \lambda^2 \chi(\omega_{1} + \omega_{2} + \omega_{3}, \lambda)\chi(\omega_{2} + \omega_{3}, \lambda) \chi(\omega_{2}, \lambda)\chi(\omega_{3}, \lambda) \Big>_{\lambda} \nonumber
    \\
    & - \frac{3\pi(1-\alpha)(1-\beta)^{2}}{175\eta^{*}(\omega_2 + \omega_3)} \Big< \lambda \chi(\omega_{1} + \omega_{2} + \omega_{3}, \lambda)\chi(\omega_2 + \omega_3, \lambda)  \Big>_{\lambda}  \Big< \lambda \chi(\omega_2 + \omega_3, \lambda)\chi(\omega_3, \lambda)\Big>_{\lambda} \nonumber
    \\ 
    &  - \frac{1548\pi(1-\beta)}{25025} \Big< \lambda^2  \chi(\omega_{1} + \omega_{2} + \omega_{3}, \lambda) \chi(\omega_{2} + \omega_{3}, \lambda) \chi(\omega_{3}, \lambda) \Big>_{\lambda} \Bigg].
\end{align}
From this same calculation we can deduce that the leading order nonlinearity in the the dimensionless steady-state force is: 
\begin{equation}
    \frac{\tilde{F}^{(3)} }{\eta_0 a V} =  \Wi^2 (1-\beta )\pi \left[ \alpha^2 \left( \frac{5652}{2275} - \frac{6(1-\beta)}{175} \right) - \alpha \left( \frac{8439}{2275} + \frac{9(1-\beta)}{175} \right) - \frac{1977}{25025} - \frac{3(1-\beta)}{175} \right].
\end{equation}
\endgroup
\end{document}